\documentclass{aa}  

\usepackage{graphicx}	
\usepackage{amsmath}	
\usepackage{multicol}        
\usepackage{multirow}        
\usepackage{bm}		
\usepackage{pdflscape}	
\usepackage{bigints} 
\usepackage[euler]{textgreek}
\usepackage{txfonts}
\usepackage{newtxtext,newtxmath}
\usepackage{multicol}
\usepackage{subcaption}
\usepackage[T1]{fontenc}
\usepackage[colorlinks = true,
            linkcolor = blue,
            urlcolor  = blue,
            citecolor = blue,
            anchorcolor = blue]{hyperref}

\newcommand{\ergs}{\,erg~s$^{-1}$}	
\newcommand{\angstrom}{\textup{\AA}}
\newcommand{\Alike}{1987A-like} 
\newcommand{\Ni}{$^{56}$Ni~} 
\newcommand{\Co}{$^{56}$Co~} 
\newcommand{\Fe}{$^{56}$Fe~} 

\defcitealias{pumo23}{Paper I}
\defcitealias{PC2025}{Paper II}

\begin{document} 
\nolinenumbers

   \title{Physical properties of long-rising type II supernovae}
   \subtitle{Bayesian analytic modeling and spectrophotometric correlations}

   \author{S.~P.~Cosentino\inst{1,2} 
   \thanks{Contact~\email{\href{mailto:stefano.cosentino@dfa.unict.it}{stefano.cosentino@dfa.unict.it}}}
          \and
          C.~Inserra\inst{3}
          \and
          M.~L.~Pumo\inst{1,2,4}
          }

   \institute{Dipartimento di Fisica e Astronomia ``Ettore Majorana'', Università degli Studi di Catania, Via Santa Sofia 64, 95123 Catania, Italy
         \and
         	INAF - Osservatorio Astrofisico di Catania, Via Santa Sofia 78, 95123 Catania, Italy
         \and
         	Cardiff Hub for Astrophysics Research and Technology, School of Physics \& Astronomy, Cardiff University, Queens Buildings, The Parade, Cardiff, CF24 3AA, UK  
        	 \and
        	 	INFN - Laboratori Nazionali del Sud, Via S. Sofia 62, 95125 Catania, Italy
             }

   \date{Received 2026 January 21; accepted 2026 February 7; published ...;}

 \abstract
   {The supernova (SN) 1987A, with its long-rising ($\gtrsim$40~days) light curve, defines a rare subclass of type II SNe known as \Alike~events. Representing only $\sim$1–3\% of all core-collapse SNe and often found in low-metallicity environments, their large diversity suggests a wide range of progenitor and explosion properties.}
   {Our aim with this study is to improve the understanding of \Alike~SNe by characterizing their explosion parameters, including kinetic energy, ejected mass, progenitor radius at the explosion, and synthesized \Ni mass. Additionally, we seek to identify systematic trends in both the physical properties and the observed spectrophotometric features of these peculiar events.}
   {A new Bayesian parameter estimation method based on our $^{56}$Ni-dependent analytical model for hydrogen-rich SNe is applied to derive explosion parameters from the light curves and expansion velocities of one of the largest and most comprehensive \Alike~SN samples to date. These data are measured through a consistent analysis of observations available in the literature.}
   {The analysis reveals a heterogeneous population that nevertheless clusters into two main groups: (i) lower-energy explosions with modest \Ni\ yields ($\sim$0.07~M$_\odot$), similar to SN~1987A, and (ii) more energetic events (up to $\sim$5~foe) with larger nickel production and, in some cases, unusually extended progenitors. We confirm a robust correlation between \Ni\ mass, peak luminosity, and explosion energy, as well as between ejecta mass and the recombination timescale. 
An anticorrelation between Ba~II line strength and photospheric velocity indicates that stronger Ba~II absorptions in \Alike~SNe arise from more compact, slowly expanding ejecta.  
}
   {Our findings indicate that 87A-like SNe populate a continuous distribution of explosion energies and progenitor radii. The study underscores the need to extend analytical frameworks to include additional power sources that will enable scalable and accurate modeling of the growing number of peculiar transients that will be discovered by current and upcoming surveys (e.g., ZTF and LSST).}
   \keywords{supernovae: general --
             supernovae: individual: \Alike~SNe -- Methods: analytical -- Methods: statistical             }

   \maketitle
%

\nolinenumbers
\section{Introduction}

The explosion of Supernova (SN) 1987A, located only $\approx50\,$kpc from Earth in the direction of the Large Magellanic Cloud (LMC), was an extraordinary event that significantly advanced our understanding of core-collapse (CC) SN events and progenitor stars \citep[e.g.,][]{burrows90,woosley02}.
The unusual shape of its long-rising ($\approx80\,$days) light curve (LC) is consistent with the explosion of a compact blue supergiant (BSG) progenitor \citep{arnett89,kleiser11,orlando15}.
The slow brightness rise and the LC secondary peak, characteristic features of SN 1987A, make it the prototype of a peculiar subclass of H-rich SN events, referred to as long-rising (typically $\gtrsim 40\,$days) type II SNe (SNe II), or simply \Alike~events \citep[e.g.,][hereafter \citetalias{pumo23}]{turatto07,pasto12,taddia16,pumo23}.

Notably, \Alike~SNe are intrinsically rare events, with observational estimates suggesting they account for only $\approx1\%-3\%$ of all CC-SNe \citep[][who recently reported a volumetric rate of about $1.37\times 10^{-6}\,$Mpc$^{-3}\,$yr$^{-1}$]{smartt09,pasto12,taddia16,sit_2023}.
Their rarity reflects the limited evolutionary pathways through which a massive star \citep[with a zero age main sequence mass of $M_{\rm ZAMS} \gtrsim 8-10\,$M$_\odot$; e.g.,][]{Pumo_2009} can become a BSG prior to its explosion as a CC-SN. 

Although the sample of long-rising SNe II is statistically limited, their bolometric LCs exhibit significant variation in rise time ($t_{\rm M}$) and the maximum luminosity ($L_{\rm M}$) of the secondary peak \citepalias{pumo23}. This peak, taking place during the hydrogen recombination phase, is commonly interpreted as the result of the interplay between the cooling driven by the SN expansion rate and heating from the radioactive decay of isotopes, primarily $^{56}$Ni, synthesized during the explosion \citep[see, e.g.,][hereafter \citetalias{PC2025}]{popov93,UC11,PZ11,PC2025}.

Modeling the LC peak and analysing its main features allow us to infer key physical properties of the ejected material (hereafter the ejecta), including the explosion energy ($E$), progenitor radius ($R_0$), total ejecta mass ($M_{\rm ej}$), and the mass of radioactive \Ni\ ($M_{\rm Ni}$). These quantities, in turn, provide crucial insights into the progenitor’s nature and the explosion mechanism \citep[e.g.,][]{arnett96,KW09,PZ13,Fang_2025}.

Previous studies about long-rising SNe have identified several clones of SN 1987A \citep[e.g., SN 2009E, SN 2009mw, and SN 2018hna; see also][]{pasto12,takats16,singh19}, which are characterized by peak luminosities below $2 \times 10^{42}$\ergs and rise times of approximately 70–90 days. 
These events are generally interpreted as CC-explosions with energies in the range of $\sim0.5-5$~foe (1~foe~$\equiv 10^{51}\,$erg), originating from relatively compact ($R_0\sim 30-100\,\text{R}_\odot$) and massive ($M_{\rm ej}\sim 15-20\,\text{M}_\odot$) progenitors \citepalias[see][and references therein]{pumo23}. The \Ni masses powering their tail luminosities typically range from $M_{\rm Ni} \sim 0.05-0.1\,\text{M}_\odot$, higher than those of standard type II-plateau (IIP) SNe \citep[see, e.g.,][]{pasto12,Muller_2017,2021MNRAS.505.1742R}.

There is also evidence that similar long-rising LCs can originate from non-BSG progenitors ($R_0\gtrsim 300\,\text{R}_\odot$) capable of synthesizing substantial amounts of \Ni ($M_{\rm Ni} \gtrsim 0.1\,\text{M}_\odot$). Examples include SN 2004ek and SN 2004em, which have been described by \citet{taddia16} as intermediate events bridging the gap between the clones of SN 1987A and typical SNe IIP. The discovery of brighter \Alike~objects, including PTF12kso and PTF12gcx \citep{taddia16}, seems to indicate the existence of Ni-rich ($\gtrsim 0.1-0.2\,$M$_\odot$) and high-energy ($\gtrsim 5-10\,$foe) events forming a luminous tail of \Alike~events, which could be characterized by a non-conventional explosion \citepalias{pumo23}.
Among these events, OGLE-2014-SN-073 (hereafter OGLE-14) stands out as the brightest long-rising SN \citep{terreran17}. With a redshift of $z \simeq 0.12$, it is also the most distant of \Alike~SNe, second only to the gravitationally lensed SN Refsdal \citep[$z \simeq 1.49$, see][]{kelly16}.
OGLE-14 is characterized by a peak bolometric luminosity of approximately $10^{43}\,$\ergs and a rise time of about 100 days. Its extraordinary LC challenges explanation within the framework of the conventional neutrino-driven CC paradigm and has been linked to pair-instability scenarios \citep{terreran17}.
An alternative explanation for this exceptional event involves the presence of a prompt injection of energy from a compact internal engine (e.g., magnetar) following the explosion. Such a mechanism has also been proposed for more recent and nearby analogous events, including SN 2020faa \citep{Yang2024,Salmaso2023} and SN 2021aatd \citep{Szalai2024}. 
The influence of additional energy sources, such as \Ni-\Co decay, magnetar spin-down, or accretion onto a compact object, can extend the long-rising phase and alter the late-time LC \citep[e.g.,][and further comments in \citetalias{PC2025}]{Kasen_2010,Inserra_13,Dexter_2013,Khatami_2019,Matsumoto_2025}. These additional energy sources can give rise to long-lived events such as SN DES16C3cje \citep{gutierrez20}, which exhibited a rise time of approximately 140 days, followed by a decline consistent with fall-back accretion power after 300 days.
 
Despite their significance, fundamental questions persist regarding \Alike~SNe, primarily due to the absence of sufficiently accurate and efficient methods for characterizing their physical properties \citepalias{pumo23,PC2025}. 
Moreover, the available information about the physical properties of the studied sample remains too limited and uncertain to draw definitive conclusions \citepalias{pumo23}.
Form 2018 June to 2021 December, the sample of \Alike~SNe has nearly doubled thanks to the systematic study by \citet{sit_2023} as part of the Census of the Local Universe \citep[CLU; ][]{Cook_2019} experiment conducted with the Zwicky Transient Facility \citep[ZTF; ][]{Bellm_2019,Graham_2019} on galaxies at less than 200 Mpc ($z\lesssim 0.05$). This study presented the LCs and spectra of 13 additional long-rising SNe. However, a complete physical characterization of their progenitors and explosion mechanisms is still missing.

With the aim of improving our understanding of long-rising type II SN explosions, we present a new Bayesian parameter estimation method based on the LC analytical model developed in \citetalias{PC2025}. This approach enables us to infer the physical properties of SN progenitors at the explosion, including $E$, $M_{\rm ej}$, $R_0$, and $M_{\rm Ni}$.
In this paper, after presenting this method and validating its accuracy (Section \ref{Sec:SuperBAM}), we apply it to one of the largest and most comprehensive samples of well-observed \Alike~objects (Section \ref{Sec:Single}). Finally, we conduct a global study inside \Alike~class of their explosion parameters in relation to the measured spectrophotometric features (Section \ref{Sec:Modeling}). The key results are outlined in Section \ref{Sec:Conclusion}, and we highlight their significance in interpreting the nature of these peculiar events.

\section{``Fast'' SN modeling procedures}\label{Sec:SuperBAM}
The physical characterization of long-rising SNe II can be approached through various modeling methods, each balancing computational complexity and accuracy. 
The most detailed analyses employ full-numerical hydrodynamic simulations, which incorporate a wide range of physical phenomena, including relativistic hydrodynamics and nuclear heating processes \citep[e.g.,][]{PZ11,Bersten_2011}. 
Although the numerical hydrodynamic modeling (HM) is highly accurate, it is computationally expensive and less practical for statistical studies of a heterogeneous sample of peculiar events, such as \Alike~SNe. A more rapid but less precise alternative is the use of scaling relations, which connect the SN physical parameters to observables through simple  proportionality relations. However, the accuracy of this method diminishes when the observed SNe differ significantly from the reference event \citepalias[e.g.,][and references therein]{pumo23}.
Intermediate approaches, such as analytical or semi-analytical models, stand out as the most practical solution, as they significantly reduce the complexity of full numerical simulations, minimizing computational time (thus being ``faster'') while still capturing the main physical processes involved in the SN post-explosive evolution \citepalias[see, e.g.,][and references therein]{PC2025}.

Among the most well-known and widely used analytic models for H-rich SNe, including \Alike~events, are those developed by \citet{Arnett80,arnett96} and \citet{arnett89}, which describe the behavior of the bolometric LC dominated by the diffusion of photons through the expanding SN ejecta. These models are particularly effective during the early post-explosive phases, when hydrogen recombination does not yet significantly influence the cooling emission. For the later stages, when hydrogen recombination becomes important, the standard analytical framework is provided by \citet{popov93}. Furthermore, this model excludes the presence of additional heating effects, such as those from radioactive decay (e.g., \Ni-\Co elements), ejecta interactions with the circumstellar medium (CSM), or energy injection from a central compact object. Subsequent analytical approaches have included these additional energy sources retaining the diffusive approximation of Arnett's models \citep[see, e.g.,][]{Chatzopoulos_2012,Inserra_13, Nicholl_2017, Villar_2017}. More recently, similar mechanisms have been analytically incorporated into models for H-rich SNe, allowing for the evaluation of heating source effects on the SN LC even during phases where hydrogen recombination dominates \citep[e.g.,][and \citetalias{PC2025}]{Dexter_2013,Matsumoto_2025}. 
Specifically, \citetalias{PC2025}'s model is the first analytical description to account for different \Ni distributions and their effect on radioactive heating at the recombination front. These features make it a valuable description for developing new modeling procedures aimed at accurately and efficiently determining the explosion and progenitor parameters of \Alike~events.
In the following, we present our implementation of this framework (Sect. \ref{SubSec:SuperBAM}) and assess its performance by comparing results for a set of well-studied \Alike~SNe against HM benchmarks (Sect. \ref{SubSec:test}).

\subsection{Supernova bayesian analytic modeling -- \textsc{SuperBAM}}\label{SubSec:SuperBAM}
``Supernova Bayesian Analytic Modeling'' (\textsc{SuperBAM}) is a novel modeling approach based on Bayesian statistics and the analytical model presented in \citetalias{PC2025}. It characterizes key physical properties of H-rich SNe, including $M_{\rm Ni}$, $E$, $M_{\rm ej}$, and $R_0$, through the analysis of the bolometric LC shape during the recombination phase \citep[e.g.,][]{PZ11}.
To achieve this, \textsc{SuperBAM} follows a three-phase process:
\begin{enumerate}
\item Prior information – \textsc{SuperBAM} extracts key spectrophotometric features from SN observations to construct prior probability distributions using appropriate scaling equations.
\item Likelihood definition – A likelihood function for the modeling parameters is defined by comparing analytical model predictions with observed data.
\item Posterior exploration – Priors and likelihoods are combined to obtain the posterior distribution, which is then analysed to determine the best-fit SN parameters.
\end{enumerate}
These steps are detailed in the following subsections.

\subsubsection{Prior information}\label{SubSubSec:Prior}
\textsc{SuperBAM} employs physically motivated priors based on empirical scaling relations that link observable quantities, such as LC morphology and photospheric velocity, to the SN explosion parameters. 
This approach naturally narrows the parameter space without enforcing arbitrary boundaries, thus improving convergence efficiency and reducing the risk of bias near the edges of the physically plausible domain \citep[see, e.g.,][]{Silva_2024}.
In contrast to a blind, uniform search over the full multidimensional parameter space ($E$, $M_{\rm ej}$, $R_0$, $M_{\rm Ni}$), this method enables a faster and more robust determination of the most probable physical configuration for each SN.

\begin{figure}[t]
\includegraphics[width=\columnwidth]{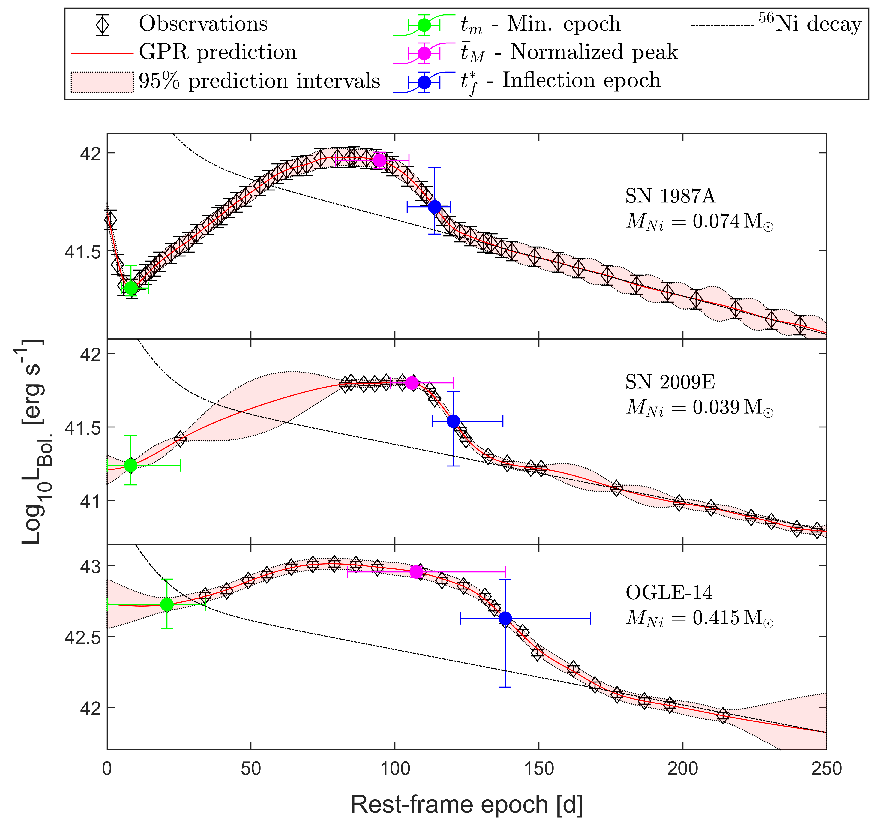} 
\caption{Application of GPR to the bolometric LCs of SN 1987A, SN 2009E, and OGLE-14 taken from the literature (see ref. in Sect. \ref{SubSec:test})
 to identify the main LC features. The adopted GPR employs a constant mean function combined with a Matérn 3/2 kernel. 
For each SN LC, the key epochs $t_{\rm m}$, $\bar{t}_{\rm M}$, and $t_{\rm f}^*$ are highlighted. The dashed line shows the luminosity contribution $L_{\rm Ni}$ due to \Ni-\Co radioactive decay with $M_{\rm Ni}$ value reported within the panel of each SN.
\label{Fig:GP}}
\end{figure}

To automatically identify the LC characteristics necessary for defining prior probability, \textsc{SuperBAM} employs a gaussian process regression\footnote{The GPR is a nonparametric supervised learning method used to solve regression problems. The GPR assumes that the N observed variables $L=(L_1,...,L_{\rm N})^T$ are randomly drawn from a multivariate Gaussian distribution $f(m,K)$, where $m=(\mu_1,...\mu_{\rm N})^T$ is the mean vector for each variable and $K$ is their covariance matrix (NxN), also called kernel \citep{roberts13}. A GPR is hence specified by its mean function and kernel \citep[see, e.g.,][]{Inserra_2018}.} (GPR), which permits interpolation of the LC, facilitating both its analysis and the computation of its derivative \citepalias[see, e.g., Fig. 13 in ][]{PC2025}.
To select the best combination of mean function and kernel\footnote{For a complete list of explored kernels, see also \href{https://it.mathworks.com/help/stats/gaussian-process-regression-models.html}{GPR MATLAB Documentation}.} for reproducing light and velocity curves, we tested several configurations using SN 1987A data. In these tests, the curves were artificially down-sampled by randomly removing individual points or entire observing periods of up to 30 days. By comparing the reconstructed values from the GPR against the original data via the mean square deviation, we identified the Matérn 3/2 kernel with a constant mean function as the most effective regression method for this type of data. The Matérn 3/2 kernel models functions with continuous derivatives up to the second order \citep[e.g.,][]{roberts13}, enabling the reconstruction of smooth profiles even when significant variations in the derivative or changes in concavity are present, as commonly observed in the \Alike~LCs (see Fig. \ref{Fig:GP}). Since the LC behavior of SN 1987A is representative of the class, the same GPR setup is adopted for all other \Alike~events.

Once the entire LC is reconstructed through GPR, three characteristic epochs ($t_{\rm m}$, $\bar{t}_{\rm M}$, and $t_{\rm f}^*$) are identified to estimate prior constrains.
The first one is $t_{\rm m}$, defined as the epoch of minimum bolometric luminosity after the SN breakout but preceding the maximum of the secondary peak, and it is generally associated with the beginning of the hydrogen recombination  \citepalias[i.e., $t_{\rm i}$; see also][]{pumo23}.
The other two epochs, $\bar{t}_{\rm M}$ and $t_{\rm f}^*$, were already introduced in \citetalias{PC2025} and are defined in terms of the bolometric magnitude (i.e., $M_{\rm bol} \equiv -2.5\log_{10} L_{\rm bol} + 88.7$). Specifically, $\bar{t}_{\rm M}$ corresponds to the normalized peak's maximum where the derivative of $M_{\rm bol}$ equals to $-2.5\log_{10} e /(111\,{\rm day})$, i.e., the slope of the \Co decay tail \citepalias[$\tau_{\rm ^{56}Co}\simeq 111\,{\rm day}$; cf. Eq. 22 in][]{PC2025}, while $t_{\rm f}^*$ is the epoch of maximum derivative, corresponding to the inflection point of $M_{\rm bol}$ that marks the end of the recombination phase.
The uncertainties on these epochs depend both on the GPR reconstruction, strongly linked to the number of observations available in the corresponding LC phase, and on the LC morphology (see Sect. \ref{SubSec:test} for further details). Moreover, the $t_{\rm f}^*$ positive error is constrained by the first epoch after the inflection where $\dot{M}_{\rm bol}\simeq-2.5\log_{10} e /(111\,{\rm d})$, which defines the beginning of the radioactive tail.

Considering all observations taken after the upper limit of $t_{\rm f}^*$, the \Ni mass can be directly estimated by fitting these LC points with the \Ni–\Co radioactive luminosity, $L_{\rm Ni}(t) = M_{\rm Ni}\times \epsilon(t)$, where $\epsilon(t)$ is the specific energy release rate (erg g$^{-1}$ s$^{-1}$) from the \Ni $\rightarrow$ \Co $\rightarrow$ \Fe decay chain \citepalias[see Eq.~22 in][]{PC2025}.
In \Alike~SNe, the trapping of $\gamma$-rays from \Ni –\Co decay remains efficient even beyond $300-400$ days \citepalias[see][and references therein]{PC2025}. Since $t_{\rm f}^* \gg \tau_{\rm ^{56}Ni} \simeq 8.8$d, the \Ni–\Co radioactive luminosity for $t>t_{\rm f}^*$ can be expressed as:
\begin{equation}\label{Eq:Ni_tail}
L_{\rm Ni}(t)\simeq M_{\rm Ni}\times \frac{\epsilon_{\rm ^{56}Co}\,\tau_{\rm ^{56}Ni}}{\tau_{\rm ^{56}Co}-\tau_{\rm ^{56}Ni}}\times \exp{\left(-t/\tau_{\rm ^{56}Co}\right)}.
\end{equation}
Thus, $M_{\rm Ni}$ is estimated by performing a linear fit to the logarithm of the bolometric luminosity during the radioactive tail ($t>t_{\rm f}^*$):
\begin{equation}\label{Eq:Ni_fit}
	\log_{10}L_{\rm Bol.}[{\rm erg/s}]= 43.18 +\log_{10}M_{\rm Ni}[{\rm M}_\odot]-\frac{t[{\rm d}]}{255.6}.
\end{equation}
This provides a direct measurement of $M_{\rm Ni}$, one of the four key physical parameters of these SNe, effectively reducing the dimensionality of the inference problem from four free parameters to three.

To constrain the remaining parameters—$E$, $M_{\rm ej}$, and $R_0$—we generally adopted the following system of scaling relations\footnote{The first two scaling equations in Eq. \ref{Eq:Scaling} are the same as in Eq. (4) of \citetalias{pumo23}, while the third replaces Eq. (15) of the same work. Instead of relying on $L_{\rm min}$, $t_{\rm M}$, and $v_{\rm M}$, it directly relates the progenitor radius to $t_{\rm m}$, based on the proportionality derived in their Eq. (9). This assumes that $t_i$ (the onset of recombination) coincides with the pre-peak luminosity minimum. Although \citetalias{PC2025} has shown that this is not always strictly valid (particularly in cases where a thin shell or CSM interaction modifies the early luminosity) this relation is nonetheless simple and robust enough to provide useful prior information on $R_0$, while allowing sufficient flexibility for posterior inference.}:
 \begin{equation}\label{Eq:Scaling}
 \begin{cases}
 E\,&=(1.3\,\text{foe})\times \left(\frac{\bar{t}_{\rm M}}{94\,\text{d}}\right)^2\times\left(\frac{v_{\rm M}}{\rm 2145\,km/s}\right)^3\\
 M_{\rm ej}\,&=(16\,\text{M}_\odot)\times \left(\frac{\bar{t}_{\rm M}}{94\,\text{d}}\right)^2\times\left(\frac{v_{\rm M}}{\rm 2145\,km/s}\right)\\
  R_0\,&=(3\times 10^{12}\,\text{cm})\times \left(\frac{t_{\rm m}}{8.7\,\text{d}}\right)^2
 \end{cases}.
 \end{equation}
 This system relies on two characteristic times ($t_{\rm m}$ and $\bar{t}_{\rm M}$) linked to the photometric evolution, and on the velocity $v_{\rm M}$, which can be measured from at least one spectrum around peak luminosity (close to $\bar{t}_{\rm M}$) or from multiple spectra obtained near that epoch. Following \citetalias{pumo23}, $v_{\rm M}$ can be derived from the blueshift of the Fe II 5619\angstrom~P-Cygni absorption, which remains approximately constant around the LC peak. However, the scaling relations of system~(\ref{Eq:Scaling}) can only be applied when at least one spectroscopic measurement of the velocity around maximum light is available. Since this is not always the case, in order to characterize SNe lacking spectroscopic coverage, \textsc{SuperBAM} also allows the use of purely photometric scaling relations to define the priors, under the assumption that $v_{\rm M} \propto L_{\rm M}^{1/2}/\bar{t}_{\rm M}$ \citepalias[similarly as seen in][]{pumo23}, where $L_{\rm M}=L_{\rm Bol.}(\bar{t}_{\rm M})$ is the bolometric luminosity at $\bar{t}_{\rm M}$.
This assumption is specifically valid for type~II SNe whose emission during the recombination phase can be approximated by that of a blackbody with a nearly constant temperature equal to the hydrogen recombination one \citepalias[see][and references]{pumo23,PC2025}. Under this hypothesis, the first two equations of system~(\ref{Eq:Scaling}) can be rewritten as:
 \begin{equation}\label{Eq:Scaling_new}
 \begin{cases}
 E\,&=(1.3\,\text{foe})\times \left(\frac{\bar{t}_{\rm M}}{94\,\text{d}}\right)^{-1}\times\left(\frac{L_{\rm M}}{8.6\times 10^{42}\text{\ergs}}\right)^{3/2}\\
 M_{\rm ej}\,&=(16\,\text{M}_\odot)\times \left(\frac{\bar{t}_{\rm M}}{94\,\text{d}}\right)\times\left(\frac{L_{\rm M}}{8.6\times 10^{42}\text{\ergs}}\right)^{1/2}
 \end{cases}.
 \end{equation}
  These alternative relations, although having the advantage of not relying on spectroscopic data, are typically affected by larger uncertainties due to the relative errors on luminosity (generally higher than those on velocity), as well as by the assumption of a common recombination temperature among different SNe. Nevertheless, they represent a valid and practical option for the construction of priors when spectroscopic information is unavailable.

In both cases, to incorporate this prior information on the physical parameters into the Bayesian framework, we define a probability distribution function (PDF) for each parameter. For $M_{\rm Ni}$, we adopt a Gaussian PDF with mean $\langle M_{\rm Ni}\rangle$ equal to the value obtained from the best-fit of the radioactive tail, and a standard deviation $\delta M_{\rm Ni}$ derived from the weighted fit uncertainties. For the remaining parameters ($E$, $M_{\rm ej}$, and $R_0$), the prior PDFs are obtained by combining skew-normal distributions \citep[e.g.,][]{ASHOUR2010341}, modified to ensure support on positive values. These distributions are constructed from the measured features ($t_{\rm m}, \bar{t}_{\rm M}, v_{\rm M}, L_{\rm M}$) and their uncertainties, and subsequently mapped into the physical parameter space through the scaling relations of Eq.~\ref{Eq:Scaling} (or \ref{Eq:Scaling_new}). The full procedure is presented in Appendix~\ref{App:PDF}.

\subsubsection{Likelihood definition}\label{SubSubSec:Likelihood}
\textsc{SuperBAM}'s likelihood function is designed to compare the observed bolometric LC data ($t^{\rm Obs.},\,L_{\rm Bol.}^{\rm Obs.},\,\pm \Delta L^{\rm Obs.}_{\rm Bol.}$) with the luminosity predicted by the analytical model of \citetalias{PC2025} ($L_{\rm SN}$). In particular, this likelihood function considers only LC observations taken later than 30 days after explosion (i.e., $t^{\rm Obs.}>30\,$d), so that the inference of the main physical parameters relies exclusively on the morphology of the secondary peak, without being significantly affected by early-time luminosity contributions \citepalias[e.g., optical thin shell or CSM interaction, see also][]{PC2025}.

To improve the efficiency of the model–data comparison, \textsc{SuperBAM}'s likelihood works with dimensionless functions and variables. For this reason, the observed luminosity is normalized to the radioactive luminosity from $^{56}$Ni decay expressed by Eq.~(\ref{Eq:Ni_tail}). For each LC point, the normalized luminosity is defined as:
\begin{equation}
\bar{L}^{\rm Obs.} = \frac{L_{\rm Bol.}^{\rm Obs.}}{L_{\rm Ni}(t^{\rm Obs.})},
\end{equation}
where $L_{\rm Ni}(t)$ is computed assuming an ejected nickel mass equals to $\langle M_{\rm Ni}\rangle$, as estimated from the prior constraints (Sect.~\ref{SubSubSec:Prior}).

On the modeling side, this normalization has the key advantage of removing the explicit dependence on the $^{56}$Ni mass, thereby reducing the number of free parameters in the normalized luminosity predicted by the model. The latter, $\bar{L}=L_{\rm SN}/L_{\rm Ni}$, is defined as the ratio between Eq.~(58) of \citetalias{PC2025} and Eq.~\ref{Eq:Ni_tail}. When expressed in terms of the dimensionless time coordinate $y=t/\tau_{^{56}{\rm Co}}$, $\bar{L}$ takes the following form:
\begin{equation}\label{Eq:L_mod}
\bar{L}(y)=\frac{[(y-y_i)\times H(y,y_i)+y_i]^2\times e^{y+k_2}}{\lambda\,k_1^2\,(e^{k_2}-1)}+\frac{e^{k_2\,(1-z^3)}-1}{e^{k_2}-1},
\end{equation}
where ($y_i,\,\lambda,\,k_1,\,k_2$) are the set of dimensionless model parameters and $z(\equiv x_i)$ is the comoving coordinate of the wave-front of cooling and recombination (WCR). Here, $H(y,y_i)$ denotes the Heaviside function, equal to one if $y>y_i$ and zero otherwise.
Unlike the variable change adopted in \citetalias{PC2025} for Eq.~(58), the time coordinate used here ($y$) is independent of the SN timescale $t_a$, leading to a direct dependency of $y_i=t_i/\tau_{^{56}{\rm Co}}$ from progenitor radius.
So the dimensionless parameters can be related to the physical SN quantities through the following scaling relations:
\begin{equation}\label{Eq:SuperBAM_param}
\begin{cases}
y_{i}\,&= 8.15\times 10^{-2}\times \left(\frac{R_0}{3\times 10^{12}\, {\rm cm}}\right)^{1/2}\\[6pt]
\lambda\,&= 5.02\times\frac{M_{\rm Ni}}{7.4\,{\rm M_\odot}}\times \left(\frac{M_{\rm ej}}{16\,{\rm M_\odot}}\right)^{-1/2}\times \left(\frac{E}{1.3\,{\rm foe}}\right)^{-1/2}\\[6pt]
k_1\,&= 1.35\times \left(\frac{E}{1.3\,{\rm foe}}\right)^{-1/4}\times \left(\frac{M_{\rm ej}}{16\,{\rm M_\odot}}\right)^{3/4}
\end{cases}.
\end{equation}

The remaining parameter $k_2$ depends only on the mixing of $^{56}$Ni. In this work, we adopt the fixed value $k_2=32.87$ for all \Alike~SNe, corresponding to a configuration in which 95\% of the nickel mass is confined within 45\% of the ejecta radius, assuming an exponential distribution (see \citetalias{PC2025}). The effects of varying this parameter and a comparison 
with other models are discussed in Appendix~\ref{App:mixing}.
With this formulation, the WCR evolution is governed by the following differential equation [cf. Eq.~(54) in \citetalias{PC2025}]:
\begin{equation}\label{Eq:WCR_evolution}
\frac{dz^4}{dy}=-\frac{2}{y}\left[z^4+\frac{y^2}{k_1^2}z^2-\lambda\times e^{-y}\times\left(1-e^{-k_2z^3}\right)\right],
\end{equation}
with boundary condition $z(y_i)=1$.  

Hence, once the boundary conditions are specified, the model luminosity $\bar{L}_{\rm Mod.}=\bar{L}(y;y_i,\lambda,k_1)$ depends on the three parameters ($y_i,\,\lambda,\,k_1$). These, together with the variance $\sigma$ that accounts for intrinsic model uncertainties, define the parameter set $\theta$ of the likelihood function:
\begin{align}\label{Eq:likelihood}
P(\bar{L}|\theta)=&\frac{1}{\sqrt{2\pi}}\times \prod_{\rm Obs.}\frac{1}{\sqrt{(\sigma^{\rm Obs.})^2+\sigma^2}}\times\\
&\exp{\left\{-\frac{\left[\log \bar{L}^{\rm Obs.}-\log \bar{L}_{\rm Mod.}(y^{\rm Obs.};k_1,\lambda,y_i)\right]^2}{2[(\sigma^{\rm Obs.})^2+\sigma^2]}\right\}},\nonumber
\end{align}
where $y^{\rm Obs.}=t^{\rm Obs.}/\tau_{^{56}{\rm Co}}$, $\sigma^{\rm Obs.}=\Delta \log{\bar{L}^{\rm Obs.}}=\Delta L^{\rm Obs.}/L^{\rm Obs.}$ is the error deviation linked to the observations, and $\sigma$ represents a systematic uncertainty term that is used to capture extra variance in the data, arising from unmodeled variability or other sources of systematic error in the model.
The likelihood function thus provides the statistical framework to compare observations and model predictions in terms of normalized luminosity. To achieve a consistent inference on the progenitor and explosion properties, it is then combined with the prior distributions introduced in Sect.~\ref{SubSubSec:Prior}, leading to the definition of the posterior probability distribution discussed in the next section.

\subsubsection{Posterior exploration}\label{SubSubSec:Posterior}
The likelihood function is defined in terms of the parameter set $\theta = (y_i, \lambda, k_1, \sigma)$. 
To express the posterior consistently in this parameter space, the priors—originally formulated for the physical quantities $(E, M_{\rm ej}, R_0)$—must be transformed accordingly.
The use of $\theta$ instead of the direct physical parameters has the practical advantage that each of the three dimensionless parameters $(y_i, \lambda, k_1)$ affects the WCR evolution in an independent way (cf. Eq.~\ref{Eq:WCR_evolution}). As a result, the LC model depends on them without strong internal degeneracies, unlike the case of the physical triplet $E$, $M_{\rm ej}$ and $R_0$ \citepalias[see also][for a detailed discussion of degeneracy in 87A-like events]{PC2025}.  
Following the procedure described in Appendix~\ref{App:PDF}, the prior PDFs (${\rm PDF}^{\rm Pr.}$) for the dimensionless parameters can be constructed from those of $(E, M_{\rm ej}, R_0)$ through the combination of Eqs.~\ref{Eq:Scaling} (or \ref{Eq:Scaling_new}) and \ref{Eq:SuperBAM_param}. This yields:
\begin{align}
\label{Eq:Prior_param}
y_{i}\propto  t_{\rm m},\qquad&\,
\lambda\propto M_{\rm Ni}\times t_{\rm M}^{-5/2}\times v_{\rm M}^{-1},\qquad
k_1\propto  t_{\rm M}^{3/2}\nonumber\\
\text{or}\qquad &\,
\lambda\propto M_{\rm Ni}\times t_{\rm M}^{-3/2}\times L_{\rm M}^{-1/2}
.
\end{align}
The prior distribution for the modeling parameter set can then be written as:
\begin{equation}
P(\theta)= {\rm PDF}^{\rm Pr.}(y_i)\times {\rm PDF}^{\rm Pr.}(\lambda)\times {\rm PDF}^{\rm Pr.}(k_1)\times {\rm PDF}^{\rm Pr.}(\sigma),
\end{equation} 
where ${\rm PDF}^{\rm Pr.}(\sigma)$ is taken as a normal distribution with mean and variance estimated from the observed scatter $\sigma^{\rm Obs.}$.  
Finally, the posterior distribution is obtained as the product of likelihood and prior, up to a normalization factor $N_{\rm Norm}$:
\begin{equation}
P(\theta|\bar{L})=N_{\rm Norm.}\times P(\bar{L}|\theta)\times P(\theta).
\end{equation}

\begin{figure}[t]
\includegraphics[width=\columnwidth]{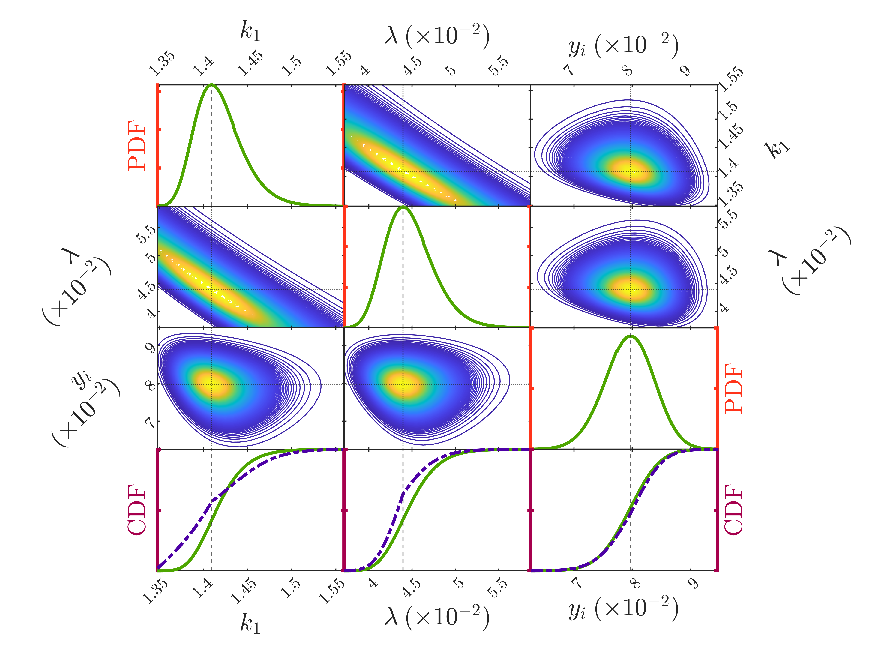} 
\caption{Mosaic of $P(\theta|\bar{L})$ plots for SN 1987A. The color density maps show 2D slices of the log-posterior distribution with $\sigma$ fixed, while two modeling parameters vary at a time. The diagonal panels display the posterior PDFs along each parameter axis, keeping the other parameters fixed at their best-fit values. The lower row shows the cumulative distribution function (CDF) for each parameter axis (solid line), compared to the CDF integrated on the entire space (dotted line).
\label{Fig:Posterior}}
\end{figure}

The combination of Bayesian statistics with the efficiency of the analytic approach allows us to explore nearly the entire support of the posterior probability, thus distinguishing local maxima from the global one (see Fig.~\ref{Fig:Posterior}).
\textsc{SuperBAM} implements an exploratory algorithm based on the Hamiltonian Monte Carlo (HMC) Sampler\footnote{The HMC algorithm is a Markov Chain Monte Carlo technique that generates random samples to efficiently explore and maximize a multi-parametric distribution. See also the \href{https://it.mathworks.com/help/stats/bayesian-linear-regression-using-hamiltonian-monte-carlo.html}{MatLab documentation} for details.}, which identifies the parameter vector $\theta^M=(y_i^M,\lambda^M,k_1^M,\sigma^M)$ that maximizes the posterior $P(\theta|\bar{L})$. By varying one parameter at a time, the posterior PDFs (${\rm PDF}^{\rm Po.}$) can be derived around $\theta^M$ through renormalization (see the diagonal panels in Fig.~\ref{Fig:Posterior}):
\begin{equation}
 {\rm PDF}^{\rm Po.}(\theta_i)=K_i \times P\!\left(\theta_i,\theta^M_{j\neq i}\mid\bar{L}\right),
\end{equation}
where $K_i^{-1}=\int_0^{+\infty} {\rm PDF}^{\rm Po.}(\theta_i)\, d\theta_i$ is the normalization constant. 
Once the posterior PDFs for $(k_1^M,\lambda^M,y_i^M)$ are obtained, the corresponding PDFs of the physical explosion parameters $(E, M_{\rm ej}, R_0)$ can be reconstructed by applying the inverse transformations of Eq.~\ref{Eq:SuperBAM_param}, as described in Appendix~\ref{App:PDF}.

It should be noted, however, that these posterior PDFs are derived in a univariate way (i.e., along single axes). This approach may underestimate modeling uncertainties, since off-axis correlations between parameters are not fully captured. For instance, inspection of the $k_1$--$\lambda$ plane reveals a non-negligible covariance, intrinsic to the model due to their product entering the normalized luminosity in Eq.~\ref{Eq:L_mod}. As illustrated by the difference between the CDFs along the axes and over the full space (Fig.~\ref{Fig:Posterior}), the broader multidimensional distributions highlight the need to account for covariance-driven uncertainties.
To address this limitation, \textsc{SuperBAM} employs a Monte Carlo (MC) procedure that randomly generates $10^4$ triplets of modeling parameters ($y_i,\lambda,k_1$) around the HMC best-fit solution $\theta^M$, fixing $\sigma= \sigma^M$. For each triplet, the posterior probability value is computed and used as a statistical weight, naturally decreasing with distance from the optimum. From this weighted sample of parameter triplets, expectation values and variances are derived for both the modeling parameters and, through the inverse transformations of Eq.~\ref{Eq:SuperBAM_param}, for the physical parameters $(E, M_{\rm ej}, R_0)$.
 
 \subsection{Validation test}\label{SubSec:test}
 To validate our approach, we apply \textsc{SuperBAM} to the previously published LCs of SN 1987A \citep{catchpole1987}, SN 2009E \citep{pasto12}, and OGLE-14 \citep{terreran17}, shown in Fig. \ref{Fig:GP}, and compare the results with the HM from \citet{PZ11}'s post-explosive model (see Table 1 of \citetalias{PC2025} and references therein). As a consequence, the temporal sampling of the bolometric LCs reflects the reconstruction methods used in the original works and may slightly differ from that obtained with our own bolometric reconstruction technique (see also Sect. \ref{SubSec:Bolo_lum}).
Based on the procedure outlined in Section~\ref{SubSec:SuperBAM}, \textsc{SuperBAM} provides an initial estimate from the priors for $E$, $M_{\rm ej}$, $R_0$, and $M_{\rm Ni}$, as well as two posterior estimates of $E$, $M_{\rm ej}$, and $R_0$ obtained through different approaches. The first, referred to as the posterior best model, corresponds to the parameter configuration that maximizes the posterior probability (i.e., the mode of the distribution), derived using the HMC sampler; the associated uncertainty is estimated from a univariate analysis of the posterior. The second, instead, is obtained through a MC simulation, yielding the mean values and standard deviations of the physical parameters over the multidimensional posterior support.
These two posterior estimates are consistent with each other within their respective error bars, demonstrating \textsc{SuperBAM}’s ability to converge toward coherent and more precise modeling results compared to those obtained from the priors through the scaling relations (see Table~\ref{Tab:Validation}).
\begin{table}
\centering
\scriptsize
\caption{Results of the validation test.}
\begin{tabular}{clccc}
\hline
Par. & Method & SN~1987A & SN~2009E & OGLE-14 \\
\hline
\multirow{2}{*}{} 
 & Pr.       & $1.4\pm0.9$ & $0.8\pm0.2$ & $18\pm11$ \\
 E & Po.       & $1.6\pm0.2$ & $1.1\pm0.1$ & $20\pm5$ \\
 $[{\rm foe}]$& MC     & $1.7\pm0.4$ & $1.2\pm0.2$ & $21\pm7$ \\
\multirow{2}{*}{} & HM  & $1.3\pm0.1$ & $0.6	\pm0.2$ & $12^{+13}_{-6}$ \\
\hline
\multirow{2}{*}{} 
 & Pr.       & $16\pm5$     & $15\pm2$  & $43\pm16$ \\
 M$_{\rm ej}$ & Po.       & $18\pm1$ & $23\pm1$  & $46\pm5$ \\
 $[{\rm M}_\odot]$& MC     & $19\pm3$ & $24\pm2$  & $48\pm8$ \\
\multirow{2}{*}{} & HM  & $16\pm1$ & $19	\pm3$ & $60^{+42}_{-16}$ \\
\hline
\multirow{2}{*}{} 
 & Pr.       & $4\pm3$     & $4\pm4$ & $25\pm21$ \\
R$_0$ & Po.       & $2.9\pm0.3$ & $4.6\pm0.2$ & $25\pm3$ \\
 $[10^{12}\,{\rm cm}]$ & MC     & $2.8\pm0.3$ & $4.5\pm0.5$ & $27\pm4$ \\
\multirow{2}{*}{} & HM  & $3.0\pm0.9$ & $7\pm1$ & $38^{+8}_{-10}$ \\
\hline
 M$_{\rm Ni}$ & Pr.       & $7.4\pm0.9$     & $3.9\pm0.2$ & $42\pm13$ \\
$[10^{-2}\, {\rm M}_\odot]$ & HM 		   & $7\pm1$     & $4\pm1$     & $47\pm2$ \\
\hline
\end{tabular}\label{Tab:Validation}
\tablefoot{Comparison of priors (Pr.), best model posterior (Po.), and MC results with those from numerical HM for SN~1987A, SN~2009E, and OGLE-14.
} 
\end{table}
As the output of \textsc{SuperBAM}, we adopt the best-fit model obtained with the HMC-Sampler, while assigning uncertainties derived from the MC approach. This choice combines the advantage of a model that best reproduces the observed data with error estimates that account for multidimensional parameter correlations.

When comparing the \textsc{SuperBAM} results with the HM, we find that the results are fully consistent within the errors, with the only exception of the energy for SN 2009E
(about a factor two higher). This offset in the explosion energy is present in all SNe and it is especially evident for SN 2009E and OGLE-14. This systematic discrepancy can be traced back to the intrinsic assumptions of semi-analytic models. Indeed, unlike hydrodynamical simulations, our model adopts a uniform density profile for the ejecta. As already discussed in previous works, this simplification leads to an overestimate of the energy budget because more mass is effectively concentrated in the fast-moving outer layers, increasing the kinetic energy content.
This behavior was noted for OGLE-14 by \citet{terreran17}, where the semi-analytic model of \citet{zampieri03} yielded an explosion energy $\sim$70\% larger than the HM estimate, precisely because the early acceleration phase of the ejecta is not accounted for. A similar result was reported for SN 2009E in \citet{pasto12}, where the same semi-analytic approach produced an $M_{\rm ej}\sim$26 ${\rm M}_\odot$ and an $E\sim$1.3 foe, again exceeding HM estimates due to the uniform density assumption.
\textsc{SuperBAM} mitigates part of this bias, as it accounts for the effect of $^{56}$Ni heating on the recombination process. As noted by \citet{pasto12}, neglecting the non-uniform distribution of $^{56}$Ni leads to an underestimate of its contribution at late photospheric phases, forcing the model to invoke a more massive envelope to sustain the recombination. By incorporating this effect, \textsc{SuperBAM} achieves results that converge more closely to HM values, especially for the ejecta mass.
Smaller differences are found in the initial radius. In our framework, the uniform-density assumption tends to confine the ejecta more tightly at early times, leading to slightly smaller radii than those inferred from HM. This discrepancy increases with ejecta mass, but in the case of OGLE-14 the results remain consistent within the error bars.

Finally, it is important to emphasize that the $^{56}$Ni mass derived by \textsc{SuperBAM} from the LC tail (cf. Eq.~\ref{Eq:Ni_fit}), corresponds to the amount of nickel present in the ejecta at late times. This represents a lower limit compared to HM results, which include the fraction of nickel that falls back onto the compact remnant. Consequently, our $^{56}$Ni masses are systematically lower than those obtained with HM.
In summary, \textsc{SuperBAM} offers a fast and physically consistent way to characterize \Alike~SNe, providing results comparable to HM, while significantly reducing the computational time. This makes \textsc{SuperBAM} particularly well suited for the analysis of large SN samples.

\begin{table*}
\caption{\label{Tab:Info_SNe87A} Sample of \Alike~SNe and key information used for the spectrophotometric analysis. 
}
\centering
\scriptsize
\begin{tabular}{l c c c c c c c}
\hline\hline
\multirow{2}{*}{SN Name} & \multirow{1}{*}{Explosion Epoch} & \multirow{1}{*}{Host Galaxy}& \multirow{1}{*}{Redshift} & \multirow{1}{*}{m-M} & \multirow{1}{*}{$A_V$(tot)} & \multirow{2}{*}{12+log(O/H)\tablefootmark{a}} & \multirow{2}{*}{Ref.} \\
 & [MJD] & Name & $(z)$ & [mag] & [mag] & &\\
\hline
SN 1987A	&	46849.8	&	LMC	&	0.0009	& $18.5\pm0.1$ &	0.600	&	$8.37\pm0.06$ & 1,2 \\
SN 1998A	&	50763.3	&	IC 2627	&	0.0072 & $32.2\pm0.3$	&	0.383	&	$8.68\pm0.06$ &3 \\
SN 2000cb	&	51651.1	&	IC 1158	&	0.0064 & $32.7$	&	0.353	&	$8.45\pm0.06$ &4 \\
SN 2004ek	&	53250.5	&	UGC 724	&	0.0173 &	$34.4\pm 0.3$ &	0.530	&	$8.59$ & 5 \\
SN 2004em	&	53263.3	&	IC 1303	&	0.0149	& $33.9\pm0.3$ &	0.298	&	$8.56 \pm 0.11$ & 5 \\
SN 2005ci	&	53512.4	&	NGC 5682	&	0.0076	& $32.8$ &	0.089	& $8.31\pm 0.04$	&5 \\
SN 2006V	&	53748.0	&	UGC 6510	&	0.0158	& $34.4\pm0.8$ &	0.090	&$8.35 \pm 0.11$&6 \\
SN 2006au	&	53794.0	&	UGC 11057	&	0.0099	& $ 33.4\pm0.8$ &	0.970	&$8.49 \pm 0.12$&6 \\
SN 2009E	&	54824.5	&	NGC 4141	&	0.0063	& $32.4\pm0.4$ &	0.124	&$8.22 \pm 0.08$&7 \\
SN 2009mw	&	55174.5	&	ESO 499-G005	&	0.0143	& $33.5\pm0.2$&	0.167	& $8.32$ &8 \\
PTF12gcx	&	56081.3	&	SDSS J154417.02+095743.8	& 0.0450 & $36.5\pm0.2$	&	0.140	&	$8.52 \pm 0.18$ &5 \\
PTF12kso	&	56175.0	&	Anon.	&	0.0300	& $35.6\pm0.2$ &	0.184	&	$\lesssim8.04$&5 \\
OGLE-14	&	56861.1	&	Anon.	&	0.1225	& $38.8$ &	0.170	&	 $8.36\pm0.10$&9 \\
SN Refsdal	&	56968.9	&	Anon.-Lensed\tablefootmark{b}	&	1.4910 & $45.2\pm0.1$	&	0.310	 &	$8.3\pm0.1$&10 \\
DES16C3cje	&	57670.2	&	PGC324331	&	0.0618 & $37.2$ &	0.527	&	$<8.19\pm0.02$ &11 \\
SN 2018cub	&	58216.3	&	WISEA J150541.98+604751.4	&	0.0440 & $36.5$ &	0.047	&	$8.43$ &12 \\
SN 2018ego	&	58248.3	&	2MASX J15525218+1958107	&	0.0375 & $36.2$ &	0.144	& $7.60$ &12 \\
SN 2018imj	&	58380.0	&	IC0454	&	0.0132 & $33.7\pm0.2$ 	&	0.627	&	$8.19$ &11 \\
SN 2018hna	&	58411.3	&	UGC 07534	&	0.0024 & $30.5\pm	0.3$ &	0.028	&	$8.14\pm0.02$ &13 \\
SN 2019bsw	&	58482.7	&	WISEA J100506.20-162425.1	&	0.0295 & $35.6$	&	0.140	& NaN & 12 \\
SN 2020faa	&	58926.0	&	WISEA J144709.05+724415.5	&	0.0411 & $36.2\pm0.2$ &	0.067	&	$8.40\pm0.1$ &14,15 \\
SN 2020oem	&	58982.8	&	WISEA J152729.72+034646.8	&	0.0424 & $36.4$	&	0.139	&	$8.34$ &12 \\
SN 2020abah	&	59175.5	&	CGCG 127-002	&	0.0301 & $35.7$	&	0.065	&	$8.42$ &12 \\
SN 2021zj\tablefootmark{c}	&	59224.4	&	SDSS J111632.91+290546.5	&	0.0460 & $36.6\pm0.1$	&	0.038	& NaN &12,16 \\
SN 2021mju	&	59287.5	&	WISEA J164148.29+192203.6	&	0.0283 & $35.5$ &	0.210	&	8.14 &12 \\
SN 2021skm	&	59389.4	&	2MASX J16165615+2148359	&	0.0315 & NaN &	0.222	&	8.30 &12 \\
SN 2021wun	&	59425.2	&	SDSS J154631.94+252545.5	&	0.0228 & $35.1$ &	0.128	&	8.17 &12 \\
SN 2021aatd	&	59493.5	&	GALEXASCJ005904.40-001210.0	&	0.0152 & $34.2$	&	0.078	&	NaN &17 \\
\hline
\end{tabular}
\tablebib{(1)~\citet{1995ApJS...99..223P};
(2)~\citet{catchpole1987}; (3)~\citet{pasto05}; (4) \citet{kleiser11}; (5) \citet{taddia16};
(6) \citet{taddia12}; (7) \citet{pasto12}; (8) \citet{takats16};
(9) \citet{terreran17}; (10) \citet{kelly16}; (11) \citet{gutierrez20};
(12) \citet{sit_2023}; (13) \citet{singh19}; (14) \citet{Yang2024}; (15) \citet{Salmaso2023};
(16) \citet{JacobsonG2024}; (17) \citet{Szalai2024}.
}

\tablefoot{Column~2 lists the estimated explosion epoch, computed as the midpoint between the last non-detection and the discovery date. 
The host galaxy name is given in Column~3 (with “Anon.” indicating anonymous hosts), followed by the redshift ($z$ in Col.~4) and the distance modulus (m-M in Col.~5; when not reported in the cited references, values are taken from the \href{https://ned.ipac.caltech.edu/}{NASA/IPAC Extragalactic Database}). 
Columns~6 and~7 report the total visual extinction, $A_V$ \citep[Milky Way + host contribution, following][]{2011ApJ...737..103S}, and the oxygen abundance at the SN site [12+log(O/H)], respectively. 
The final column lists the main references from which these quantities, as well as the multi-band photometric data and spectra used in this work, are collected. 
Entries marked as “NaN” indicate that the corresponding information is not available in the literature.
\tablefoottext{a}{
Unless otherwise specified in the cited references, the oxygen abundances are derived using the N2 diagnostic \cite[see][]{Taddia13}.}\\
\tablefoottext{b}{SN Refsdal, discovered in the MACS J1149+2223 galaxy cluster, was observed as a strongly lensed SN with multiple resolved images \citep{kelly16}. Among the four images, this work focuses on S2, assuming a magnification factor of $\mu=15$ \citep{2016ApJ...822...78G}.}\\
\tablefoottext{c}{No public spectra are available for SN 2021zj during the secondary peak phase. Classified as a young SN II by \citet{2021TNSAN..14....1S}, it showed flash ionization signatures in the early spectra.
}
} 

\end{table*}
\section{The sample of 87A-like SNe}\label{Sec:Single}
Over the last two years, the number of published 87A-like SNe has nearly doubled, largely thanks to the efforts of wide-field surveys such as ZTF/CLU \citep{sit_2023}. In \citetalias{pumo23}, a sample of 14 SNe \Alike~was analyzed using the approach of scaling relations, relying on bolometric LCs and velocity data already available in the literature. In this work, we expand this sample to nearly double its size by homogeneously reconstructing the bolometric LCs, estimating line velocities and their pseudo-equivalent widths (pEWs) from peak spectra, and applying the \textsc{SuperBAM} procedure to derive the physical properties at the explosion.

To this aim, we performed a systematic search through the literature of the past three decades, selecting the best-observed long-rising type II SNe according to the following criteria:
\begin{itemize}
\item Multi-band photometric coverage (at least three filters) starting no later than $\bar{t}_{\rm M}-40\,$d and extending through the end of the recombination phase ($t_{\rm f}^*$) with at least one point on the $^{56}$Ni tail. This allows us to reconstruct the bolometric LC and constrain priors such as the $^{56}$Ni mass.
\item Reliable estimates of distance (or redshift) and extinction are required. It is necessary to correct both spectra and photometry before deriving bolometric LCs and velocities.
\end{itemize}
Both \textsc{SuperBAM} and alternative modeling approaches require these conditions to be applied; otherwise, the analysis cannot be considered reliable \citep[e.g.,][]{MB19}.
Following these criteria, we selected a sample of 28 long-rising type II SNe published before 2025. The list reported in Table \ref{Tab:Info_SNe87A} includes, for each of these events, the estimated explosion epoch (which is not a free parameter of the fit, but is fixed as the midpoint between the last non-detection and the discovery date), the host galaxy with its distance modulus, the adopted redshift, the total visual extinction, and the environmental metallicity in terms of oxygen abundance [12+log(O/H)], along with the main bibliographic references from which this information is taken.

The collection of multi-band photometric data for each SN enables the reconstruction of their bolometric LCs. 
In addition to the datasets published in the references listed in Table~\ref{Tab:Info_SNe87A}, we also include, where available, photometry from the ATLAS \citep[Asteroid Terrestrial-impact Last Alert System; see, e.g.,][]{Tonry_2018}. 
Spectroscopic data are gathered from the same cited works, selecting spectra obtained near the luminosity peak (see Tab. \ref{Tab:Spectra_feature} for acquisition dates and epochs of the spectra considered for each SN). 
These spectra are used to measure photospheric velocities and line equivalent widths, providing complementary constraints on the ejecta composition and kinematics. 
 Although, no public spectra are available for SN 2021zj during the peak stage, its modeling was carried out by \textsc{SuperBAM} using the full photometric scaling relations system in Eq. \ref{Eq:Scaling_new}. However, this object is not included in the spectroscopic analysis made for the other \Alike~SNe in the sample.

\begin{figure*}[t]
\centering
\includegraphics[width=\textwidth]{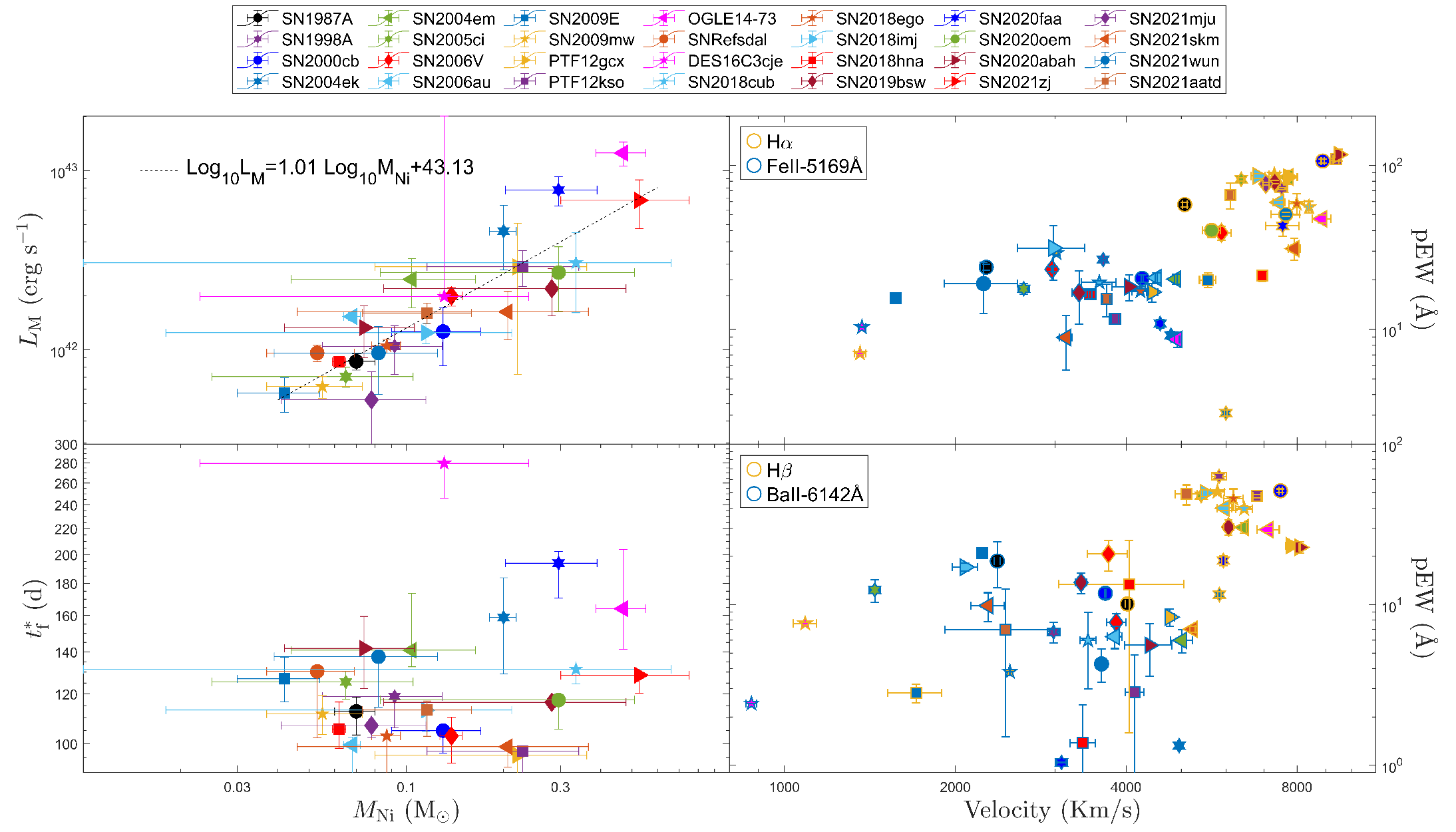}
\caption{Observational properties derived from the bolometric LCs and spectra of the SN sample. 
\textit{Top left:} Distribution of the peak luminosity $L_{\rm M}$ as a function of the $^{56}$Ni mass estimated from the radioactive tail fit. 
The trend line is obtained through a linear regression $\log_{10}L_{\rm M}$-$\log_{10}M_{\mathrm{Ni}}$; the correlation is highly significant ($p$-value$\ll 0.01$ and Pearson coefficient of $0.89$). 
\textit{Bottom left:} Distribution of $t_f^*$ versus the $^{56}$Ni mass for the same SN sample. 
\textit{Top right:} pEWs as a function of expansion velocities for the H$\alpha$ and Fe\,II lines, for the SNe with available data. 
The color and shape of the markers follow the top legend, while the marker border indicates the line type according to the internal legend. 
\textit{Bottom right:} Same as the top-right panel, but for the H$\beta$ and Ba\,II lines.}
\label{Fig:Obs_features}
\end{figure*}

\subsection{Bolometric luminosity}\label{SubSec:Bolo_lum}
For the computation of bolometric luminosity, all available multi-band photometry from the literature has been re-analysed in a homogeneous way, ensuring a consistent procedure for the construction of bolometric LCs. In particular, we apply \textsc{Superbol} procedure for the bolometric integration of the spectral energy distribution (SED) described by \citet{Nicholl_2018}. 
In addition,
the multi-band LCs are first reconstructed in time using GPR interpolation, and extrapolated under the constant-color assumption\footnote{
The constant-color refers to a temporal extrapolation technique in which missing-band fluxes are estimated by preserving the color difference measured at the nearest epoch with multi-band coverage \citep[see][]{Nicholl_2018}.}.
The use of GPR provides synthetic photometric points at missing epochs, effectively increasing the number of multi-band observations and allowing for virtually simultaneous coverage across filters\footnote{In wide-field surveys (e.g., LSST), multi-band photometry is typically obtained on different nights, depending on the observing strategy \citep{Bianco_2022}. In such cases, GPR is particularly effective.}. This interpolation yields a denser bolometric sampling while preserving consistency with curves derived from real data within the uncertainties\footnote{See also Fig. 4.2 in \citet{Cosentino2024}.}.
In addition to photometry, information on distance and extinction is required. 
For SNe with $z > 0.01$, distances are derived from the redshift, assuming $\Omega_\Lambda=0.685$, $\Omega_M= 0.315 $ and the Hubble-Lemaıtre constant equals to $H_0 = 67.4$ km s$^{-1}$ Mpc$^{-1}$ \citep[e.g.,][]{2020A&A...641A...6P}.
For nearby SNe, where peculiar motions can dominate over the Hubble flow, we adopt published distance moduli (Tab. \ref{Tab:Info_SNe87A}). In the case of the gravitational lensed SN Refsdal, we also correct the observed flux for the magnification lens factor $\mu=15$ \citep[e.g.,][and references therein]{2016ApJ...822...78G}.
The reddening correction is applied assuming the \citet{Cardelli} extinction law with $R_V = 3.1$, appropriate for the diffuse interstellar medium. The color excess $E(B-V)$, computed as the difference between observed and intrinsic colors, is then estimated from the total extinction as $E(B-V)=A_V/R_V$.
Finally, no UV cut-off is applied to the SEDs. {Although the early post-explosion phases ($<10$–20 d) might be affected by a significant drop in the UV flux, this effect is negligible for SNe II because of the hydrogen-rich, metal-poor nature of their ejecta \citep[e.g.,][]{2008ApJ...683L.131G,Bufano_2009}, and is irrelevant for our analysis, which does not rely on the earliest epochs. However, the SED fitting is potentially affected by line blanketing\footnote{As in normal Type~II SNe, the SEDs of \Alike~events can be affected by 
line blanketing in the blue bands \citep[e.g.,][]{KW09,pasto12}. We verified that recomputing the bolometric LC of SN~1987A with and without the U and B bands leads to differences of $\sim 5$--$20\%$, comparable to the intrinsic uncertainties of the blackbody fitting procedure. Bolometric LCs obtained by integrating all available bands and extrapolating to missing regions using a blak-body fit provide robust estimates of the total emergent luminosity \citep[e.g.,][]{2014MNRAS.437.3848L,taddia16}.}.

The bolometric LCs obtained in this way are then used by \textsc{SuperBAM} to constrain the physical parameters of the explosions (see Sect. \ref{Sec:Modeling}). As a preparatory step, however, the procedure also measures key morphological features of the LCs, reported in Table \ref{Tab:Prior_features}, such as $M_{\rm Ni}$, $t_{\rm f}^*$ and $L_{\rm M}$ (see left panels of Fig. \ref{Fig:Obs_features}).
The distribution of nickel masses, directly inferred from the radioactive tails, spans nearly an order of magnitude, from the $\sim0.04$~M$_\odot$ of SN~2009E up to the $0.4$--$0.5$~M$_\odot$ measured for OGLE-14 and SN~2021zj. 
Grouping the sample according to $M_{\rm Ni}$ highlights three main subsets. 
Events with $M_{\rm Ni}<0.1$~M$_\odot$, including SN~1987A and SN~Refsdal, cluster around peak luminosities of $\sim10^{42}$ erg s$^{-1}$ and recombination end times $t_{\rm f}^*\approx100$--150~d. 
SNe with intermediate nickel masses ($0.1$--$0.2$~M$_\odot$) reach higher peak luminosities, still consistent with the observed scaling, but in some cases show broader LCs, as for DES16C3cje, whose $t_{\rm f}^*$ exceeds 200~d. 
Finally, the most nickel-rich explosions ($M_{\rm Ni}>0.2$~M$_\odot$) attain luminosities above $10^{43}$ erg s$^{-1}$, with OGLE14-73 representing the event with the brightest and broadest peak of the class. 
SN~2021zj and SN~2020faa also belong to this group, though they display a slow pre-maximum rise, with $t_{\rm m}\sim40$--60~d, unlike OGLE14-73 where no early coverage is available.
Across the entire sample, $L_{\rm M}$ shows a linear increase with $M_{\rm Ni}$, as illustrated in the left-up panel of Fig.~\ref{Fig:Obs_features}. 
This trend has also been reported in broader samples of H-rich SNe \citep[e.g.,][]{Martinez22}, and becomes particularly clear for \Alike~SNe,  in which the radioactive Ni-heating is more powerful than the ejecta recombination-cooling (i.e., where the $\Lambda$-term dominates; see Eqs.~67–68 in \citetalias{PC2025}). However, events such as OGLE-14 and SN~2020faa exhibit luminosities significantly above the predicted trend, suggesting the contribution of additional energy sources that enhance the overall brightness.
On the contrary, the same clear  dependency on $M_{\rm Ni}$ is not observed for $t_{\rm f}^*$, which is more influenced by other properties of the explosion, deriving from its link with the diffusion and ejecta expansion rates \citepalias[$t_{\rm f}^*\propto M_{\rm Ni}^{1/3}E^{-1/2}M_{\rm ej}^{5/6}$; cf. Eq. 76 of][]{PC2025}.

\subsection{Spectroscopic features}\label{SubSec:spectra}
The available SN spectra have been reduced and corrected for both reddening and redshift, thus placing them in the rest frame and recovering their intrinsic continuum shape. These corrections, applied consistently across the sample, allow a direct comparison of line profiles and velocities, providing essential constraints on the ejecta properties. 

For this purpose, the \textit{deredden} and \textit{dopcor} tasks within the Image Reduction and Analysis Facility (IRAF; \citealt{IRAF_86}) are employed.
After the corrections, the main absorption features typical of SNe~\Alike\ are analyzed, with particular attention to the Balmer lines (H$\alpha$, H$\beta$) and metal lines such as Ba\,{\sc ii}-$\lambda6142$ and Fe\,{\sc ii}-$\lambda5169$. The minima of the P-Cygni absorption profiles are measured using Gaussian fits with the IRAF \textit{splot} tool, in order to infer the expansion velocities from their Doppler shifts. Simultaneously, these fits provide the pEWs used to characterize line strengths and assess the relative contribution of different ions. The results of this analysis are summarized in Tab. \ref{Tab:Spectra_feature} and illustrated in the right panels of Fig. \ref{Fig:Obs_features}.

From a spectroscopic point of view, the majority of the SNe show broad P-Cygni Balmer profiles together with Fe\,{\sc ii} and Ba\,{\sc ii} features, which are particularly suited to trace the photospheric velocity. In our procedure, the photospheric velocity at peak ($v_{\rm M}$) adopted in the priors of \textsc{SuperBAM} is taken from the Fe\,{\sc ii} velocity, which is available for the large majority of the sample. When Fe\,{\sc ii} is not detected, the Ba\,{\sc ii} velocity is used instead, and in the few cases where neither of these lines is measurable (e.g., SN Refsdal and SN 2020oem) the H$\alpha$ velocity is employed, rescaled by a factor of 2 to account for the systematic offset between hydrogen and metal lines. Indeed, Balmer lines yield expansion velocities of 6000--9000 km s$^{-1}$, systematically higher (by a factor of 1.5--2.5) than those derived from metal lines (2000--5000 km s$^{-1}$). This difference originates from the fact that Balmer transitions are mostly shaped by photoionization and recombination processes in the outer ejecta layers, where the gas is only partially ionized and expands at higher velocities. Conversely, ionized metal lines require higher temperatures to sustain the proper ionization state, and become more prominent where the abundance of heavy elements is larger, i.e., in deeper and slower-moving regions of the photosphere. The smaller velocity offset between H$\beta$ and Ba\,{\sc ii} supports this interpretation, since the H$\beta$ formation region lies deeper than that of H$\alpha$ and requires a higher excitation energy to populate the $n=4$ level. Only a few objects deviate from these general trends, such as SN2009E and SN2005ci, which display unusually large offsets between Balmer and metal velocities.

As for pEWs, the Balmer lines generally exceed the strength of metal features by factors of 5–10. A noteworthy case is DES16C3cje, whose narrower H$\alpha$ profile and weaker relative pEW suggest a non-standard post-explosion scenario \citep[see also][]{gutierrez20}. At peak phase, Fe\,{\sc ii} pEWs across the whole sample remain relatively uniform, typically in the range of 10–30\,\AA. In contrast, Ba\,{\sc ii} displays a much larger scatter, with values spanning from a few \AA\,(e.g., SN~2018hna) up to $\sim$10\,\AA\ or more (e.g., SN~1987A and SN~2009E). Previous studies had suggested the existence of two distinct subgroups of 87A-like SNe based on Ba\,{\sc ii} strength \citep[e.g.,][]{takats16}; however, in our extended sample the distribution appears more continuous, with several events occupying intermediate values.

In addition, a possible anti-correlation between the velocity of the Ba\,{\sc ii} line and its pEW is hinted at in our data. Although the statistical significance of this trend is limited (see Sect.~\ref{SubSec:correlation}), it can be interpreted in terms of the ejecta density structure. In more compact explosions, where the ejecta expand less rapidly, the higher density in the line-forming regions favors stronger Ba\,{\sc ii} absorption \citep[cf.][]{1995A&A...303..118M}. Within this framework, SNe with lower Ba\,{\sc ii} velocities during the peak phase are naturally expected to display deeper Ba\,{\sc ii} lines. 
While the spectra considered here are all collected around the recombination peak phase, variations in the location and temperature of the Ba\,{\sc ii} formation region can also affect this line strength \citep[e.g.,][]{xiang23}. 
A systematic investigation of spectrophotometric dependencies, including the role of physical parameters, is deferred to Sect.~\ref{SubSec:correlation}.

\section{Modeling long-rising SNe with \textsc{SuperBAM}}
\label{Sec:Modeling}
\begin{figure*}
\includegraphics[width=2\columnwidth]{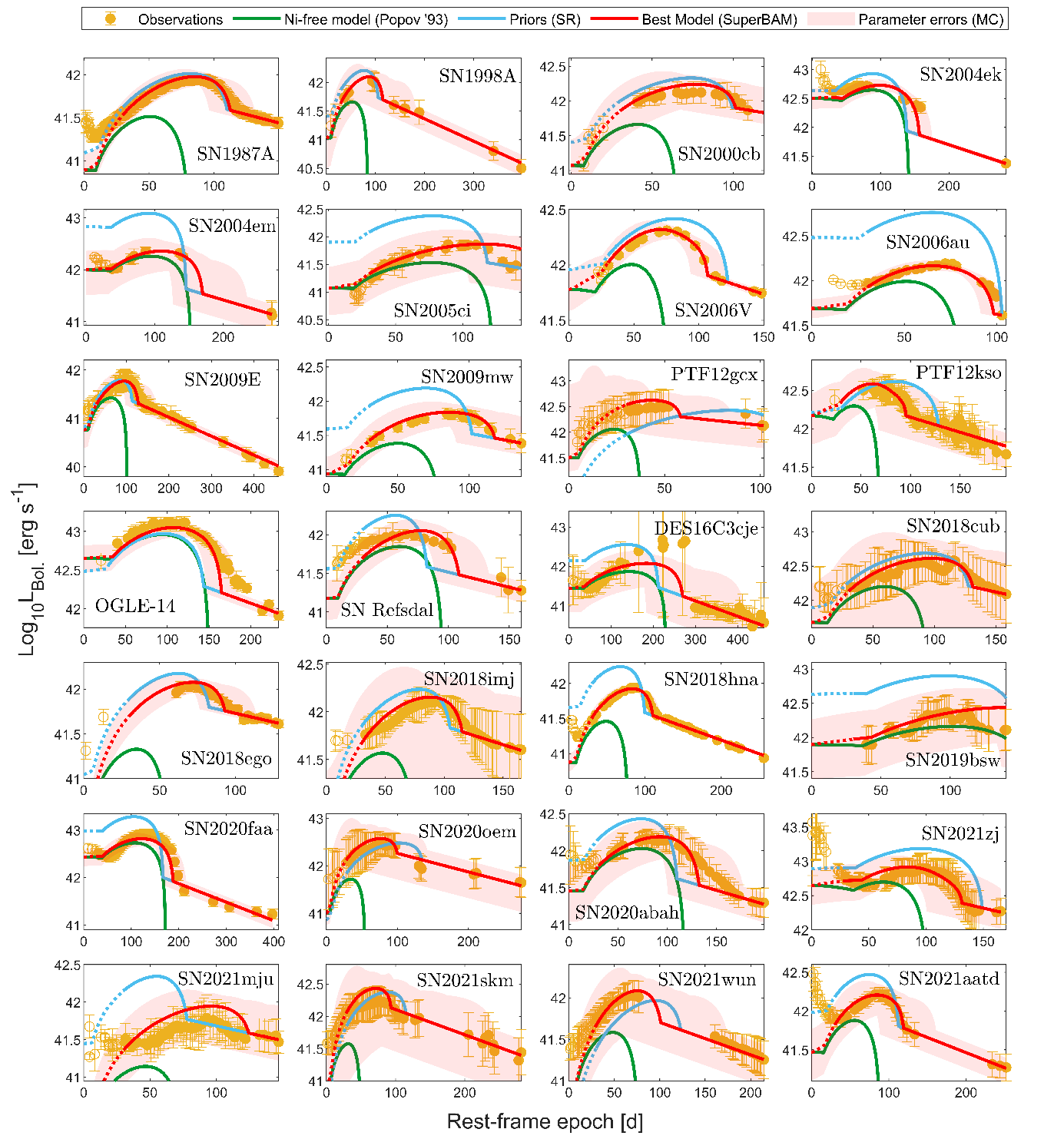} 
\caption{Bolometric LC data for each SN in our sample. The name of each SN is indicated within its panel. The adopted explosion epochs ($t=0$) are fixed from observational constraints (see Table \ref{Tab:Info_SNe87A}).
 Along with the data, we present the synthetic LCs obtained from the analytical model described in \citetalias{PC2025}. The parameters for these models were determined using the prior values given by the scaling relations, and the best-fit model obtained from \textsc{SuperBAM} analysis. The shaded region marks the LCs associated with the explored parameters within the error ranges of the MC method. Data earlier than 30 days are not included in the fitting procedure.
 For comparison, the synthetic LC of the nickel-free model by \citet{popov93} is also included, calculated for the best-model parameters.
\label{Fig:Modeling}}
\end{figure*}
To explore the physical properties of the selected \Alike~sample, we applied the \textsc{SuperBAM} procedure to the entire set of objects. As discussed in Sect. \ref{Sec:SuperBAM}, the pipeline provides estimates of the explosion energy, ejecta mass, and progenitor radius (summarized in Tab. \ref{Tab:Modeling}), as well as the synthesized $^{56}$Ni mass, the latter being directly constrained from the radioactive tail of the LC (Tab. \ref{Tab:Prior_features}). By comparing the observed LCs with the simulated ones obtained from the models presented in \citetalias{PC2025} and \citet{popov93}, and using the parameters derived by \textsc{SuperBAM} through scaling relations and HMC sampling, we can evaluate both the physical parameter inference and the capability of these models to reproduce the observed bolometric evolution (Fig.~\ref{Fig:Modeling}).

The Ni–free simulations of \citet{popov93} systematically underestimate both the peak luminosity and the end of the recombination phase, owing to the absence of additional heating from radioactive $^{56}$Ni in the ejecta. A direct comparison between the \citet{popov93} and \citetalias{PC2025} models therefore allows us to quantify the role of $^{56}$Ni in shaping the LC morphology. Interestingly, this difference becomes progressively smaller in SNe with recombination onset later than $\sim$30 days, i.e., for progenitors with significantly larger initial radii (cf. Eq.~\ref{Eq:SuperBAM_param}).
As expected, the best-fit posterior models provide a significantly better agreement with the data compared to the priors, which generally capture the onset and termination of recombination but fail to accurately reproduce the luminous intensity.

Among the 28 SNe in the sample, ten show a reduced chi-squared\footnote{The reduced chi–squared is defined as $$\chi^2_\nu =\nu^{-1} \times\sum_{\rm Obs.} \left[\left(L^{\rm Obs.}_{\rm Bol.} - L^{\rm Mod.}\right)/\Delta L^{\rm Obs.}_{\rm Bol.} \right]^2,$$ where $\nu=N^{\rm Obs.}-4$ is the freedom degree depending on the number of bolometric observations $N^{\rm Obs.}$ after 30d since explosion, the sum is extended on these observed data ($L^{\rm Obs.}_{\rm Bol.}$, $\Delta L^{\rm Obs.}_{\rm Bol.}$), and $L^{\rm Mod.}$ are the corresponding model bolometric luminosities at $t^{\rm Obs.}$.} ($\chi^2_\nu$) above unity, while the others are well fitted within the observational uncertainties.
A first group of events, including SN1987A, SN1998A, SN2000cb, SN2004em, SN2006V, SN2009E, SN2009mw, SN2018hna, and SN2021aatd, shows excellent agreement between models and observations (i.e., $\chi^2_\nu\sim 1$), indicating that the automatic procedure robustly recovers the main LC features. For SN~2006V and SN~2018hna, however, the statistical agreement is formally poorer (reduced $\chi^2_\nu \sim 2.3$–$2.7$), mainly because the exceptionally dense UV--NIR 
multi-band coverage yields very small bolometric uncertainties 
($\sim$0.02 dex). Rescaling the log10-luminosity error to more typical values (0.05 dex) would bring the $\chi^2_\nu$ below unity, confirming the adequacy of the fit.
\begin{figure*}
\includegraphics[width=2\columnwidth]{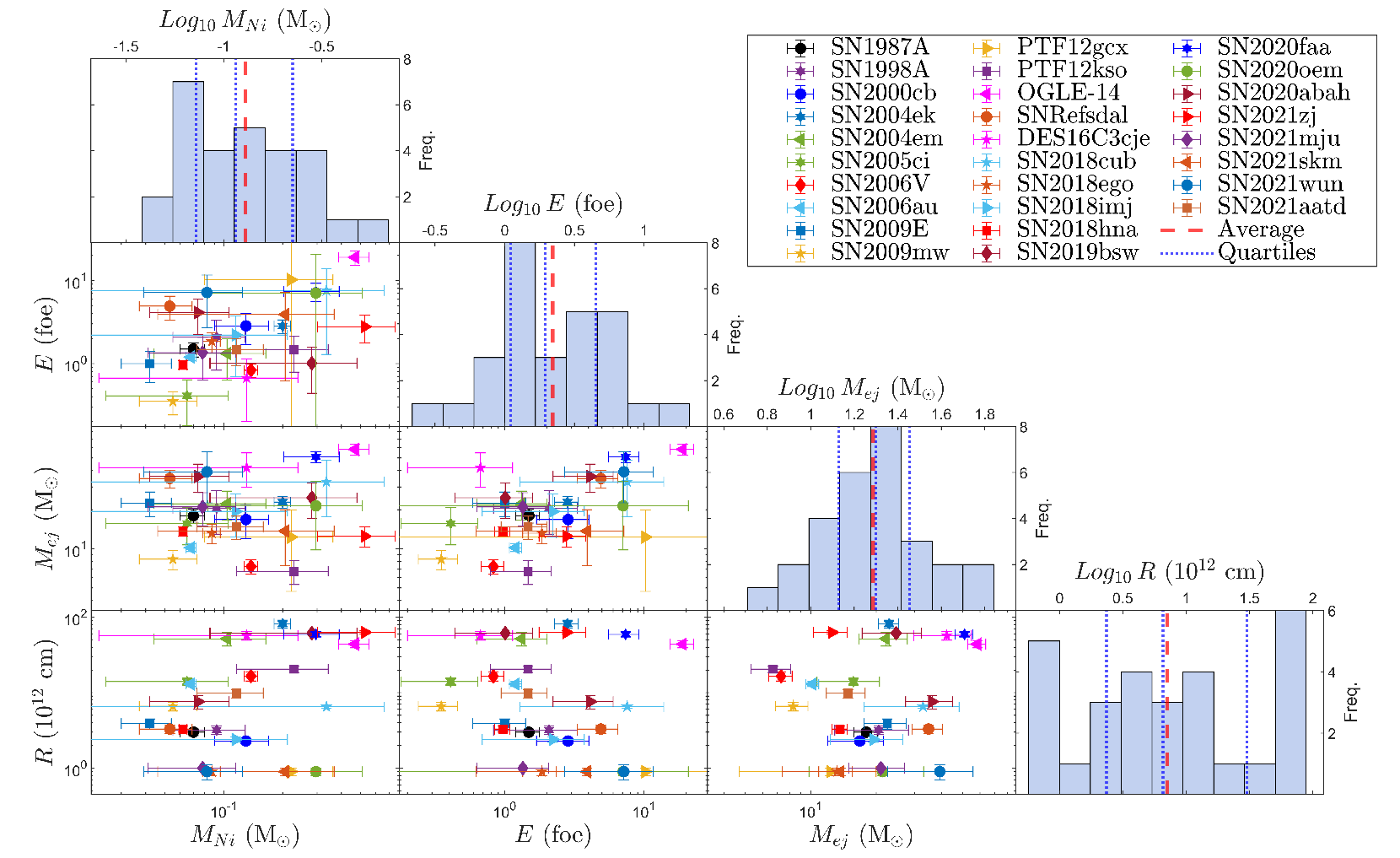} 
    \caption{Distribution of SNe physical properties. The diagonal panels show the histograms of the logarithm of each parameter ($M_{\rm Ni}$, $E$, $M_{\rm ej}$, $R$), with the mean value (dashed line) and quartiles (25\%, 50\%, and 75\%; see dotted lines) indicated.  The lower triangular panels show scatter plots of the parameter pairs for each SN in the sample.}
    \label{Fig:Parameters}
\end{figure*}

A second group, made by PTF12gcx, PTF12kso, DES16C3cje, SN~2018cub, SN~2018imj, SN~2020oem, SN~2020abah, SN~2021zj, SN~2021skm, and SN~2021wun, also yields apparently excellent fits. However, in some of these cases the error bars on the reconstructed bolometric flux are significantly large, that reduce the $\chi^2_\nu\lesssim 0.5$, limiting so the statistical robustness of the LC comparison and parameters inference.

Other SNe display discrepancies mainly due to limited temporal coverage or underestimated photometric errors. For instance, SN~2004ek, SN~2006au, and SN~2018ego suffer from sparse data and very small luminosity uncertainties, which drive the $\chi^2_\nu$ to high values (grater than 3) despite an overall model-consistent evolution. SN~Refsdal is fairly well reproduced around peak luminosity, but the lack of late–time data prevents a reliable constraint on the end of recombination.
In some cases, the models clearly overfit or misinterpret parts of the LC. SN~2005ci shows spurious overfitting of the early rise, while SN2019bsw and SN~2021mju suffer from noisy data that lead the procedure to confuse the late plateau with the onset of the tail, introducing large uncertainties in all explosive parameters.

The two most luminous events in the sample, OGLE-14 and SN~2020faa ($L_{\rm M}\sim10^{43}$ erg s$^{-1}$), both return reduced $\chi^2_\nu$ values well above unity, highlighting the difficulty of reproducing the entire peak solely through the contribution of radioactive $^{56}$Co. In OGLE-14, the higher $\chi^2_\nu$ value is mainly due to the deviation seen at the end of recombination, which could be explained by a different ejecta density structure or an additional energy source placed inside. For SN~2020faa, the deviation is further supported by a change of slope in the nebular tail, suggesting the presence of an additional power source. A plausible interpretation is energy input from a hidden shock–powered mechanism, possibly magnetar–driven \citep[see also][]{Salmaso2023}, contributing to the luminosity at nebular phases. Interestingly, DES16C3cje also deviates from the expected $^{56}$Co decline rate and is instead compatible with the shallower slope expected from accretion–powered emission \citep{gutierrez20}.

Finally, several SNe show an early–time luminosity excess (within $\sim$30 days of explosion) relative to the adopted model, which does not include contributions from thin-shell emission \citepalias[see details in][]{PC2025} or shock interaction with a dense CSM \citep[e.g.,][]{Cosentino25}. Out of the 28 SNe analyzed, 13 exhibit this behavior, with particularly striking examples being SN2021zj and SN2021aatd. In SN2021zj, the interaction scenario is supported by the presence of narrow emission lines in the spectra \citep{JacobsonG2024}, while in SN2021aatd no narrow features were detected beyond the first two days. However, the absence of narrow lines does not rule out CSM interaction as the underlying mechanism \citep[e.g.,][and references therein]{KK24}.

\subsection{Physical properties of the 87A-like class}\label{SubSec:PhysicalProp}
The physical parameters derived from the \textsc{SuperBAM} modeling for the \Alike~sample are listed in Tab.~\ref{Tab:Modeling}. When considered as a whole, they reveal a rather heterogeneous population, with asymmetric distributions in explosion energy and $^{56}$Ni mass (see Fig.~\ref{Fig:Parameters}). In contrast, the distribution of ejecta masses is nearly symmetric, with the mean, median, and mode all close to $\sim$19\,M$_\odot$. Explosion energies and $^{56}$Ni masses, however, display a mode shifted toward lower values, with 8 SNe clustering below the sample averages. This points to the presence of two broad subgroups of 87A-like events: one characterized by modest energies ($\sim$1.2 foe) and $^{56}$Ni masses around 0.07\,M$_\odot$ (including SN~1987A, SN~2009E, etc.), and another, more energetic, group peaking around 4 foe and $^{56}$Ni masses above 0.11\,M$_\odot$. About one quarter of the sample exhibits explosion energies exceeding 5 foe, reaching values that are difficult to reconcile with standard CC mechanisms (e.g., OGLE-14 and SN~2020faa). 

By mapping energy, ejecta mass, and $^{56}$Ni together, the extended sample allows us to better outline the parameter space of \Alike~SNe. The resulting distribution appears continuous, but still allows events to be broadly grouped into two categories: a high-$E$, high-$^{56}$Ni subgroup with ejecta masses $\gtrsim$20\,M$_\odot$ (e.g., OGLE-14), and another, more similar to SN~1987A, with lower ejecta masses ($<$20\,M$_\odot$) and explosion energies spanning a wide range from $\sim$0.3 foe up to $\sim$2 foe, in line with the trends already reported in \citetalias{pumo23}.

Similarly to ejecta masses, progenitor radii at the explosion time show a roughly symmetric distribution centered around $(6-7)\times 10^{12}$\,cm, but with two secondary tails: one at smaller radii of order $10^{12}$\,cm, and another at larger radii, beyond $3\times 10^{13}$\,cm. The low-radius tail includes events such as SN~2018ego, SN~2020oem, PTF12gcx, SN~2021mju, SN~2021skm, and SN~2021wun. However, these objects generally suffer from poor or noisy coverage in the early phases or near the end of recombination, which likely makes the radius estimates uncertain and possibly underestimated. Conversely, the high-radius tail contains several well-sampled and robustly modeled events (e.g., SN~2004em, SN~2004ek, OGLE-14, and SN~2021zj).
Interestingly, the highest-radius events are also those with the largest explosion energies and $^{56}$Ni masses. Moreover, up to $\sim7\times10^{12}$\,cm the $^{56}$Ni mass tends to decrease with increasing radius, reaching a minimum for SN~2009E, and then rises again for the most extended progenitors. This trend is consistent with the intrinsic nature of \Alike~SN LC, where the slow rise of the second peak depends sensitively on the initial radius and on the $^{56}$Ni yield. This behavior further distinguishes the group with lower radii and energy from the more energetic and extended one, which would otherwise resemble standard SNe~II were it not for their unusually high $^{56}$Ni content
\citep[average $^{56}$Ni mass for SN~II population is $\sim0.037 \pm 0.005\,\mathrm{{\rm M}_\odot}$;][]{2021MNRAS.505.1742R}.

The extended parameter space inferred from our modeling also enables a qualitative comparison with recent stellar-evolution predictions for BSG progenitors. In particular, the inferred explosion energies and $^{56}$Ni masses follow the exponential trend expected from neutrino-driven explosion models \citep[e.g.,][]{2025A&A...700A.253S}, while the broad range of ejecta masses and progenitor radii overlaps with those predicted by recent grids of pre-SN supergiant models \citep{2024A&A...686A..45S}. However, most single-star models at solar metallicity struggle to produce sufficiently compact and massive blue supergiants, confirming that either lower metallicity evolution \citep[e.g.,][]{1988ApJ...330..218W} or binary interaction and merger channels \citep[e.g.,][]{2025arXiv250821116T} may be required to explain the \Alike~SNe. Moreover, the most energetic and $^{56}$Ni-rich events in our sample may also invite comparison with pair-instability progenitor models \citep[e.g.,][]{terreran17}, which naturally predict large explosion energies and nickel yields \citep[e.g.,][]{Kasen_2011,Dessart_2012}.

Finally, note that the explosion parameters derived by \textsc{SuperBAM} are consistent with those obtained in other previous works (see Table~3 of \citetalias{pumo23} and references therein), confirming the reliability of our approach (see also Sect. \ref{SubSec:test}).
In Sect.~\ref{SubSec:correlation}, a detailed analysis of correlations among these physical parameters, as well as between the physical and spectrophotometric properties of the sample, is presented. 

\begin{figure}
\includegraphics[width=1\columnwidth]{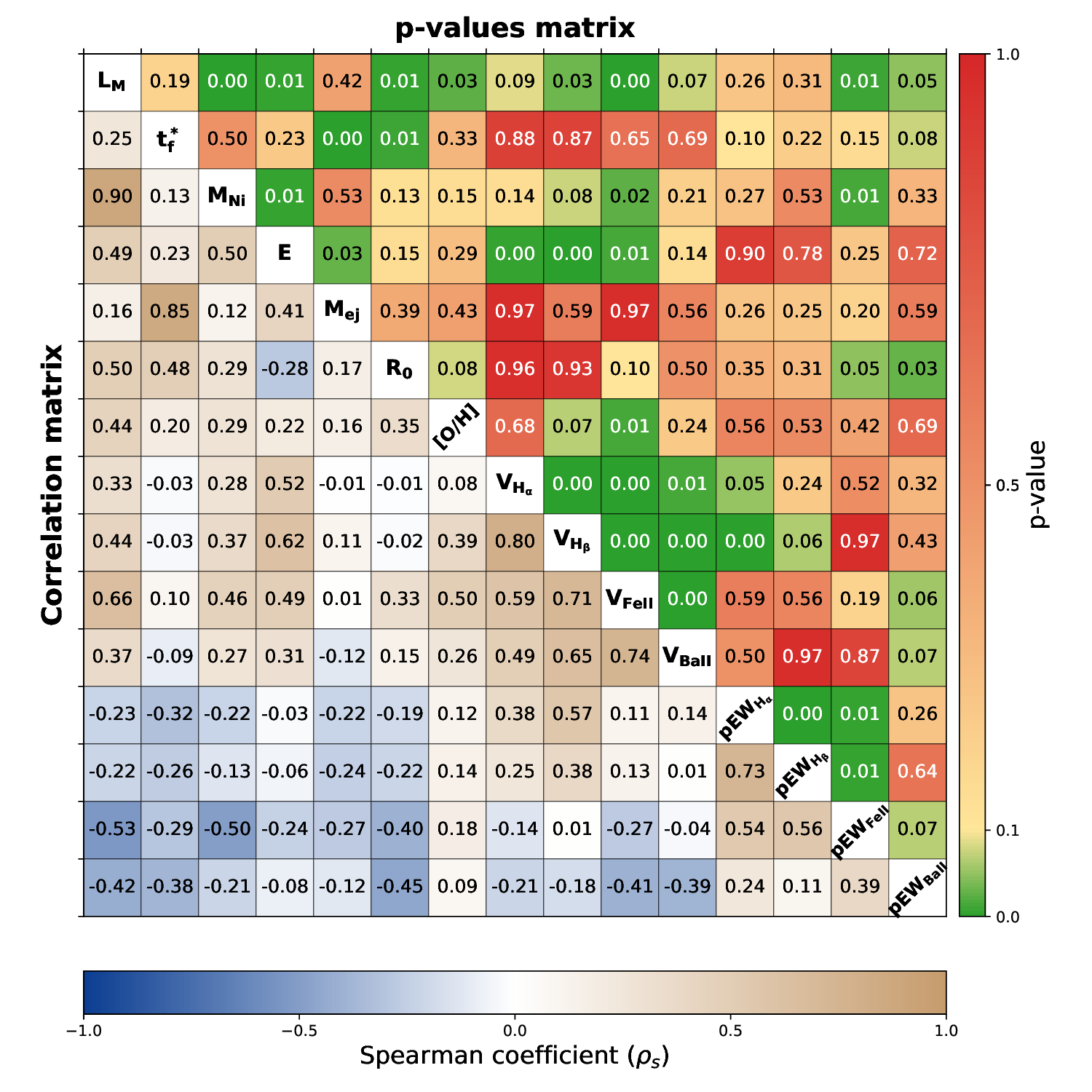}
    \caption{Spearman correlation matrix for the physical and spectroscopic parameters considered. The lower-left half of squares show the Spearman correlation coefficient ($\rho_c$), color-coded from –1 (dark blue) to +1 (light brown), with the numerical values reported inside each cell. The upper-right half of squares show the associated p-values, color-coded according to significance level (from green for high significance to red for non–significant values). The diagonal reports the labels of the parameters. In addition to the parameters discussed in the previous sections, we also include [O/H]=
12+log(O/H) listed in Table \ref{Tab:Info_SNe87A}.}
    \label{Fig:Correlations}
\end{figure}
\subsection{Correlations}\label{SubSec:correlation}

To investigate possible statistical links between the explosion properties and the spectrophotometric features of the \Alike~SN sample, we computed the Spearman correlation coefficient ($\rho_c$) and evaluated its significance with a $t$-test for each parameter combination (see Fig.~\ref{Fig:Correlations}).
As already discussed in Sect.~\ref{SubSec:Bolo_lum}, the $^{56}$Ni mass correlates positively ($\rho_c>0$) with the peak luminosity $L_{\rm M}$, with very high significance ($p \ll 0.01$). $M_{\rm Ni}$ also shows a positive correlation with the explosion energy ($p=0.01$), in agreement with theoretical expectations that link larger nickel yields to more energetic explosions \citep[e.g.,][]{2025A&A...700A.253S}. From the spectroscopic side, $M_{\rm Ni}$ appears correlated with the Fe\,{\sc ii} line, both in velocity ($p=0.02$) and in pEW ($p=0.02$). Interestingly, higher nickel masses are associated with weaker Fe\,{\sc ii} absorption but faster line velocities. A possible interpretation is that larger amounts of $^{56}$Ni keep the recombination front and the photosphere at larger radii in the ejecta \citepalias[see, e.g.,][]{PC2025}, so that Fe\,{\sc ii} lines form in faster-moving layers where the Fe abundance is intrinsically lower. This effect would explain the counter-intuitive result that Ni-rich explosions show weaker Fe\,{\sc ii} features, not because of a reduced Fe content, but because the absorption originates in more external and less Fe-rich regions. Such behavior suggests that scaling relations for line velocities may need to explicitly account for the nickel mass.
By contrast, the ejecta mass is strongly correlated with the recombination timescale $t_{\rm f}^*$, as expected \citepalias{PC2025}, and shows a moderately significant trend with explosion energy ($p=0.03$), but no direct correlation with $^{56}$Ni mass. Spectroscopically, $M_{\rm ej}$ does not display connections with line properties.

As anticipated in Sect.~\ref{SubSec:spectra}, line velocities are mutually correlated, as are velocity widths, with the exception of Ba\,{\sc ii}, which does not follow the same trends. In fact, Ba\,{\sc ii} shows a particularly peculiar behavior. An anti-correlation emerges between its velocity and pEW, which becomes statistically significant ($p < 0.05$) once clear outliers (e.g., DES16C3cje) are excluded. This trend is consistent with the idea that more compact explosions, with less extended ejecta, retain higher densities in the line-forming regions and therefore favor stronger Ba\,{\sc ii} absorption \citep[e.g.,][]{1995A&A...303..118M}. In support of this interpretation, the modeling results indicate an additional correlation with the progenitor radius at explosion, which appears to be inversely related to the Ba\,{\sc ii}'s pEW. On the other hand, no correlation was found with other explosive parameters such as $E$ and $M_{\rm Ni}$. This suggests that the key factor regulating the intensity of the Ba\,{\sc ii} lines should be the compactness of the progenitors and ejecta, rather than differences in their nucleosynthetic production.

Progenitor radius also exhibits weak correlations with other parameters. In particular, an expected correlation between $R_0$ and the oxygen abundance [O/H] of the host environment is present in our analysis \citep[e.g.,][]{Taddia13,taddia16}. Curiously, metallicity itself shows possible links with both peak luminosity and expansion velocity. SNe in more metal-rich environments tend to expand faster and peak at higher luminosities. Although the significance is low, these trend could provide insights into the role of metallicity in shaping the properties of \Alike~events.

However we note that correlation does not imply causation. Moreover, although the present sample is significantly larger than in previous works, it could not constitute a fully unbiased representation of the \Alike~population. Selection effects linked to discovery methods, survey strategies, and spectroscopic follow-up may introduce statistical biases that affect both distributions and correlations. These results should therefore be used with caution, although they will gain robustness as the number of well-observed \Alike~SNe increases in the coming years.

\section{Summary and further consideration}\label{Sec:Conclusion}
In this work we introduced the novel procedure \textsc{SuperBAM} (Supernova Bayesian Analytic Modeling), a fast-modeling framework designed to infer the main physical properties of H-rich SNe from their bolometric LCs and spectroscopic information during the recombination phase. \textsc{SuperBAM} relies on Bayesian statistics and analytical prescriptions, providing posterior distributions for explosion energy, ejecta mass, progenitor radius, and synthesized $^{56}$Ni mass. The method was validated against well-studied benchmarks and shown to reproduce the results of detailed numerical hydrodynamic simulations at a fraction of their computational cost.

We then applied \textsc{SuperBAM} to the most comprehensive sample of 87A-like SNe assembled so far (28 events), reconstructing their bolometric LCs in a homogeneous way and measuring spectroscopic features at maximum. The analysis reveals a heterogeneous population that nevertheless clusters into two broad groups: (i) lower-energy explosions with modest $^{56}$Ni yields, reminiscent of SN~1987A, and (ii) more energetic events (sometimes exceeding 5 foe) with higher nickel production and, in some cases, unusually large progenitor radii. The parameter space, however, appears continuous, suggesting a spectrum of possible progenitor/explosion configurations rather than discrete categories.

From the statistical comparison of physical properties and spectrophotometric observables, we confirm the robust correlation between $^{56}$Ni mass and both peak luminosity and explosion energy, as well as the strong link between ejecta mass and recombination timescale. We also identify a peculiar behavior of Ba\,{\sc ii}, whose line strength anti-correlates with velocity and appears inversely related to progenitor radius. This suggests that the prominent Ba absorption observed in many 87A-like events arises primarily from progenitor compactness and ejecta density, rather than from enhanced nucleosynthetic production. Notably, for progenitors with radii approaching $\sim 10^{13}$ cm, the Ba\,{\sc ii} strength converges toward that of standard SNe~II \citep[pEW$_{\rm Ba\,{\sc II}}\sim 5\,$\AA; cf. ][]{2017ApJ...850...90G}, supporting the interpretation that compactness, and not abundance anomalies, could be the key driver. Future progress on this front will require detailed spectral synthesis calculations that account for the thermodynamical state of the ejecta (temperature and density), in order to disentangle abundance from excitation effects.

Finally, several events display luminosity components not fully reproduced by standard recombination plus $^{56}$Ni heating. The brightest SNe in our sample (e.g., OGLE-14 and SN~2020faa) require additional power sources such as magnetar spin-down or accretion, while others show early-time excesses consistent with interaction with CSM (e.g., SN 2021zj and SN 2021aatd). This highlights the need to extend fast-modeling frameworks similar to \textsc{SuperBAM} to incorporate non-radioactive energy sources and CSM interaction, in order to capture the full diversity of 87A-like explosions. In particular, the inclusion of CSM-related components would allow us to constrain not only the explosion parameters but also the physical properties of the surrounding material. Combined with detailed spectroscopic analysis, this approach would provide key insights into the progenitor mass-loss history and the pre-explosion evolution leading to events such as SN~2021zj. A dedicated study of this object, including unpublished spectroscopic observations, is planned for future work.

The development of such efficient, Bayesian-based procedures is particularly timely in view of ongoing and upcoming large surveys (e.g., ZTF and LSST), which will discover thousands of peculiar SNe. \textsc{SuperBAM} represents a step toward scalable analysis pipelines, able to rapidly characterize large samples with improved accuracy over simple scaling relations and at a fraction of the cost of full numerical simulations. Moreover, our framework has proven effective even in the absence of spectroscopic information, providing physically meaningful estimates based solely on photometric data. This feature is particularly valuable for next-generation surveys, which will deliver extensive multi-band photometry but limited spectroscopic coverage.
In this way, this work opens the path for systematic population studies of rare transients, bridging the gap between survey data and detailed physical modeling.

\begin{acknowledgements}
    
We acknowledge support from the Piano di Ricerca di Ateneo UNICT – Linea 2 PIA.CE.RI. 2024–2026 of Catania University (project AstroCosmo, P.I. A. Lanzafame). CI gratefully acknowledges the support received from the MERAC Foundation.
SPC thanks Cardiff University for hosting him during the research activities that contributed to this work.
      
\end{acknowledgements}

\bibliographystyle{aa}
\bibliography{biblio} 

@ARTICLE{1988ApJ...330..218W,
       author = {{Woosley}, S.~E.},
        title = "{SN 1987A: After the Peak}",
      journal = {\apj},
         year = 1988,
        month = jul,
       volume = {330},
        pages = {218},
          doi = {10.1086/166468},
       adsurl = {https://ui.adsabs.harvard.edu/abs/1988ApJ...330..218W}
}

@article{Dessart_2012,
    author = {Dessart, Luc and Waldman, Roni and Livne, Eli and Hillier, D. John and Blondin, Stéphane},
    title = {Radiative properties of pair-instability supernova explosions},
    journal = {\mnras},
    volume = {428},
    number = {4},
    pages = {3227-3251},
    year = {2012},
    month = {11},
    issn = {0035-8711},
    doi = {10.1093/mnras/sts269},
    url = {https://doi.org/10.1093/mnras/sts269},
    eprint = {https://academic.oup.com/mnras/article-pdf/428/4/3227/18463652/sts269.pdf},
}

@article{Kasen_2011,
doi = {10.1088/0004-637X/734/2/102},
url = {https://doi.org/10.1088/0004-637X/734/2/102},
year = {2011},
month = {jun},
publisher = {The American Astronomical Society},
volume = {734},
number = {2},
pages = {102},
author = {Kasen, Daniel and Woosley, S. E. and Heger, Alexander},
title = {PAIR INSTABILITY SUPERNOVAE: LIGHT CURVES, SPECTRA, AND SHOCK BREAKOUT},
journal = {\apj}
}

@ARTICLE{2025arXiv250821116T,
       author = {{Tsuna}, Daichi and {Fuller}, Jim and {Lu}, Wenbin},
        title = "{Fates of Rotating Supergiants from Stellar Mergers and the Landscape of Transients upon Core-collapse}",
      journal = {arXiv e-prints},
         year = 2025,
        month = aug,
          eid = {arXiv:2508.21116},
        pages = {arXiv:2508.21116},
          doi = {10.48550/arXiv.2508.21116},
archivePrefix = {arXiv},
       eprint = {2508.21116},
 primaryClass = {astro-ph.HE},
       adsurl = {https://ui.adsabs.harvard.edu/abs/2025arXiv250821116T}
}

@ARTICLE{2024A&A...686A..45S,
       author = {{Schneider}, F.~R.~N. and {Podsiadlowski}, Ph. and {Laplace}, E.},
        title = "{Pre-supernova evolution and final fate of stellar mergers and accretors of binary mass transfer}",
      journal = {\aap},
         year = 2024,
        month = jun,
       volume = {686},
          eid = {A45},
        pages = {A45},
          doi = {10.1051/0004-6361/202347854},
archivePrefix = {arXiv},
       eprint = {2403.03984},
 primaryClass = {astro-ph.SR},
       adsurl = {https://ui.adsabs.harvard.edu/abs/2024A&A...686A..45S}
}

@ARTICLE{2025A&A...700A.253S,
       author = {{Schneider}, F.~R.~N. and {Laplace}, E. and {Podsiadlowski}, Ph.},
        title = "{Supernovae from stellar mergers and accretors of binary mass transfer: Implications for Type IIP, 1987A-like and interacting supernovae}",
      journal = {\aap},
         year = 2025,
        month = aug,
       volume = {700},
          eid = {A253},
        pages = {A253},
          doi = {10.1051/0004-6361/202554362},
archivePrefix = {arXiv},
       eprint = {2507.06391},
 primaryClass = {astro-ph.HE},
       adsurl = {https://ui.adsabs.harvard.edu/abs/2025A&A...700A.253S}
}

@ARTICLE{1994ApJ...427..874C,
       author = {{Colgan}, Sean W.~J. and {Haas}, Michael R. and {Erickson}, Edwin F. and {Lord}, Steven D. and {Hollenbach}, David J.},
        title = "{Day 640 Infrared Line and Continuum Measurements: Dust Formation in SN 1987A}",
      journal = {\apj},
         year = 1994,
        month = jun,
       volume = {427},
        pages = {874},
          doi = {10.1086/174193},
       adsurl = {https://ui.adsabs.harvard.edu/abs/1994ApJ...427..874C}
}

@ARTICLE{Martinez22,
       author = {{Martinez}, L. and {Anderson}, J.~P. and {Bersten}, M.~C. and {Hamuy}, M. and {Gonz{\'a}lez-Gait{\'a}n}, S. and {Orellana}, M. and {Stritzinger}, M. and {Phillips}, M.~M. and {Guti{\'e}rrez}, C.~P. and {Burns}, C. and {de Jaeger}, T. and {Ertini}, K. and {Folatelli}, G. and {F{\"o}rster}, F. and {Galbany}, L. and {Hoeflich}, P. and {Hsiao}, E.~Y. and {Morrell}, N. and {Pessi}, P.~J. and {Suntzeff}, N.~B.},
        title = "{Type II supernovae from the Carnegie Supernova Project-I. III. Understanding SN II diversity through correlations between physical and observed properties}",
      journal = {\aap},
         year = 2022,
        month = apr,
       volume = {660},
          eid = {A42},
        pages = {A42},
          doi = {10.1051/0004-6361/202142555},
archivePrefix = {arXiv},
       eprint = {2202.11220},
 primaryClass = {astro-ph.SR},
       adsurl = {https://ui.adsabs.harvard.edu/abs/2022A&A...660A..42M}
}

@ARTICLE{2014MNRAS.437.3848L,
       author = {{Lyman}, J.~D. and {Bersier}, D. and {James}, P.~A.},
        title = "{Bolometric corrections for optical light curves of core-collapse supernovae}",
      journal = {\mnras},
         year = 2014,
        month = feb,
       volume = {437},
       number = {4},
        pages = {3848-3862},
          doi = {10.1093/mnras/stt2187},
archivePrefix = {arXiv},
       eprint = {1311.1946},
 primaryClass = {astro-ph.HE},
       adsurl = {https://ui.adsabs.harvard.edu/abs/2014MNRAS.437.3848L}
}

@article{Young_2004,
doi = {10.1086/425675},
url = {https://doi.org/10.1086/425675},
year = {2004},
month = {dec},
publisher = {},
volume = {617},
number = {2},
pages = {1233},
author = {Young, Timothy R.},
title = {A Parameter Study of Type II Supernova Light Curves Using 6 M☉ He Cores},
journal = {\apj}
}

@article{Utrobin_2021,
doi = {10.3847/1538-4357/abf4c5},
url = {https://doi.org/10.3847/1538-4357/abf4c5},
year = {2021},
month = {jun},
publisher = {The American Astronomical Society},
volume = {914},
number = {1},
pages = {4},
author = {Utrobin, V. P. and Wongwathanarat, A. and Janka, H.-Th. and Müller, E. and Ertl, T. and Menon, A. and Heger, A.},
title = {Supernova 1987A: 3D Mixing and Light Curves for Explosion Models Based on Binary-merger Progenitors},
journal = {\apj}
}

@ARTICLE{2021MNRAS.505.1742R,
       author = {{Rodr{\'\i}guez}, {\'O}. and {Meza}, N. and {Pineda-Garc{\'\i}a}, J. and {Ramirez}, M.},
        title = "{The iron yield of normal Type II supernovae}",
      journal = {\mnras},
         year = 2021,
        month = aug,
       volume = {505},
       number = {2},
        pages = {1742-1774},
          doi = {10.1093/mnras/stab1335},
archivePrefix = {arXiv},
       eprint = {2105.03268},
 primaryClass = {astro-ph.SR},
       adsurl = {https://ui.adsabs.harvard.edu/abs/2021MNRAS.505.1742R}
}

@ARTICLE{2021TNSAN..14....1S,
       author = {{Smith}, K.~W. and {Fulton}, M. and {Gillanders}, J. and {Srivastav}, S. and {Barbarino}, C. and {Yang}, S. and {Chen}, T.~W. and {Arcavi}, I. and {Gutierrez}, C. and {Anderson}, J. and {Bravo}, T.~M. and {Gromadzki}, M. and {Inserra}, C. and {Kankare}, E. and {Nicholl}, M. and {Yaron}, O. and {Manulis}, I. and {Young}, D. and {Tonry}, J. and {Denneau}, L. and {Heinze}, A. and {Weiland}, H. and {Stalder}, B. and {Rest}, A. and {Smartt}, S.~J. and {McBrien}, O.},
        title = "{ePESSTO+ spectroscopic classification of optical transients}",
      journal = {TNS AstroNote},
         year = 2021,
        month = jan,
       volume = {14},
        pages = {1-14},
       adsurl = {https://ui.adsabs.harvard.edu/abs/2021TNSAN..14....1S}
}

@article{Tonry_2018,
doi = {10.1088/1538-3873/aabadf},
url = {https://doi.org/10.1088/1538-3873/aabadf},
year = {2018},
month = {may},
publisher = {The Astronomical Society of the Pacific},
volume = {130},
number = {988},
pages = {064505},
author = {Tonry, J. L. and Denneau, L. and Heinze, A. N. and Stalder, B. and Smith, K. W. and Smartt, S. J. and Stubbs, C. W. and Weiland, H. J. and Rest, A.},
title = {ATLAS: A High-cadence All-sky Survey System},
journal = {\pasp}
}

@ARTICLE{2017ApJ...850...90G,
       author = {{Guti{\'e}rrez}, Claudia P. and {Anderson}, Joseph P. and {Hamuy}, Mario and {Gonz{\'a}lez-Gaitan}, Santiago and {Galbany}, Lluis and {Dessart}, Luc and {Stritzinger}, Maximilian D. and {Phillips}, Mark M. and {Morrell}, Nidia and {Folatelli}, Gast{\'o}n},
        title = "{Type II Supernova Spectral Diversity. II. Spectroscopic and Photometric Correlations}",
      journal = {\apj},
         year = 2017,
        month = nov,
       volume = {850},
       number = {1},
          eid = {90},
        pages = {90},
          doi = {10.3847/1538-4357/aa8f42},
archivePrefix = {arXiv},
       eprint = {1709.02799},
 primaryClass = {astro-ph.HE},
       adsurl = {https://ui.adsabs.harvard.edu/abs/2017ApJ...850...90G}
}

@ARTICLE{1995A&A...303..118M,
       author = {{Mazzali}, P.~A. and {Chugai}, N.~N.},
        title = "{Barium in SN 1987A and SNe II-P.}",
      journal = {\aap},
         year = 1995,
        month = nov,
       volume = {303},
        pages = {118},
       adsurl = {https://ui.adsabs.harvard.edu/abs/1995A&A...303..118M}
}

@ARTICLE{KK24,
       author = {{Khatami}, David K. and {Kasen}, Daniel N.},
        title = "{The Landscape of Thermal Transients from Supernovae Interacting with a Circumstellar Medium}",
      journal = {\apj},
         year = 2024,
        month = sep,
       volume = {972},
       number = {2},
          eid = {140},
        pages = {140},
          doi = {10.3847/1538-4357/ad60c0},
archivePrefix = {arXiv},
       eprint = {2304.03360},
 primaryClass = {astro-ph.HE},
       adsurl = {https://ui.adsabs.harvard.edu/abs/2024ApJ...972..140K}
}

@INPROCEEDINGS{IRAF_86,
       author = {{Tody}, Doug},
        title = "{The IRAF Data Reduction and Analysis System}",
    booktitle = {Instrumentation in astronomy VI},
         year = 1986,
       editor = {{Crawford}, David L.},
       series = {Society of Photo-Optical Instrumentation Engineers (SPIE) Conference Series},
       volume = {627},
        month = jan,
        pages = {733},
          doi = {10.1117/12.968154},
       adsurl = {https://ui.adsabs.harvard.edu/abs/1986SPIE..627..733T}
}

@ARTICLE{2020A&A...641A...6P,
       author = {{Planck Collaboration} and {Aghanim}, N. and {Akrami}, Y. and {Ashdown}, M. and {Aumont}, J. and {Baccigalupi}, C. and {Ballardini}, M. and {Banday}, A.~J. and {Barreiro}, R.~B. and {Bartolo}, N. and {Basak}, S. and {Battye}, R. and {Benabed}, K. and {Bernard}, J. -P. and {Bersanelli}, M. and {Bielewicz}, P. and {Bock}, J.~J. and {Bond}, J.~R. and {Borrill}, J. and {Bouchet}, F.~R. and {Boulanger}, F. and {Bucher}, M. and {Burigana}, C. and {Butler}, R.~C. and {Calabrese}, E. and {Cardoso}, J. -F. and {Carron}, J. and {Challinor}, A. and {Chiang}, H.~C. and {Chluba}, J. and {Colombo}, L.~P.~L. and {Combet}, C. and {Contreras}, D. and {Crill}, B.~P. and {Cuttaia}, F. and {de Bernardis}, P. and {de Zotti}, G. and {Delabrouille}, J. and {Delouis}, J. -M. and {Di Valentino}, E. and {Diego}, J.~M. and {Dor{\'e}}, O. and {Douspis}, M. and {Ducout}, A. and {Dupac}, X. and {Dusini}, S. and {Efstathiou}, G. and {Elsner}, F. and {En{\ss}lin}, T.~A. and {Eriksen}, H.~K. and {Fantaye}, Y. and {Farhang}, M. and {Fergusson}, J. and {Fernandez-Cobos}, R. and {Finelli}, F. and {Forastieri}, F. and {Frailis}, M. and {Fraisse}, A.~A. and {Franceschi}, E. and {Frolov}, A. and {Galeotta}, S. and {Galli}, S. and {Ganga}, K. and {G{\'e}nova-Santos}, R.~T. and {Gerbino}, M. and {Ghosh}, T. and {Gonz{\'a}lez-Nuevo}, J. and {G{\'o}rski}, K.~M. and {Gratton}, S. and {Gruppuso}, A. and {Gudmundsson}, J.~E. and {Hamann}, J. and {Handley}, W. and {Hansen}, F.~K. and {Herranz}, D. and {Hildebrandt}, S.~R. and {Hivon}, E. and {Huang}, Z. and {Jaffe}, A.~H. and {Jones}, W.~C. and {Karakci}, A. and {Keih{\"a}nen}, E. and {Keskitalo}, R. and {Kiiveri}, K. and {Kim}, J. and {Kisner}, T.~S. and {Knox}, L. and {Krachmalnicoff}, N. and {Kunz}, M. and {Kurki-Suonio}, H. and {Lagache}, G. and {Lamarre}, J. -M. and {Lasenby}, A. and {Lattanzi}, M. and {Lawrence}, C.~R. and {Le Jeune}, M. and {Lemos}, P. and {Lesgourgues}, J. and {Levrier}, F. and {Lewis}, A. and {Liguori}, M. and {Lilje}, P.~B. and {Lilley}, M. and {Lindholm}, V. and {L{\'o}pez-Caniego}, M. and {Lubin}, P.~M. and {Ma}, Y. -Z. and {Mac{\'\i}as-P{\'e}rez}, J.~F. and {Maggio}, G. and {Maino}, D. and {Mandolesi}, N. and {Mangilli}, A. and {Marcos-Caballero}, A. and {Maris}, M. and {Martin}, P.~G. and {Martinelli}, M. and {Mart{\'\i}nez-Gonz{\'a}lez}, E. and {Matarrese}, S. and {Mauri}, N. and {McEwen}, J.~D. and {Meinhold}, P.~R. and {Melchiorri}, A. and {Mennella}, A. and {Migliaccio}, M. and {Millea}, M. and {Mitra}, S. and {Miville-Desch{\^e}nes}, M. -A. and {Molinari}, D. and {Montier}, L. and {Morgante}, G. and {Moss}, A. and {Natoli}, P. and {N{\o}rgaard-Nielsen}, H.~U. and {Pagano}, L. and {Paoletti}, D. and {Partridge}, B. and {Patanchon}, G. and {Peiris}, H.~V. and {Perrotta}, F. and {Pettorino}, V. and {Piacentini}, F. and {Polastri}, L. and {Polenta}, G. and {Puget}, J. -L. and {Rachen}, J.~P. and {Reinecke}, M. and {Remazeilles}, M. and {Renzi}, A. and {Rocha}, G. and {Rosset}, C. and {Roudier}, G. and {Rubi{\~n}o-Mart{\'\i}n}, J.~A. and {Ruiz-Granados}, B. and {Salvati}, L. and {Sandri}, M. and {Savelainen}, M. and {Scott}, D. and {Shellard}, E.~P.~S. and {Sirignano}, C. and {Sirri}, G. and {Spencer}, L.~D. and {Sunyaev}, R. and {Suur-Uski}, A. -S. and {Tauber}, J.~A. and {Tavagnacco}, D. and {Tenti}, M. and {Toffolatti}, L. and {Tomasi}, M. and {Trombetti}, T. and {Valenziano}, L. and {Valiviita}, J. and {Van Tent}, B. and {Vibert}, L. and {Vielva}, P. and {Villa}, F. and {Vittorio}, N. and {Wandelt}, B.~D. and {Wehus}, I.~K. and {White}, M. and {White}, S.~D.~M. and {Zacchei}, A. and {Zonca}, A.},
        title = "{Planck 2018 results. VI. Cosmological parameters}",
      journal = {\aap},
         year = 2020,
        month = sep,
       volume = {641},
          eid = {A6},
        pages = {A6},
          doi = {10.1051/0004-6361/201833910},
archivePrefix = {arXiv},
       eprint = {1807.06209},
 primaryClass = {astro-ph.CO},
       adsurl = {https://ui.adsabs.harvard.edu/abs/2020A&A...641A...6P}
}

@article{Nicholl_2018,
doi = {10.3847/2515-5172/aaf799},
url = {https://dx.doi.org/10.3847/2515-5172/aaf799},
year = {2018},
month = {dec},
publisher = {The American Astronomical Society},
volume = {2},
number = {4},
pages = {230},
author = {Matt Nicholl},
title = {SuperBol: A User-friendly Python Routine for Bolometric Light Curves},
journal = {Research Notes of the AAS}
}

@article{Bianco_2022,
doi = {10.3847/1538-4365/ac3e72},
url = {https://dx.doi.org/10.3847/1538-4365/ac3e72},
year = {2021},
month = {dec},
publisher = {The American Astronomical Society},
volume = {258},
number = {1},
pages = {1},
author = {Federica B. Bianco and Željko Ivezić and R. Lynne Jones and Melissa L. Graham and Phil Marshall and Abhijit Saha and Michael A. Strauss and Peter Yoachim and Tiago Ribeiro and Timo Anguita and A. E. Bauer and Franz E. Bauer and Eric C. Bellm and Robert D. Blum and William N. Brandt and Sarah Brough and Márcio Catelan and William I. Clarkson and Andrew J. Connolly and Eric Gawiser and John E. Gizis and Renée Hložek and Sugata Kaviraj and Charles T. Liu and Michelle Lochner and Ashish A. Mahabal and Rachel Mandelbaum and Peregrine McGehee and Eric H. Neilsen Jr. and Knut A. G. Olsen and Hiranya V. Peiris and Jason Rhodes and Gordon T. Richards and Stephen Ridgway and Megan E. Schwamb and Dan Scolnic and Ohad Shemmer and Colin T. Slater and Anže Slosar and Stephen J. Smartt and Jay Strader and Rachel Street and David E. Trilling and Aprajita Verma and A. K. Vivas and Risa H. Wechsler and Beth Willman},
title = {Optimization of the Observing Cadence for the Rubin Observatory Legacy Survey of Space and Time: A Pioneering Process of Community-focused Experimental Design},
journal = {\apjs}
}

@phdthesis{Cosentino2024,
  author       = {Cosentino, Stefano Pio},
  title        = {Procedure di modellizzazione rapide per la caratterizzazione di Supernovae Core Collapse e di transienti similari},
  school       = {Università degli Studi di Catania},
  year         = {2024},
  month        = dec,
  type         = {PhD thesis},
  url          = {https://hdl.handle.net/20.500.11769/658009},
}

@ARTICLE{MB19,
       author = {{Martinez}, Laureano and {Bersten}, Melina C.},
        title = "{Mass discrepancy analysis for a select sample of Type II-Plateau supernovae}",
      journal = {A\&A},
         year = 2019,
        month = sep,
       volume = {629},
          eid = {A124},
        pages = {A124},
          doi = {10.1051/0004-6361/201834818},
archivePrefix = {arXiv},
       eprint = {1908.01828},
 primaryClass = {astro-ph.SR},
       adsurl = {https://ui.adsabs.harvard.edu/abs/2019A&A...629A.124M}
}

@article{zampieri03,
    author = {Zampieri, L. and Pastorello, A. and Turatto, M. and Cappellaro, E. and Benetti, S. and Altavilla, G. and Mazzali, P. and Hamuy, M.},
    title = {Peculiar, low-luminosity Type II supernovae: low-energy explosions in massive progenitors?},
    journal = {\mnras},
    volume = {338},
    number = {3},
    pages = {711-716},
    year = {2003},
    month = {01},
    issn = {0035-8711},
    doi = {10.1046/j.1365-8711.2003.06082.x},
    url = {https://doi.org/10.1046/j.1365-8711.2003.06082.x},
    eprint = {https://academic.oup.com/mnras/article-pdf/338/3/711/4269997/338-3-711.pdf},
}

@article{ASHOUR2010341,
title = {Approximate skew normal distribution},
journal = {Journal of Advanced Research},
volume = {1},
number = {4},
pages = {341-350},
year = {2010},
issn = {2090-1232},
doi = {https://doi.org/10.1016/j.jare.2010.06.004},
url = {https://www.sciencedirect.com/science/article/pii/S209012321000069X},
author = {Samir K. Ashour and Mahmood A. Abdel-hameed}
}

@article{Silva_2024,
doi = {10.3847/1538-4357/ad402a},
url = {https://dx.doi.org/10.3847/1538-4357/ad402a},
year = {2024},
month = {jun},
publisher = {The American Astronomical Society},
volume = {969},
number = {1},
pages = {57},
author = {Silva-Farfán, Javier and Förster, Francisco and Moriya, Takashi J. and Hernández-García, L. and Muñoz Arancibia, A. M. and Sánchez-Sáez, P. and Anderson, Joseph P. and Tonry, John L. and Clocchiatti, Alejandro},
title = {Physical Properties of Type II Supernovae Inferred from ZTF and ATLAS Photometric Data},
journal = {\apj}
}

@article{Inserra_2018,
doi = {10.3847/1538-4357/aaaaaa},
url = {https://dx.doi.org/10.3847/1538-4357/aaaaaa},
year = {2018},
month = {feb},
publisher = {The American Astronomical Society},
volume = {854},
number = {2},
pages = {175},
author = {Inserra, C. and Prajs, S. and Gutierrez, C. P. and Angus, C. and Smith, M. and Sullivan, M.},
title = {A Statistical Approach to Identify Superluminous Supernovae and Probe Their Diversity},
journal = {\apj}
}

@article{roberts13,
author = {Roberts, S.  and Osborne, M.  and Ebden, M.  and Reece, S.  and Gibson, N.  and Aigrain, S. },
title = {Gaussian processes for time-series modelling},
journal = {Philosophical Transactions of the Royal Society A: Mathematical, Physical and Engineering Sciences},
volume = {371},
number = {1984},
pages = {20110550},
year = {2013},
doi = {10.1098/rsta.2011.0550},

URL = {https://royalsocietypublishing.org/doi/abs/10.1098/rsta.2011.0550},
eprint = {https://royalsocietypublishing.org/doi/pdf/10.1098/rsta.2011.0550}
}

@article{catchpole1987,
  title={Spectroscopic and photometric observations of SN 1987a--II. Days 51 to 134},
  author={Catchpole, RM and Menzies, JW and Monk, AS and Wargau, WF and Pollaco, D and Carter, BS and Whitelock, PA and Marang, F and Laney, CD and Balona, LA and others},
  journal={\mnras},
  volume={229},
  number={1},
  pages={15P--25P},
  year={1987},
  publisher={Oxford University Press Oxford, UK}
}

@ARTICLE{Cosentino25,
    author = {Cosentino, Stefano P and Pumo, Maria L and Cherubini, Silvio},
    title = {High-energy neutrinos by hydrogen-rich supernovae interacting with low-massive circumstellar medium: the case of SN 2023ixf},
    journal = {\mnras},
    volume = {540},
    number = {4},
    pages = {2894-2913},
    year = {2025},
    month = {05},
    issn = {0035-8711},
    doi = {10.1093/mnras/staf861},
    url = {https://doi.org/10.1093/mnras/staf861},
    eprint = {https://academic.oup.com/mnras/article-pdf/540/4/2894/63373064/staf861.pdf},
}

@ARTICLE{2008ApJ...683L.131G,
       author = {{Gezari}, Suvi and {Dessart}, Luc and {Basa}, St{\'e}phane and {Martin}, D. Chris and {Neill}, James D. and {Woosley}, S.~E. and {Hillier}, D. John and {Bazin}, Gurvan and {Forster}, Karl and {Friedman}, Peter G. and {Le Du}, J{\'e}r{\'e}my and {Mazure}, Alain and {Morrissey}, Patrick and {Neff}, Susan G. and {Schiminovich}, David and {Wyder}, Ted K.},
        title = "{Probing Shock Breakout with Serendipitous GALEX Detections of Two SNLS Type II-P Supernovae}",
      journal = {\apjl},
         year = 2008,
        month = aug,
       volume = {683},
       number = {2},
        pages = {L131},
          doi = {10.1086/591647},
archivePrefix = {arXiv},
       eprint = {0804.1123},
 primaryClass = {astro-ph},
       adsurl = {https://ui.adsabs.harvard.edu/abs/2008ApJ...683L.131G}
}

@article{Bufano_2009,
doi = {10.1088/0004-637X/700/2/1456},
url = {https://dx.doi.org/10.1088/0004-637X/700/2/1456},
year = {2009},
month = {jul},
publisher = {The American Astronomical Society},
volume = {700},
number = {2},
pages = {1456},
author = {Bufano, F. and Immler, S. and Turatto, M. and Landsman, W. and Brown, P. and Benetti, S. and Cappellaro, E. and Holland, S. T. and Mazzali, P. and Milne, P. and Panagia, N. and Pian, E. and Roming, P. and Zampieri, L. and Breeveld, A. A. and Gehrels, N.},
title = {ULTRAVIOLET SPECTROSCOPY OF SUPERNOVAE: THE FIRST TWO YEARS OF SWIFT OBSERVATIONS},
journal = {\apj}
}

@article{Villar_2017,
doi = {10.3847/2041-8213/aa9c84},
url = {https://dx.doi.org/10.3847/2041-8213/aa9c84},
year = {2017},
month = {dec},
publisher = {The American Astronomical Society},
volume = {851},
number = {1},
pages = {L21},
author = {V. A. Villar and J. Guillochon and E. Berger and B. D. Metzger and P. S. Cowperthwaite and M. Nicholl and K. D. Alexander and P. K. Blanchard and R. Chornock and T. Eftekhari and W. Fong and R. Margutti and P. K. G. Williams},
title = {The Combined Ultraviolet, Optical, and Near-infrared Light Curves of the Kilonova Associated with the Binary Neutron Star Merger GW170817: Unified Data Set, Analytic Models, and Physical Implications},
journal = {\apjl}
}

@article{Nicholl_2017,
doi = {10.3847/1538-4357/aa9334},
url = {https://dx.doi.org/10.3847/1538-4357/aa9334},
year = {2017},
month = {nov},
publisher = {The American Astronomical Society},
volume = {850},
number = {1},
pages = {55},
author = {Matt Nicholl and James Guillochon and Edo Berger},
title = {The Magnetar Model for Type I Superluminous Supernovae. I. Bayesian Analysis of the Full Multicolor Light-curve Sample with MOSFiT},
journal = {\apj}
}

@ARTICLE{2011ApJ...737..103S,
       author = {{Schlafly}, Edward F. and {Finkbeiner}, Douglas P.},
        title = "{Measuring Reddening with Sloan Digital Sky Survey Stellar Spectra and Recalibrating SFD}",
      journal = {ApJ},
         year = 2011,
        month = aug,
       volume = {737},
       number = {2},
          eid = {103},
        pages = {103},
          doi = {10.1088/0004-637X/737/2/103},
archivePrefix = {arXiv},
       eprint = {1012.4804},
 primaryClass = {astro-ph.GA},
       adsurl = {https://ui.adsabs.harvard.edu/abs/2011ApJ...737..103S}
}

@ARTICLE{2016ApJ...822...78G,
       author = {{Grillo}, C. and {Karman}, W. and {Suyu}, S.~H. and {Rosati}, P. and {Balestra}, I. and {Mercurio}, A. and {Lombardi}, M. and {Treu}, T. and {Caminha}, G.~B. and {Halkola}, A. and {Rodney}, S.~A. and {Gavazzi}, R. and {Caputi}, K.~I.},
        title = "{The Story of Supernova Refsdal Told by Muse}",
      journal = {ApJ},
         year = 2016,
        month = may,
       volume = {822},
       number = {2},
          eid = {78},
        pages = {78},
          doi = {10.3847/0004-637X/822/2/78},
archivePrefix = {arXiv},
       eprint = {1511.04093},
 primaryClass = {astro-ph.GA},
       adsurl = {https://ui.adsabs.harvard.edu/abs/2016ApJ...822...78G}
}

@ARTICLE{1995ApJS...99..223P,
       author = {{Pun}, Chun S.~J. and {Kirshner}, Robert P. and {Sonneborn}, George and {Challis}, Peter and {Nassiopoulos}, George and {Arquilla}, Richard and {Crenshaw}, D. Michael and {Shrader}, Chris and {Teays}, Terry and {Cassatella}, Angelo and {Gilmozzi}, Roberto and {Talavera}, Antonio and {Wamsteker}, Willem and {Fransson}, Claes and {Panagia}, Nino},
        title = "{Ultraviolet Observations of SN 1987A with the IUE Satellite}",
      journal = {ApJS},
         year = 1995,
        month = jul,
       volume = {99},
        pages = {223},
          doi = {10.1086/192185},
       adsurl = {https://ui.adsabs.harvard.edu/abs/1995ApJS...99..223P}
}

@ARTICLE{Matsumoto_2025,
       author = {{Matsumoto}, Tatsuya and {Metzger}, Brian D. and {Goldberg}, Jared A.},
        title = "{Long Plateau Doth So: How Internal Heating Sources Affect Hydrogen-rich Supernova Light Curves}",
      journal = {\apj},
         year = 2025,
        month = jan,
       volume = {978},
       number = {1},
          eid = {56},
        pages = {56},
          doi = {10.3847/1538-4357/ad93a9},
archivePrefix = {arXiv},
       eprint = {2401.13731},
 primaryClass = {astro-ph.HE},
       adsurl = {https://ui.adsabs.harvard.edu/abs/2025ApJ...978...56M}
}

@article{Kasen_2010,
doi = {10.1088/0004-637X/717/1/245},
url = {https://dx.doi.org/10.1088/0004-637X/717/1/245},
year = {2010},
month = {jun},
publisher = {The American Astronomical Society},
volume = {717},
number = {1},
pages = {245},
author = {Daniel Kasen and Lars Bildsten},
title = {SUPERNOVA LIGHT CURVES POWERED BY YOUNG MAGNETARS},
journal = {\apj}
}

@article{Chatzopoulos_2012,
doi = {10.1088/0004-637X/746/2/121},
url = {https://dx.doi.org/10.1088/0004-637X/746/2/121},
year = {2012},
month = {jan},
publisher = {The American Astronomical Society},
volume = {746},
number = {2},
pages = {121},
author = {E. Chatzopoulos and J. Craig Wheeler and J. Vinko},
title = {GENERALIZED SEMI-ANALYTICAL MODELS OF SUPERNOVA LIGHT CURVES},
journal = {\apj}
}

@article{Khatami_2019,
doi = {10.3847/1538-4357/ab1f09},
url = {https://dx.doi.org/10.3847/1538-4357/ab1f09},
year = {2019},
month = {jun},
publisher = {The American Astronomical Society},
volume = {878},
number = {1},
pages = {56},
author = {David K. Khatami and Daniel N. Kasen},
title = {Physics of Luminous Transient Light Curves: A New Relation between Peak Time and Luminosity},
journal = {\apj}
}

@ARTICLE{Inserra_13,
       author = {{Inserra}, C. and {Smartt}, S.~J. and {Jerkstrand}, A. and {Valenti}, S. and {Fraser}, M. and {Wright}, D. and {Smith}, K. and {Chen}, T. -W. and {Kotak}, R. and {Pastorello}, A. and {Nicholl}, M. and {Bresolin}, F. and {Kudritzki}, R.~P. and {Benetti}, S. and {Botticella}, M.~T. and {Burgett}, W.~S. and {Chambers}, K.~C. and {Ergon}, M. and {Flewelling}, H. and {Fynbo}, J.~P.~U. and {Geier}, S. and {Hodapp}, K.~W. and {Howell}, D.~A. and {Huber}, M. and {Kaiser}, N. and {Leloudas}, G. and {Magill}, L. and {Magnier}, E.~A. and {McCrum}, M.~G. and {Metcalfe}, N. and {Price}, P.~A. and {Rest}, A. and {Sollerman}, J. and {Sweeney}, W. and {Taddia}, F. and {Taubenberger}, S. and {Tonry}, J.~L. and {Wainscoat}, R.~J. and {Waters}, C. and {Young}, D.},
        title = "{Super-luminous Type Ic Supernovae: Catching a Magnetar by the Tail}",
      journal = {ApJ},
         year = 2013,
        month = jun,
       volume = {770},
       number = {2},
          eid = {128},
        pages = {128},
          doi = {10.1088/0004-637X/770/2/128},
archivePrefix = {arXiv},
       eprint = {1304.3320},
 primaryClass = {astro-ph.HE},
       adsurl = {https://ui.adsabs.harvard.edu/abs/2013ApJ...770..128I}
}

@article{Dexter_2013,
doi = {10.1088/0004-637X/772/1/30},
url = {https://dx.doi.org/10.1088/0004-637X/772/1/30},
year = {2013},
month = {jul},
publisher = {The American Astronomical Society},
volume = {772},
number = {1},
pages = {30},
author = {Jason Dexter and Daniel Kasen},
title = {SUPERNOVA LIGHT CURVES POWERED BY FALLBACK ACCRETION},
journal = {\apj}
}

@article{Bersten_2011,
doi = {10.1088/0004-637X/729/1/61},
url = {https://dx.doi.org/10.1088/0004-637X/729/1/61},
year = {2011},
month = {feb},
publisher = {The American Astronomical Society},
volume = {729},
number = {1},
pages = {61},
author = {Melina C. Bersten and Omar Benvenuto and Mario Hamuy},
title = {HYDRODYNAMICAL MODELS OF TYPE II PLATEAU SUPERNOVAE},
journal = {\apj}
}

@article{Cook_2019,
doi = {10.3847/1538-4357/ab2131},
url = {https://dx.doi.org/10.3847/1538-4357/ab2131},
year = {2019},
month = {jul},
publisher = {The American Astronomical Society},
volume = {880},
number = {1},
pages = {7},
author = {David O. Cook and Mansi M. Kasliwal and Angela Van Sistine and David L. Kaplan and Jessica S. Sutter and Thomas Kupfer and David L. Shupe and Russ R. Laher and Frank J. Masci and Daniel A. Dale and Branimir Sesar and Patrick R. Brady and Lin Yan and Eran O. Ofek and David H. Reitze and Shrinivas R. Kulkarni},
title = {Census of the Local Universe (CLU) Narrowband Survey. I. Galaxy Catalogs from Preliminary Fields},
journal = {\apj}
}

@article{Bellm_2019,
doi = {10.1088/1538-3873/aaecbe},
url = {https://dx.doi.org/10.1088/1538-3873/aaecbe},
year = {2018},
month = {dec},
publisher = {The Astronomical Society of the Pacific},
volume = {131},
number = {995},
pages = {018002},
author = {Eric C. Bellm and Shrinivas R. Kulkarni and Matthew J. Graham and Richard Dekany and Roger M. Smith and Reed Riddle and Frank J. Masci and George Helou and Thomas A. Prince and Scott M. Adams and C. Barbarino and Tom Barlow and James Bauer and Ron Beck and Justin Belicki and Rahul Biswas and Nadejda Blagorodnova and Dennis Bodewits and Bryce Bolin and Valery Brinnel and Tim Brooke and Brian Bue and Mattia Bulla and Rick Burruss and S. Bradley Cenko and Chan-Kao Chang and Andrew Connolly and Michael Coughlin and John Cromer and Virginia Cunningham and Kishalay De and Alex Delacroix and Vandana Desai and Dmitry A. Duev and Gwendolyn Eadie and Tony L. Farnham and Michael Feeney and Ulrich Feindt and David Flynn and Anna Franckowiak and S. Frederick and C. Fremling and Avishay Gal-Yam and Suvi Gezari and Matteo Giomi and Daniel A. Goldstein and V. Zach Golkhou and Ariel Goobar and Steven Groom and Eugean Hacopians and David Hale and John Henning and Anna Y. Q. Ho and David Hover and Justin Howell and Tiara Hung and Daniela Huppenkothen and David Imel and Wing-Huen Ip and Željko Ivezić and Edward Jackson and Lynne Jones and Mario Juric and Mansi M. Kasliwal and S. Kaspi and Stephen Kaye and Michael S. P. Kelley and Marek Kowalski and Emily Kramer and Thomas Kupfer and Walter Landry and Russ R. Laher and Chien-De Lee and Hsing Wen Lin and Zhong-Yi Lin and Ragnhild Lunnan and Matteo Giomi and Ashish Mahabal and Peter Mao and Adam A. Miller and Serge Monkewitz and Patrick Murphy and Chow-Choong Ngeow and Jakob Nordin and Peter Nugent and Eran Ofek and Maria T. Patterson and Bryan Penprase and Michael Porter and Ludwig Rauch and Umaa Rebbapragada and Dan Reiley and Mickael Rigault and Hector Rodriguez and Jan van Roestel and Ben Rusholme and Jakob van Santen and S. Schulze and David L. Shupe and Leo P. Singer and Maayane T. Soumagnac and Robert Stein and Jason Surace and Jesper Sollerman and Paula Szkody and F. Taddia and Scott Terek and Angela Van Sistine and Sjoert van Velzen and W. Thomas Vestrand and Richard Walters and Charlotte Ward and Quan-Zhi Ye and Po-Chieh Yu and Lin Yan and Jeffry Zolkower},
title = {The Zwicky Transient Facility: System Overview, Performance, and First Results},
journal = {\pasp}
}

@article{Graham_2019,
doi = {10.1088/1538-3873/ab006c},
url = {https://dx.doi.org/10.1088/1538-3873/ab006c},
year = {2019},
month = {may},
publisher = {The Astronomical Society of the Pacific},
volume = {131},
number = {1001},
pages = {078001},
author = {Matthew J. Graham and S. R. Kulkarni and Eric C. Bellm and Scott M. Adams and Cristina Barbarino and Nadejda Blagorodnova and Dennis Bodewits and Bryce Bolin and Patrick R. Brady and S. Bradley Cenko and Chan-Kao Chang and Michael W. Coughlin and Kishalay De and Gwendolyn Eadie and Tony L. Farnham and Ulrich Feindt and Anna Franckowiak and Christoffer Fremling and Suvi Gezari and Shaon Ghosh and Daniel A. Goldstein and V. Zach Golkhou and Ariel Goobar and Anna Y. Q. Ho and Daniela Huppenkothen and Željko Ivezić and R. Lynne Jones and Mario Juric and David L. Kaplan and Mansi M. Kasliwal and Michael S. P. Kelley and Thomas Kupfer and Chien-De Lee and Hsing Wen Lin and Ragnhild Lunnan and Ashish A. Mahabal and Adam A. Miller and Chow-Choong Ngeow and Peter Nugent and Eran O. Ofek and Thomas A. Prince and Ludwig Rauch and Jan van Roestel and Steve Schulze and Leo P. Singer and Jesper Sollerman and Francesco Taddia and Lin Yan and Quan-Zhi Ye and Po-Chieh Yu and Tom Barlow and James Bauer and Ron Beck and Justin Belicki and Rahul Biswas and Valery Brinnel and Tim Brooke and Brian Bue and Mattia Bulla and Rick Burruss and Andrew Connolly and John Cromer and Virginia Cunningham and Richard Dekany and Alex Delacroix and Vandana Desai and Dmitry A. Duev and Michael Feeney and David Flynn and Sara Frederick and Avishay Gal-Yam and Matteo Giomi and Steven Groom and Eugean Hacopians and David Hale and George Helou and John Henning and David Hover and Lynne A. Hillenbrand and Justin Howell and Tiara Hung and David Imel and Wing-Huen Ip and Edward Jackson and Shai Kaspi and Stephen Kaye and Marek Kowalski and Emily Kramer and Michael Kuhn and Walter Landry and Russ R. Laher and Peter Mao and Frank J. Masci and Serge Monkewitz and Patrick Murphy and Jakob Nordin and Maria T. Patterson and Bryan Penprase and Michael Porter and Umaa Rebbapragada and Dan Reiley and Reed Riddle and Mickael Rigault and Hector Rodriguez and Ben Rusholme and Jakob van Santen and David L. Shupe and Roger M. Smith and Maayane T. Soumagnac and Robert Stein and Jason Surace and Paula Szkody and Scott Terek and Angela Van Sistine and Sjoert van Velzen and W. Thomas Vestrand and Richard Walters and Charlotte Ward and Chaoran Zhang and Jeffry Zolkower},
title = {The Zwicky Transient Facility: Science Objectives},
journal = {\pasp}
}

@article{Fang_2025,
doi = {10.3847/1538-4357/ad8b19},
url = {https://dx.doi.org/10.3847/1538-4357/ad8b19},
year = {2024},
month = {dec},
publisher = {The American Astronomical Society},
volume = {978},
number = {1},
pages = {35},
author = {Qiliang Fang and Keiichi Maeda and Haonan Ye and Takashi J. Moriya and Tatsuya Matsumoto},
title = {Diversity in Hydrogen-rich Envelope Mass of Type II Supernovae. I. Plateau Phase Light-curve Modeling},
journal = {\apj}
}

@article{Muller_2017,
doi = {10.3847/1538-4357/aa72f1},
url = {https://dx.doi.org/10.3847/1538-4357/aa72f1},
year = {2017},
month = {jun},
publisher = {The American Astronomical Society},
volume = {841},
number = {2},
pages = {127},
author = {Tomás Müller and José L. Prieto and Ondřej Pejcha and Alejandro Clocchiatti},
title = {The Nickel Mass Distribution of Normal Type II Supernovae},
journal = {\apj}
}

@ARTICLE{Taddia13,
       author = {{Taddia}, F. and {Sollerman}, J. and {Razza}, A. and {Gafton}, E. and {Pastorello}, A. and {Fransson}, C. and {Stritzinger}, M.~D. and {Leloudas}, G. and {Ergon}, M.},
        title = "{A metallicity study of 1987A-like supernova host galaxies}",
      journal = {\aap},
         year = 2013,
        month = oct,
       volume = {558},
          eid = {A143},
        pages = {A143},
          doi = {10.1051/0004-6361/201322276},
archivePrefix = {arXiv},
       eprint = {1308.5545},
 primaryClass = {astro-ph.CO},
       adsurl = {https://ui.adsabs.harvard.edu/abs/2013A&A...558A.143T}
}

@ARTICLE{Salmaso2023,
       author = {{Salmaso}, I. and {Cappellaro}, E. and {Tartaglia}, L. and {Benetti}, S. and {Botticella}, M.~T. and {Elias-Rosa}, N. and {Pastorello}, A. and {Patat}, F. and {Reguitti}, A. and {Tomasella}, L. and {Valerin}, G. and {Yang}, S.},
        title = "{Hidden shock powering the peak of SN 2020faa}",
      journal = {\aap},
         year = 2023,
        month = may,
       volume = {673},
          eid = {A127},
        pages = {A127},
          doi = {10.1051/0004-6361/202245781},
archivePrefix = {arXiv},
       eprint = {2302.12527},
 primaryClass = {astro-ph.HE},
       adsurl = {https://ui.adsabs.harvard.edu/abs/2023A&A...673A.127S}
}

@ARTICLE{Yang2024,
       author = {{Yang}, S. and {Sollerman}, J. and {Chen}, T. -W. and {Kool}, E.~C. and {Lunnan}, R. and {Schulze}, S. and {Strotjohann}, N. and {Horesh}, A. and {Kasliwal}, M. and {Kupfer}, T. and {Mahabal}, A.~A. and {Masci}, F.~J. and {Nugent}, P. and {Perley}, D.~A. and {Riddle}, R. and {Rusholme}, B. and {Sharma}, Y.},
        title = "{Is supernova SN 2020faa an iPTF14hls look-alike?}",
      journal = {\aap},
         year = 2021,
        month = feb,
       volume = {646},
          eid = {A22},
        pages = {A22},
          doi = {10.1051/0004-6361/202039440},
archivePrefix = {arXiv},
       eprint = {2009.07270},
 primaryClass = {astro-ph.HE},
       adsurl = {https://ui.adsabs.harvard.edu/abs/2021A&A...646A..22Y}
}

@ARTICLE{Szalai2024,
       author = {{Szalai}, T. and {K{\"o}nyves-T{\'o}th}, R. and {Nagy}, A.~P. and {Hiramatsu}, D. and {Arcavi}, I. and {Bostroem}, A. and {Howell}, D.~A. and {Farah}, J. and {McCully}, C. and {Newsome}, M. and {Padilla Gonzalez}, E. and {Pellegrino}, C. and {Terreran}, G. and {Berger}, E. and {Blanchard}, P. and {Gomez}, S. and {Sz{\'e}kely}, P. and {B{\'a}nhidi}, D. and {B{\'\i}r{\'o}}, I.~B. and {Cs{\'a}nyi}, I. and {P{\'a}l}, A. and {Rho}, J. and {Vink{\'o}}, J.},
        title = "{The story of SN 2021aatd: A peculiar 1987A-like supernova with an early-phase luminosity excess}",
      journal = {\aap},
         year = 2024,
        month = oct,
       volume = {690},
          eid = {A17},
        pages = {A17},
          doi = {10.1051/0004-6361/202348548},
archivePrefix = {arXiv},
       eprint = {2406.02498},
 primaryClass = {astro-ph.SR},
       adsurl = {https://ui.adsabs.harvard.edu/abs/2024A&A...690A..17S}
}

@article{orlando15,
doi = {10.1088/0004-637X/810/2/168},
url = {https://dx.doi.org/10.1088/0004-637X/810/2/168},
year = {2015},
month = {sep},
publisher = {The American Astronomical Society},
volume = {810},
number = {2},
pages = {168},
author = {S. Orlando and M. Miceli and M. L. Pumo and F. Bocchino},
title = {SUPERNOVA 1987A: A TEMPLATE TO LINK SUPERNOVAE TO THEIR REMNANTS},
journal = {\apj}
}

@ARTICLE{JacobsonG2024,
       author = {{Jacobson-Gal{\'a}n}, W.~V. and {Dessart}, L. and {Davis}, K.~W. and {Kilpatrick}, C.~D. and {Margutti}, R. and {Foley}, R.~J. and {Chornock}, R. and {Terreran}, G. and {Hiramatsu}, D. and {Newsome}, M. and {Padilla Gonzalez}, E. and {Pellegrino}, C. and {Howell}, D.~A. and {Filippenko}, A.~V. and {Anderson}, J.~P. and {Angus}, C.~R. and {Auchettl}, K. and {Bostroem}, K.~A. and {Brink}, T.~G. and {Cartier}, R. and {Coulter}, D.~A. and {de Boer}, T. and {Drout}, M.~R. and {Earl}, N. and {Ertini}, K. and {Farah}, J.~R. and {Farias}, D. and {Gall}, C. and {Gao}, H. and {Gerlach}, M.~A. and {Guo}, F. and {Haynie}, A. and {Hosseinzadeh}, G. and {Ibik}, A.~L. and {Jha}, S.~W. and {Jones}, D.~O. and {Langeroodi}, D. and {LeBaron}, N. and {Magnier}, E.~A. and {Piro}, A.~L. and {Raimundo}, S.~I. and {Rest}, A. and {Rest}, S. and {Rich}, R. Michael and {Rojas-Bravo}, C. and {Sears}, H. and {Taggart}, K. and {Villar}, V.~A. and {Wainscoat}, R.~J. and {Wang}, X. -F. and {Wasserman}, A.~R. and {Yan}, S. and {Yang}, Y. and {Zhang}, J. and {Zheng}, W.},
        title = "{Final Moments. II. Observational Properties and Physical Modeling of Circumstellar-material-interacting Type II Supernovae}",
      journal = {\apj},
         year = 2024,
        month = aug,
       volume = {970},
       number = {2},
          eid = {189},
        pages = {189},
          doi = {10.3847/1538-4357/ad4a2a},
archivePrefix = {arXiv},
       eprint = {2403.02382},
 primaryClass = {astro-ph.HE},
       adsurl = {https://ui.adsabs.harvard.edu/abs/2024ApJ...970..189J}
}

@ARTICLE{kelly16,
       author = {{Kelly}, P.~L. and {Brammer}, G. and {Selsing}, J. and {Foley}, R.~J. and {Hjorth}, J. and {Rodney}, S.~A. and {Christensen}, L. and {Strolger}, L. -G. and {Filippenko}, A.~V. and {Treu}, T. and {Steidel}, C.~C. and {Strom}, A. and {Riess}, A.~G. and {Zitrin}, A. and {Schmidt}, K.~B. and {Brada{\v{c}}}, M. and {Jha}, S.~W. and {Graham}, M.~L. and {McCully}, C. and {Graur}, O. and {Weiner}, B.~J. and {Silverman}, J.~M. and {Taddia}, F.},
        title = "{SN Refsdal: Classification as a Luminous and Blue SN 1987A-like Type II Supernova}",
      journal = {ApJ},
         year = 2016,
        month = nov,
       volume = {831},
       number = {2},
          eid = {205},
        pages = {205},
          doi = {10.3847/0004-637X/831/2/205},
archivePrefix = {arXiv},
       eprint = {1512.09093},
 primaryClass = {astro-ph.GA},
       adsurl = {https://ui.adsabs.harvard.edu/abs/2016ApJ...831..205K}
}

@article{gutierrez20,
    author = {Gutiérrez, C P and Sullivan, M and Martinez, L and Bersten, M C and Inserra, C and Smith, M and Anderson, J P and Pan, Y-C and Pastorello, A and Galbany, L and Nugent, P and Angus, C R and Barbarino, C and Carollo, D and Chen, T-W and Davis, T M and Della Valle, M and Foley, R J and Fraser, M and Frohmaier, C and González-Gaitán, S and Gromadzki, M and Kankare, E and Kokotanekova, R and Kollmeier, J and Lewis, G F and Magee, M R and Maguire, K and Möller, A and Morrell, N and Nicholl, M and Pursiainen, M and Sollerman, J and Sommer, N E and Swann, E and Tucker, B E and Wiseman, P and Aguena, M and Allam, S and Avila, S and Bertin, E and Brooks, D and Buckley-Geer, E and Burke, D L and Carnero Rosell, A and Carrasco Kind, M and Carretero, J and Costanzi, M and da Costa, L N and De Vicente, J and Desai, S and Diehl, H T and Doel, P and Eifler, T F and Flaugher, B and Fosalba, P and Frieman, J and García-Bellido, J and Gerdes, D W and Gruen, D and Gruendl, R A and Gschwend, J and Gutierrez, G and Hinton, S R and Hollowood, D L and Honscheid, K and James, D J and Kuehn, K and Kuropatkin, N and Lahav, O and Lima, M and Maia, M A G and March, M and Menanteau, F and Miquel, R and Morganson, E and Palmese, A and Paz-Chinchón, F and Plazas, A A and Sako, M and Sanchez, E and Scarpine, V and Schubnell, M and Serrano, S and Sevilla-Noarbe, I and Soares-Santos, M and Suchyta, E and Swanson, M E C and Tarle, G and Thomas, D and Varga, T N and Walker, A R and Wilkinson, R and (DES Collaboration)},
    title = "{DES16C3cje: A low-luminosity, long-lived supernova}",
    journal = {\mnras},
    volume = {496},
    number = {1},
    pages = {95-110},
    year = {2020},
    month = {05},
    issn = {0035-8711},
    doi = {10.1093/mnras/staa1452},
    url = {https://doi.org/10.1093/mnras/staa1452},
    eprint = {https://academic.oup.com/mnras/article-pdf/496/1/95/33395855/staa1452.pdf},
}

@ARTICLE{terreran17,
       author = {{Terreran}, G. and {Pumo}, M.~L. and {Chen}, T. -W. and {Moriya}, T.~J. and {Taddia}, F. and {Dessart}, L. and {Zampieri}, L. and {Smartt}, S.~J. and {Benetti}, S. and {Inserra}, C. and {Cappellaro}, E. and {Nicholl}, M. and {Fraser}, M. and {Wyrzykowski}, {\L}. and {Udalski}, A. and {Howell}, D.~A. and {McCully}, C. and {Valenti}, S. and {Dimitriadis}, G. and {Maguire}, K. and {Sullivan}, M. and {Smith}, K.~W. and {Yaron}, O. and {Young}, D.~R. and {Anderson}, J.~P. and {Della Valle}, M. and {Elias-Rosa}, N. and {Gal-Yam}, A. and {Jerkstrand}, A. and {Kankare}, E. and {Pastorello}, A. and {Sollerman}, J. and {Turatto}, M. and {Kostrzewa-Rutkowska}, Z. and {Koz{\l}owski}, S. and {Mr{\'o}z}, P. and {Pawlak}, M. and {Pietrukowicz}, P. and {Poleski}, R. and {Skowron}, D. and {Skowron}, J. and {Soszy{\'n}ski}, I. and {Szyma{\'n}ski}, M.~K. and {Ulaczyk}, K.},
        title = "{Hydrogen-rich supernovae beyond the neutrino-driven core-collapse paradigm}",
      journal = {\nat},
         year = 2017,
        month = sep,
       volume = {1},
        pages = {713-720},
          doi = {10.1038/s41550-017-0228-8},
archivePrefix = {arXiv},
       eprint = {1709.10475},
 primaryClass = {astro-ph.SR},
       adsurl = {https://ui.adsabs.harvard.edu/abs/2017NatAs...1..713T}
}

@article{smartt09,
   author = "Smartt, Stephen J.",
   title = "Progenitors of Core-Collapse Supernovae", 
   journal= "ARA\&A",
   year = "2009",
   volume = "47",
   number = "Volume 47, 2009",
   pages = "63-106",
   doi = "https://doi.org/10.1146/annurev-astro-082708-101737",
   url = "https://www.annualreviews.org/content/journals/10.1146/annurev-astro-082708-101737",
   publisher = "Annual Reviews",
   issn = "1545-4282",
   type = "Journal Article",
  }

@ARTICLE{taddia16,
       author = {{Taddia}, F. and {Sollerman}, J. and {Fremling}, C. and {Migotto}, K. and {Gal-Yam}, A. and {Armen}, S. and {Duggan}, G. and {Ergon}, M. and {Filippenko}, A.~V. and {Fransson}, C. and {Hosseinzadeh}, G. and {Kasliwal}, M.~M. and {Laher}, R.~R. and {Leloudas}, G. and {Leonard}, D.~C. and {Lunnan}, R. and {Masci}, F.~J. and {Moon}, D. -S. and {Silverman}, J.~M. and {Wozniak}, P.~R.},
        title = "{Long-rising Type II supernovae from Palomar Transient Factory and Caltech Core-Collapse Project}",
      journal = {\aap},
         year = 2016,
        month = apr,
       volume = {588},
          eid = {A5},
        pages = {A5},
          doi = {10.1051/0004-6361/201527811},
       adsurl = {https://ui.adsabs.harvard.edu/abs/2016A&A...588A...5T}
}

@ARTICLE{pasto12,
       author = {{Pastorello}, A. and {Pumo}, M.~L. and {Navasardyan}, H. and {Zampieri}, L. and {Turatto}, M. and {Sollerman}, J. and {Taddia}, F. and {Kankare}, E. and {Mattila}, S. and {Nicolas}, J. and {Prosperi}, E. and {San Segundo Delgado}, A. and {Taubenberger}, S. and {Boles}, T. and {Bachini}, M. and {Benetti}, S. and {Bufano}, F. and {Cappellaro}, E. and {Cason}, A.~D. and {Cetrulo}, G. and {Ergon}, M. and {Germany}, L. and {Harutyunyan}, A. and {Howerton}, S. and {Hurst}, G.~M. and {Patat}, F. and {Stritzinger}, M. and {Strolger}, L. -G. and {Wells}, W.},
        title = "{SN 2009E: a faint clone of SN 1987A}",
      journal = {\aap},
         year = 2012,
        month = jan,
       volume = {537},
          eid = {A141},
        pages = {A141},
          doi = {10.1051/0004-6361/201118112},
archivePrefix = {arXiv},
       eprint = {1111.2497},
 primaryClass = {astro-ph.SR},
       adsurl = {https://ui.adsabs.harvard.edu/abs/2012A&A...537A.141P}
}

@article{pasto05,
    author = {Pastorello, A. and Baron, E. and Branch, D. and Zampieri, L. and Turatto, M. and Ramina, M. and Benetti, S. and Cappellaro, E. and Salvo, M. and Patat, F. and Piemonte, A. and Sollerman, J. and Leibundgut, B. and Altavilla, G.},
    title = {SN 1998A: explosion of a blue supergiant},
    journal = {\mnras},
    volume = {360},
    number = {3},
    pages = {950-962},
    year = {2005},
    month = {07},
    issn = {0035-8711},
    doi = {10.1111/j.1365-2966.2005.09079.x},
    url = {https://doi.org/10.1111/j.1365-2966.2005.09079.x},
    eprint = {https://academic.oup.com/mnras/article-pdf/360/3/950/3208473/360-3-950.pdf},
}

@article{xiang23,
    author = {Xiang, Danfeng and Wang, Xiaofeng and Zhang, Xinghan and Sai, Hanna and Zhang, Jujia and Brink, Thomas G and Filippenko, Alexei V and Mo, Jun and Zhang, Tianmeng and Chen, Zhihao and Dessart, Luc and Li, Zhitong and Yan, Shengyu and Blinnikov, Sergei I and Rui, Liming and Baron, E and DerKacy, J M},
    title = {SN 2018hna: Adding a piece to the puzzles of the explosion of blue supergiants},
    journal = {\mnras},
    volume = {520},
    number = {2},
    pages = {2965-2982},
    year = {2023},
    month = {01},
    issn = {0035-8711},
    doi = {10.1093/mnras/stad340},
    url = {https://doi.org/10.1093/mnras/stad340},
    eprint = {https://academic.oup.com/mnras/article-pdf/520/2/2965/49178552/stad340.pdf},
}

@article{singh19,
doi = {10.3847/2041-8213/ab3d44},
url = {https://dx.doi.org/10.3847/2041-8213/ab3d44},
year = {2019},
month = {sep},
publisher = {The American Astronomical Society},
volume = {882},
number = {2},
pages = {L15},
author = {Avinash Singh and D. K. Sahu and G. C. Anupama and Brajesh Kumar and Harsh Kumar and Masayuki Yamanaka and Petr V. Baklanov and Nozomu Tominaga and Sergei I. Blinnikov and Keiichi Maeda and Anirban Dutta and Varun Bhalerao and Ramya M. Anche and Sudhanshu Barway and Hiroshi Akitaya and Tatsuya Nakaoka and Miho Kawabata and Koji S Kawabata and Mahito Sasada and Kengo Takagi and Hiroyuki Maehara and Keisuke Isogai and Masaru Kino and Kenta Taguchi and Takashi Nagao},
title = {SN 2018hna: 1987A-like Supernova with a Signature of Shock Breakout},
journal = {\apjl}
}

@article{takats16,
    author = {Takáts, K. and Pignata, G. and Bersten, M. and Rojas Kaufmann, M. L. and Anderson, J. P. and Folatelli, G. and Hamuy, M. and Stritzinger, M. and Haislip, J. B. and LaCluyze, A. P. and Moore, J. P. and Reichart, D.},
    title = "{Optical photometry and spectroscopy of the 1987A-like supernova 2009mw}",
    journal = {\mnras},
    volume = {460},
    number = {4},
    pages = {3447-3457},
    year = {2016},
    month = {05},
    issn = {0035-8711},
    doi = {10.1093/mnras/stw1122},
    url = {https://doi.org/10.1093/mnras/stw1122},
    eprint = {https://academic.oup.com/mnras/article-pdf/460/4/3447/8116401/stw1122.pdf},
}

@ARTICLE{taddia12,
       author = {{Taddia}, F. and {Stritzinger}, M.~D. and {Sollerman}, J. and {Phillips}, M.~M. and {Anderson}, J.~P. and {Ergon}, M. and {Folatelli}, G. and {Fransson}, C. and {Freedman}, W. and {Hamuy}, M. and {Morrell}, N. and {Pastorello}, A. and {Persson}, S.~E. and {Gonzalez}, S.},
        title = "{The Type II supernovae 2006V and 2006au: two SN 1987A-like events}",
      journal = {\aap},
         year = 2012,
        month = jan,
       volume = {537},
          eid = {A140},
        pages = {A140},
          doi = {10.1051/0004-6361/201118091},
archivePrefix = {arXiv},
       eprint = {1111.2509},
 primaryClass = {astro-ph.CO},
       adsurl = {https://ui.adsabs.harvard.edu/abs/2012A&A...537A.140T}
}

@article{kleiser11,
    author = {Kleiser, Io K. W. and Poznanski, Dovi and Kasen, Daniel and Young, Timothy R. and Chornock, Ryan and Filippenko, Alexei V. and Challis, Peter and Ganeshalingam, Mohan and Kirshner, Robert P. and Li, Weidong and Matheson, Thomas and Nugent, Peter E. and Silverman, Jeffrey M.},
    title = "{Peculiar Type II supernovae from blue supergiants}",
    journal = {\mnras},
    volume = {415},
    number = {1},
    pages = {372-382},
    year = {2011},
    month = {07},
    issn = {0035-8711},
    doi = {10.1111/j.1365-2966.2011.18708.x},
    url = {https://doi.org/10.1111/j.1365-2966.2011.18708.x},
    eprint = {https://academic.oup.com/mnras/article-pdf/415/1/372/3122533/mnras0415-0372.pdf},
}

@INPROCEEDINGS{turatto07,
       author = {{Turatto}, Massimo and {Benetti}, Stefano and {Pastorello}, Andrea},
        title = "{Supernova classes and subclasses}",
    booktitle = {Supernova 1987A: 20 Years After: Supernovae and Gamma-Ray Bursters},
         year = 2007,
       editor = {{Immler}, Stefan and {Weiler}, Kurt and {McCray}, Richard},
       series = {American Institute of Physics Conference Series},
       volume = {937},
        month = oct,
    publisher = {AIP},
        pages = {187-197},
          doi = {10.1063/1.3682902},
archivePrefix = {arXiv},
       eprint = {0706.1086},
 primaryClass = {astro-ph},
       adsurl = {https://ui.adsabs.harvard.edu/abs/2007AIPC..937..187T}
}

@ARTICLE{arnett89,
       author = {{Arnett}, W. David and {Bahcall}, John N. and {Kirshner}, Robert P. and {Woosley}, Stanford E.},
        title = "{Supernova 1987A.}",
      journal = {ARA\&A},
         year = 1989,
        month = jan,
       volume = {27},
        pages = {629-700},
          doi = {10.1146/annurev.aa.27.090189.003213},
       adsurl = {https://ui.adsabs.harvard.edu/abs/1989ARA&A..27..629A}
}

@ARTICLE{sit_2023,
       author = {{Sit}, Tawny and {Kasliwal}, Mansi M. and {Tzanidakis}, Anastasios and {De}, Kishalay and {Fremling}, Christoffer and {Sollerman}, Jesper and {Gal-Yam}, Avishay and {Miller}, Adam A. and {Adams}, Scott and {Aloisi}, Robert and {Andreoni}, Igor and {Chu}, Matthew and {Cook}, David and {Das}, Kaustav Kashyap and {Dugas}, Alison and {Groom}, Steven L. and {Ho}, Anna Y.~Q. and {Karambelkar}, Viraj and {Neill}, James D. and {Masci}, Frank J. and {Medford}, Michael S. and {Purdum}, Josiah and {Sharma}, Yashvi and {Smith}, Roger and {Stein}, Robert and {Yan}, Lin and {Yao}, Yuhan and {Zhang}, Chaoran},
        title = "{Long-rising Type II Supernovae in the Zwicky Transient Facility Census of the Local Universe}",
      journal = {\apj},
         year = 2023,
        month = dec,
       volume = {959},
       number = {2},
          eid = {142},
        pages = {142},
          doi = {10.3847/1538-4357/ad036f},
archivePrefix = {arXiv},
       eprint = {2306.01109},
 primaryClass = {astro-ph.HE},
       adsurl = {https://ui.adsabs.harvard.edu/abs/2023ApJ...959..142S}
}

@article{PC2025,
    author = {Pumo, M L and Cosentino, S P},
    title = {Long-rising Type II supernovae resembling supernova 1987A – II. A new analytical model to describe these events},
    journal = {\mnras},
    volume = {538},
    number = {1},
    pages = {223 (Paper II)},
    year = {2025},
    month = {02},
    issn = {0035-8711},
    doi = {10.1093/mnras/staf288},
    url = {https://doi.org/10.1093/mnras/staf288},
    eprint = {https://academic.oup.com/mnras/article-pdf/538/1/223/61934605/staf288.pdf},
}

@article{pumo23,
       author = {{Pumo}, M.~L. and {Cosentino}, S.~P. and {Pastorello}, A. and {Benetti}, S. and {Cherubini}, S. and {Manic{\`o}}, G. and {Zampieri}, L.},
        title = "{Long-rising Type II supernovae resembling supernova 1987A - I. A comparative study through scaling relations}",
      journal = {\mnras},
         year = 2023,
        month = jun,
       volume = {521},
       number = {4},
        pages = {4801 (Paper I)},
          doi = {10.1093/mnras/stad861},
archivePrefix = {arXiv},
       eprint = {2303.10478},
 primaryClass = {astro-ph.HE},
       adsurl = {https://ui.adsabs.harvard.edu/abs/2023MNRAS.521.4801P},
	howpublished = {(Paper I)}
}

@ARTICLE{PZ13,
       author = {{Pumo}, M.~L. and {Zampieri}, L.},
        title = "{Calibration relations for core-collapse supernovae}",
      journal = {\mnras},
         year = 2013,
        month = oct,
       volume = {434},
       number = {4},
        pages = {3445-3453},
          doi = {10.1093/mnras/stt1256},
       adsurl = {https://ui.adsabs.harvard.edu/abs/2013MNRAS.434.3445P}
}

@ARTICLE{UC11,
       author = {{Utrobin}, V.~P. and {Chugai}, N.~N.},
        title = "{Supernova 2000cb: high-energy version of SN 1987A}",
      journal = {\aap},
         year = 2011,
        month = aug,
       volume = {532},
          eid = {A100},
        pages = {A100},
          doi = {10.1051/0004-6361/201117137},
archivePrefix = {arXiv},
       eprint = {1107.2145},
 primaryClass = {astro-ph.SR},
       adsurl = {https://ui.adsabs.harvard.edu/abs/2011A&A...532A.100U}
}

@ARTICLE{KW09,
       author = {{Kasen}, Daniel and {Woosley}, S.~E.},
        title = "{Type II Supernovae: Model Light Curves and Standard Candle Relationships}",
      journal = {\apj},
         year = 2009,
        month = oct,
       volume = {703},
       number = {2},
        pages = {2205-2216},
          doi = {10.1088/0004-637X/703/2/2205},
archivePrefix = {arXiv},
       eprint = {0910.1590},
 primaryClass = {astro-ph.CO},
       adsurl = {https://ui.adsabs.harvard.edu/abs/2009ApJ...703.2205K}
}

@ARTICLE{popov93,
       author = {{Popov}, D.~V.},
        title = "{An Analytical Model for the Plateau Stage of Type II Supernovae}",
      journal = {\apj},
         year = 1993,
        month = sep,
       volume = {414},
        pages = {712},
          doi = {10.1086/173117},
       adsurl = {https://ui.adsabs.harvard.edu/abs/1993ApJ...414..712P}
}

@BOOK{arnett96,
       author = {{Arnett}, David},
        title = "{Supernovae and Nucleosynthesis: An Investigation of the History of Matter from the Big Bang to the Present}",
         year = 1996,
       adsurl = {https://ui.adsabs.harvard.edu/abs/1996snih.book.....A}
}

@article{burrows90,
   author = "Burrows, A",
   title = "Neutrinos From Supernova Explosions", 
   journal= "Annual Review of Nuclear and Particle Science",
   year = "1990",
   volume = "40",
   number = "Volume 40, 1990",
   pages = "181-212",
   doi = "https://doi.org/10.1146/annurev.ns.40.120190.001145",
   url = "https://www.annualreviews.org/content/journals/10.1146/annurev.ns.40.120190.001145",
   publisher = "Annual Reviews",
   issn = "1545-4134",
   type = "Journal Article",
  }

@ARTICLE{Cardelli,
       author = {{Cardelli}, Jason A. and {Clayton}, Geoffrey C. and {Mathis}, John S.},
        title = "{The Relationship between Infrared, Optical, and Ultraviolet Extinction}",
      journal = {ApJ},
         year = 1989,
        month = oct,
       volume = {345},
        pages = {245},
          doi = {10.1086/167900},
       adsurl = {https://ui.adsabs.harvard.edu/abs/1989ApJ...345..245C}
}

@article{woosley02,
  title = {The evolution and explosion of massive stars},
  author = {Woosley, S. E. and Heger, A. and Weaver, T. A.},
  journal = {Rev. Mod. Phys.},
  volume = {74},
  issue = {4},
  pages = {1015--1071},
  numpages = {0},
  year = {2002},
  month = {Nov},
  publisher = {American Physical Society},
  doi = {10.1103/RevModPhys.74.1015},
  url = {https://link.aps.org/doi/10.1103/RevModPhys.74.1015}
}

@article{PZ11,
       author = {{Pumo}, M.~L. and {Zampieri}, L.},
        title = "{Radiation-hydrodynamical Modeling of Core-collapse Supernovae: Light Curves and the Evolution of Photospheric Velocity and Temperature}",
      journal = {\apj},
         year = 2011,
        month = nov,
       volume = {741},
       number = {1},
          eid = {41},
        pages = {41},
          doi = {10.1088/0004-637X/741/1/41},
archivePrefix = {arXiv},
       eprint = {1108.0688},
 primaryClass = {astro-ph.SR},
       adsurl = {https://ui.adsabs.harvard.edu/abs/2011ApJ...741...41P}
}

@ARTICLE{Arnett80,
       author = {{Arnett}, W.~D.},
        title = "{Analytic solutions for light curves of supernovae of Type II}",
      journal = {ApJ},
         year = 1980,
        month = apr,
       volume = {237},
        pages = {541-549},
          doi = {10.1086/157898},
       adsurl = {https://ui.adsabs.harvard.edu/abs/1980ApJ...237..541A}
}

@article{Pumo_2009,
doi = {10.1088/0004-637X/705/2/L138},
url = {https://dx.doi.org/10.1088/0004-637X/705/2/L138},
year = {2009},
month = {oct},
publisher = {The American Astronomical Society},
volume = {705},
number = {2},
pages = {L138},
author = {M. L. Pumo and M. Turatto and M. T. Botticella and A. Pastorello and S. Valenti and L. Zampieri and S. Benetti and E. Cappellaro and F. Patat},
title = {EC-SNe FROM SUPER-ASYMPTOTIC GIANT BRANCH PROGENITORS: THEORETICAL MODELS VERSUS OBSERVATIONS},
journal = {ApJ}
}

\begin{appendix} 

\section{Probability distribution function}\label{App:PDF}
This appendix describes the procedure adopted to construct and transform the probability distributions used in \textsc{SuperBAM} for both the prior and posterior inference. This process involves two main steps: 
(1) defining the PDFs associated with the measured spectrophotometric features, and
(2) converting the distributions of both measured features and modeling parameters into PDFs for the SN physical properties through the scaling relations.

As described in Sect.~\ref{SubSubSec:Prior}, once the key observables 
$m_i = \{t_{\rm m}, t_{\rm M},L_{\rm M}, v_{\rm M}, M_{\rm Ni}\}$ are measured, each with corresponding upper ($m_i^+$) and lower ($m_i^-$) uncertainties, a continuous probability distribution must be assigned to each feature. 
To achieve this, \textsc{SuperBAM} models each observable as a random variable following a skew-normal distribution \citep[e.g.,][]{ASHOUR2010341}, which accounts for possible asymmetries in the measurement errors:
$$
f_0(x)=\frac{2}{\omega \sqrt{2\pi}} 
\exp\!\left[-\frac{(x-\xi)^2}{2\omega^2}\right] 
\int_{-\infty}^{\alpha\left(\frac{x-\xi}{\omega}\right)} 
\frac{1}{\sqrt{2\pi}}\, e^{-\frac{t^2}{2}}\, dt.
$$
The free parameters of the distribution ($\omega$, $\xi$, and $\alpha$) are determined so that the resulting mean, standard deviation, and skewness reproduce the measured central value and asymmetric uncertainties of each feature. 
In practice, we adopt the following punctual distribution assumptions based on the measured feature and its upper and lower uncertainties: 
the mode is set to $A = m_i$, the mean to 
$B = (2m_i+ m_i^++ m_i^-)/4$, 
and the standard deviation to 
$C = (m_i^+ - m_i^-)/2$. 
Under these assumptions, the shape parameter $\alpha$ is obtained by numerically solving the implicit relation \citep[e.g.,][]{ASHOUR2010341}:
$$
\frac{(4 - \pi)\times\delta^3(\alpha)}{2[1 - \delta^2(\alpha)]^{3/2}} +
\mathrm{sign}(\alpha)\,
\frac{\exp[-2\pi / |\alpha|]}{1 - \delta^2(\alpha)} = \frac{2(B-A)}{C},
$$
where $\delta^2(\alpha) = 2\alpha^2 / [\pi\,(1 + \alpha^2)]$.
Once $\alpha$ is known, the other PDF's parameters are computed as:
$$
\omega = \frac{C}{\sqrt{1 - \delta^2(\alpha)}}, 
\qquad
\xi = B - \omega\,\delta(\alpha).
$$
These relations ensure that the resulting skew-normal function accurately reproduces the observed asymmetry in the measurement uncertainties \citep[see also Fig. 3.12 in][]{Cosentino2024}.

Since all the measured features represent intrinsically positive quantities, their probability density must also be defined on a positive domain. 
However, the standard skew-normal distribution $f_0(x)$ is defined over the entire real line and thus allows non-physical (negative) values. 
To prevent this issue and ensure continuity at the origin, we introduce a corrective function that smoothly suppresses the probability below zero while preserving the original distribution shape for positive values: 
\begin{equation*}
f(x)=N\times f_0(x)\times \begin{cases}
\exp\left[-\frac{x_0^2}{(x_0-\xi)^2}(x-\xi)^2/x^2\right]\,& x>0\\
0\,& x\le 0
\end{cases},
\end{equation*} 
where $x_0$ is related to the lowest admitted value and $N$ is a normalization constant. 
This correction maintains the position of the distribution maximum and reproduces the original skew-normal behavior for $x>x_0$. 

As described in Sects.~\ref{SubSubSec:Prior} and~\ref{SubSubSec:Posterior}, the scaling relations provide a direct link between the observed features and the prior distributions of the physical parameters ($E$, $M_{\rm ej}$, $R_0$), as well as between the model parameters ($y_i$, $\lambda$, $k_1$). 
Similarly, the posterior PDFs of the model parameters are related to the physical ones by applying the inverse transformations of the system described in Eq.~\ref{Eq:SuperBAM_param}.

In both cases, these transformations rely on the combination of independent random variables associated with the measured or inferred quantities. The operations between the involved PDFs  can be reduced to two basic types:

\begin{itemize}
\item Power operations — Let $Y = X^{\alpha}$, with $k \in \mathbb{R}$ and $X>0$ a random variable defined on a positive support with PDF $f(x)$. 
Then $Y$ is also a positive random variable, whose PDF is given by
$$
g(y) = \frac{1}{|k|}\, y^{(1-k)/k}\, f(y^{1/k}).
$$

\item Product operations — Let $Z = X\,Y$, where both $X>0$ and $Y>0$ are independent random variables with PDFs $f(x)$ and $g(y)$, respectively. 
Then $Z$ is also defined on a positive support, with the following PDF:
$$
h(z) = \int_0^{+\infty} f(x)\, g(z/x)\, \frac{1}{x}\, dx.
$$
\end{itemize}
Since all variables involved in Eqs.~\ref{Eq:Scaling} (or \ref{Eq:Scaling_new}) and~\ref{Eq:Prior_param} are defined on a positive domain and independent, the combination of their PDFs can be performed using previous transformations.

\section{Effects of $^{56}$Ni mixing}
\label{App:mixing}
In this appendix, we investigate the effects of varying the $^{56}$Ni mixing 
parameter $k_2$, or equivalently the characteristic radius $x_c$, on the spatial 
distribution of radioactive material and on the resulting bolometric LC. 
This analysis is motivated by recent 3D simulations of SN~1987A
\citep[e.g.,][]{Utrobin_2021}, which predict significant outward mixing of $^{56}$Ni.

Within our analytical framework, the $^{56}$Ni distribution is described by an exponential profile characterized by the parameter $x_c$, defined such that $95\%$ of the total nickel mass is enclosed within a comoving radius $x_c R_{\rm ej}$ \citepalias[see also Eqs. 27-29 of][]{PC2025}:
\begin{equation}
x_c=\sqrt[3]{\frac{\log{\left[0.05+0.95\times \exp{(-k_2)}\right]}}{k_2}}.
\end{equation}
 The fiducial value adopted in this work, $k_2=32.87\rightarrow x_c = 0.45$, is consistent with previous hydrodynamical models used for validation \citep{PZ11,PZ13} and with classical prescriptions for nickel mixing \citep{Young_2004}.

\begin{figure}[h]
\includegraphics[width=\columnwidth]{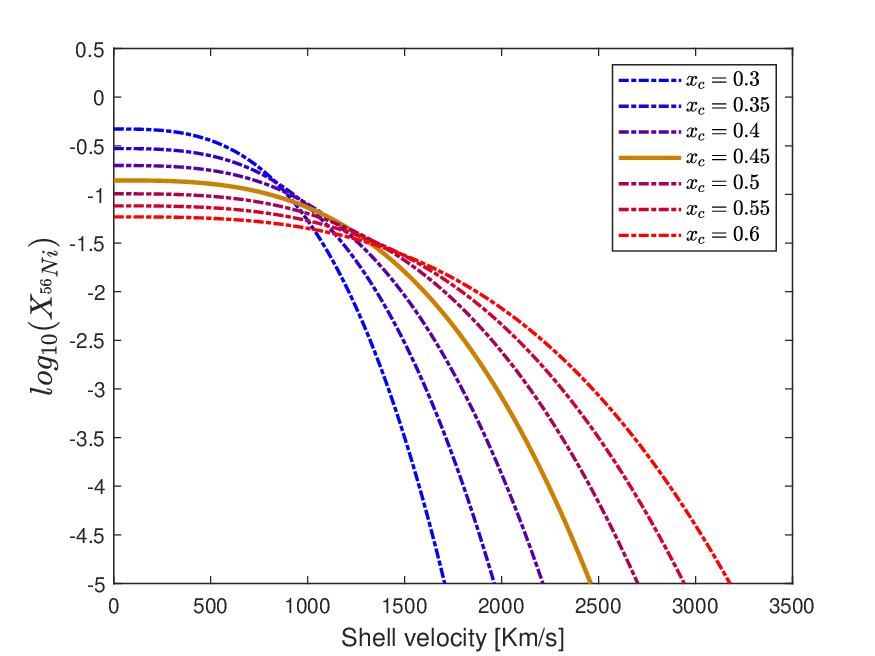} 
\caption{Abundance distribution of $^{56}$Ni inside the ejecta shells as a function of shell velocity. Different curves refers to EXP-$^{56}$Ni distribution with different confinement coefficient $x_c$ \citepalias[][]{PC2025}.
\label{Fig:Ni_mix}}
\end{figure}

Figure~\ref{Fig:Ni_mix} shows the mass fraction of $^{56}$Ni as a function of ejecta shell velocity for the SN 1987A best-fit model, computed for different values of $x_c$ and assuming comoving ejecta expansion. 
For the fiducial case ($x_c = 0.45$), less than $0.01\%$ of the nickel mass reaches velocities above $\sim 2500\,\mathrm{km\,s^{-1}}$, slightly below the maximum velocities inferred from observations \citep[e.g.,][]{1994ApJ...427..874C} and from 3D simulations \citep{Utrobin_2021}.
This difference can be partly attributed to the simplified density structure assumed in the analytical model, which adopts a uniform ejecta density, unlike hydrodynamical models where the outer layers are less dense and expand at higher velocities (cf. Section \ref{SubSec:test}). Increasing the mixing parameter to $x_c = 0.55$ results in a more extended nickel distribution with maximum velocities of around $3000\,\mathrm{km\,s^{-1}}$, comparable to those measured by \citet{1994ApJ...427..874C}.

\begin{figure}[h]
\includegraphics[width=\columnwidth]{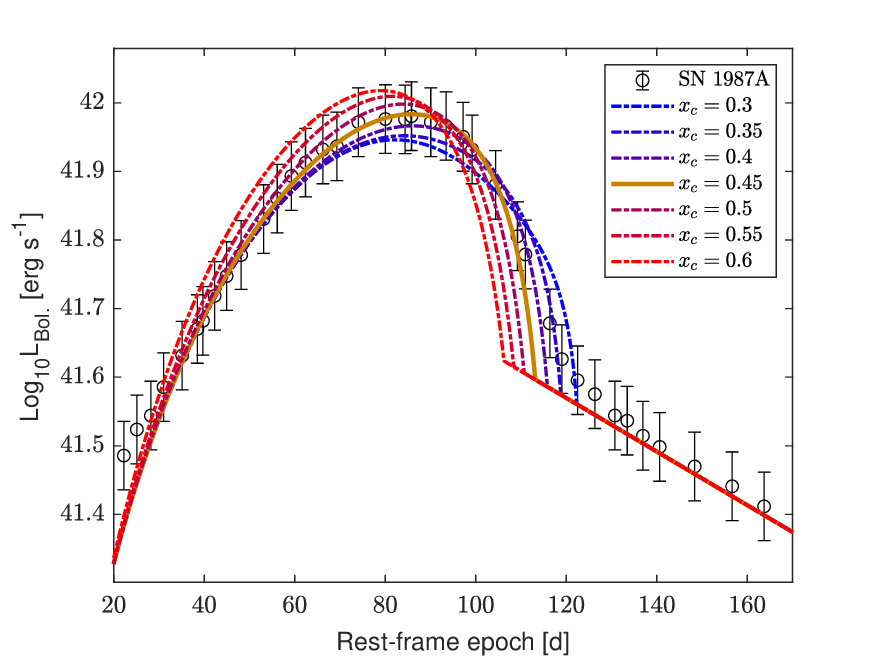} 
\caption{Effect of different $^{56}$Ni mixing on the bolometric LC for SN 1987A.
\label{Fig:Ni_mix_bol}}
\end{figure}

Figure~\ref{Fig:Ni_mix_bol} compares the bolometric LCs obtained for different values of $x_c$. A less confined nickel distribution leads to a modest anticipation of the end of the recombination phase, with $t_f$ shifting from $\sim 112$ days ($x_c = 0.45$) to $\sim 108$ days ($x_c = 0.55$), while remaining fully consistent within the observational uncertainties.
In order to evaluate the effect of $^{56}$Ni mixing on the physical parameters estimation, we further reapply the \textsc{SuperBAM} fitting procedure using $x_c = 0.55$ (corresponding to $k_2 = 18$). The resulting best-fit parameters are very similar to those obtained for the fiducial case, with all parameters remaining unchanged 
within uncertainties, except for a modest increase in the ejecta mass from $\sim 17\,{\rm M}_\odot$ to $\sim 19\,{\rm M}_\odot$. 
This confirms that variations in the $^{56}$Ni mixing parameter mainly induce second-order effects on the inferred physical properties, in agreement with what reported in \citetalias{PC2025} and previous studies \citep[e.g.,][]{PZ13}. We stress, however, that a more detailed photometric investigation of mixing effects critically relies on having accurate observations of the late recombination phase (immediately preceding the onset of the radioactive tail), where the impact of different mixing configurations becomes more pronounced.

\section{Observational features and modeling}\label{App:data}
This appendix presents the main observational and modeling information for the \Alike~SNe analysed in this work.\\ 
Tab. \ref{Tab:Prior_features} summarizes the key photometric observables derived from the bolometric LCs, followed by Tab. \ref{Tab:Spectra_feature} reporting the spectral measurements (photospheric velocities and pseudo-equivalent widths) obtained from the published spectra (see also Tab.~\ref{Tab:Info_SNe87A} and references therein). 
Tab. \ref{Tab:Modeling} contains the physical parameters inferred through our \textsc{SuperBAM} analysis.

\begin{table*}
    \caption{Bolometric LC features of the \Alike~SN sample.}
    \centering
    \scriptsize
    \begin{tabular}{l c c c c c}
        \hline
SN Name	& $t_{\rm m}$ & $\bar{t}_{\rm M}$ & $L_{\rm M}$ & $t_{\rm f}^*$ & $M_{\rm Ni}$ \\
\hline
\hline
\\
SN1987A 	&	 $8.8^{+7.2}_{-5}$ 	&	 $94^{+10.5}_{-13.8}$ 	&	 $85.95\pm9.05$ 	&	 $112.6^{+5.9}_{-9.3}$ 	&	 $0.07\pm0.01$\\	
SN1998A 	&	 $9.2^{+9.9}_{-9.2}$ 	&	 $93.2^{+25.6}_{-30.2}$ 	&	 $104.48\pm31.84$ 	&	 $118.9^{+0.1}_{-12.8}$ 	&	 $0.092\pm0.037$\\	
SN2000cb 	&	 $7.8^{+3}_{-7.8}$ 	&	 $88.5^{+16.4}_{-31}$ 	&	 $126.19\pm44.78$ 	&	 $104.9^{+0.1}_{-8.2}$ 	&	 $0.13\pm0.04$\\	
SN2004ek 	&	 $33.2^{+6.5}_{-13.2}$ 	&	 $99.8^{+33.5}_{-20.5}$ 	&	 $459\pm180.56$ 	&	 $158.9^{+24.9}_{-29.6}$ 	&	 $0.20\pm0.02$\\	
SN2004em 	&	 $39.8^{+0}_{-18}$ 	&	 $124.2^{+14.2}_{-27.8}$ 	&	 $246.85\pm75.43$ 	&	 $140.9^{+32.6}_{-8.3}$ 	&	 $0.10\pm0.06$\\	
SN2005ci 	&	 $19.3^{+1}_{-19.3}$ 	&	 $110.2^{+10}_{-19.8}$ 	&	 $70.62\pm9.06$ 	&	 $125.4^{+4.9}_{-7.6}$ 	&	 $0.065\pm0.040$\\	
SN2006V 	&	 $22.8^{+1.9}_{-22.8}$ 	&	 $83.2^{+11.5}_{-13.8}$ 	&	 $199.29\pm24.31$ 	&	 $102.9^{+7.4}_{-9.8}$ 	&	 $0.138\pm0.011$\\	
SN2006au 	&	 $20.5^{+15.8}_{-20.5}$ 	&	 $75.2^{+9.5}_{-5.2}$ 	&	 $153.01\pm7.87$ 	&	 $99.6^{+2.9}_{-12.2}$ 	&	 $0.068\pm0.004$\\	
SN2009E 	&	 $10^{+0.1}_{-10}$ 	&	 $106.5^{+17}_{-24.5}$ 	&	 $57.29\pm12.57$ 	&	 $126.9^{+10.4}_{-10.2}$ 	&	 $0.042\pm0.012$\\	
SN2009mw 	&	 $13.6^{+0.8}_{-13.6}$ 	&	 $95.2^{+14.2}_{-16}$ 	&	 $62.17\pm8.82$ 	&	 $111.6^{+7.9}_{-8.2}$ 	&	 $0.055\pm0.018$\\	
PTF12gcx 	&	 $5^{+0.4}_{-5}$ 	&	 $66.2^{+29.6}_{-33.5}$ 	&	 $289.83\pm217.16$ 	&	 $95.9^{+0.1}_{-14.8}$ 	&	 $0.221\pm0.141$\\	
PTF12kso 	&	 $23.3^{+29.1}_{-23.3}$ 	&	 $81^{+14}_{-32.2}$ 	&	 $291.16\pm65.57$ 	&	 $97.4^{+1.9}_{-8.2}$ 	&	 $0.229\pm0.113$\\	
OGLE-14 	&	 $20.8^{+19.6}_{-20.8}$ 	&	 $119^{+13}_{-16.5}$ 	&	 $1253.38\pm193.62$ 	&	 $164.1^{+39.6}_{-22.6}$ 	&	 $0.469\pm0.083$\\	
SNRefsdal 	&	 $8.8^{+0.8}_{-8.8}$ 	&	 $74^{+28}_{-23.5}$ 	&	 $95.77\pm9.92$ 	&	 $130.4^{+0.9}_{-28.2}$ 	&	 $0.053\pm0.016$\\	
DES16C3cje 	&	 $39.8^{+0}_{-39.8}$ 	&	 $213^{+66.6}_{-76.2}$ 	&	 $197.94\pm1825.21$ 	&	 $279.6^{+0.1}_{-33.3}$ 	&	 $0.131\pm0.108$\\	
SN2018cub 	&	 $13^{+26.8}_{-13}$ 	&	 $117.8^{+13.6}_{-47.2}$ 	&	 $305\pm144.43$ 	&	 $131.4^{+0.1}_{-6.8}$ 	&	 $0.335\pm0.325$\\	
SN2018ego 	&	 $5^{+8}_{-5}$ 	&	 $76.2^{+8.2}_{-10.5}$ 	&	 $105.12\pm9.59$ 	&	 $102.9^{+0.1}_{-13.3}$ 	&	 $0.087\pm0.009$\\	
SN2018imj 	&	 $7.9^{+2}_{-7.9}$ 	&	 $95.8^{+17.4}_{-20}$ 	&	 $124.16\pm16.6$ 	&	 $113.1^{+0.9}_{-8.7}$ 	&	 $0.115\pm0.097$\\	
SN2018hna 	&	 $12.8^{+5}_{-3.8}$ 	&	 $91.2^{+7.2}_{-7.2}$ 	&	 $85.49\pm5.83$ 	&	 $105.6^{+10.9}_{-7.2}$ 	&	 $0.062\pm0.003$\\	
SN2019bsw 	&	 $39.8^{+0}_{-39.8}$ 	&	 $111.8^{+4.6}_{-30.5}$ 	&	 $218.84\pm64.5$ 	&	 $116.4^{+0.1}_{-2.3}$ 	&	 $0.282\pm0.197$\\	
SN2020faa 	&	 $39.8^{+0}_{-39.8}$ 	&	 $147^{+27.5}_{-26.2}$ 	&	 $779.46\pm144.71$ 	&	 $193.9^{+8.4}_{-23.4}$ 	&	 $0.296\pm0.093$\\	
SN2020oem 	&	 $5^{+3.6}_{-5}$ 	&	 $93.5^{+23.9}_{-39.2}$ 	&	 $269.52\pm105.37$ 	&	 $117.4^{+0.1}_{-11.9}$ 	&	 $0.296\pm0.213$\\	
SN2020abah 	&	 $13.8^{+26}_{-13.8}$ 	&	 $103^{+30.8}_{-26.8}$ 	&	 $132.61\pm42.7$ 	&	 $141.9^{+17.6}_{-19.4}$ 	&	 $0.074\pm0.032$\\	
SN2021zj 	&	 $39.8^{+0}_{-13.2}$ 	&	 $112.2^{+14.8}_{-34.2}$ 	&	 $681.83\pm206.42$ 	&	 $128.6^{+0.1}_{-8.2}$ 	&	 $0.526\pm0.225$\\	
SN2021mju 	&	 $5^{+3.2}_{-5}$ 	&	 $98^{+8.9}_{-19.8}$ 	&	 $52.34\pm22.69$ 	&	 $106.9^{+0.1}_{-4.4}$ 	&	 $0.078\pm0.037$\\	
SN2021skm 	&	 $5^{+0.3}_{-5}$ 	&	 $84.8^{+14.1}_{-29.2}$ 	&	 $162.48\pm48.75$ 	&	 $98.9^{+0.1}_{-7.1}$ 	&	 $0.21\pm0.16$\\	
SN2021wun 	&	 $5^{+0.4}_{-5}$ 	&	 $91.2^{+46.4}_{-30.8}$ 	&	 $95.72\pm39.3$ 	&	 $137.6^{+0.1}_{-23.2}$ 	&	 $0.082\pm0.043$\\	
SN2021aatd 	&	 $16.3^{+21.7}_{-16.3}$ 	&	 $92.5^{+13}_{-13.8}$ 	&	 $160.35\pm21.01$ 	&	 $113.1^{+3.9}_{-10.3}$ 	&	 $0.116\pm0.043$\\
 \hline
    \end{tabular}
\tablefoot{Characteristic times ($t_{\rm m},\,\bar{t}_{\rm M},\,t_{\rm f}^*$) are expressed in days. $L_{\rm M}$ is in units of $10^{41}\,$\ergs and $M_{\rm Ni}$ in solar masses.}
    
    \label{Tab:Prior_features}
\end{table*}

\begin{table*}
\caption{Main spectroscopic features of the \Alike~SN sample.}
    \centering
    \scriptsize
    \begin{tabular}{l c c c c c c c c c c}
        \hline
        SN name & Spectra date & Rest-Frame & $V_{\rm H_\alpha}$ & pEW$_{\rm H_\alpha}$ & $V_{\rm H_\beta}$ & pEW$_{\rm H_\beta}$ & $V_{\rm Fe\,{\sc II}}$ & pEW$_{\rm Fe\,{\sc II}}$ & $V_{\rm Ba\,{\sc II}}$ & pEW$_{\rm Ba\,{\sc II}}$ \\
          & [MJD] & Epoch [d] & [km/s] & [\AA] & [km/s] & [\AA] & [km/s] & [\AA] & [km/s] & [\AA] \\
        \hline
        \hline
        \\
SN1987A	&	46934 & 84 & $5071\pm50$ 	&	 $57.5\pm1.9$ 	&	 $4019\pm38$ 	&	 $10.1\pm0.9$ 	&	 $2268\pm53$ 	&	 $23.9\pm1.0$ 	&	 $2372\pm27$ 	&	 $18.7\pm6.0$\\
SN1998A	&	 50894	& 92	&$7519\pm169$ 	&	 $72.0\pm0.8$ 	&	 $5830\pm172$ 	&	 $63.0\pm2.6$ 	&	 $3643\pm22$ 	&	 $26.6\pm0.5$ 	&	 $2979\pm75$ 	&	 $6.8\pm1.0$\\
SN2000cb	& 51722	&	71 & $8879\pm56$ 	&	 $106.1\pm2.7$ 	&	 $7479\pm48$ 	&	 $51.2\pm1.4$ 	&	 $4269\pm22$	&	 $20.5\pm1.0$ 	&	 $3675\pm7$ 	&	 $11.8\pm1.0$\\
SN2004ek	& 53328	& 76	& $5994\pm33$ 	&	 $3.1\pm0.1$ 	&	 $5838\pm45$ 	&	 $11.5\pm0.5$ 	&	 $4798\pm47$ 	&	 $9.2\pm0.5$ 	&	 $4959\pm35$ 	&	 $1.3\pm0.0$\\
SN2004em & 53328 & 64	&	 $7715\pm271$ 	&	 $84.7\pm8.3$ 	&	 $6448\pm68$ 	&	 $30.3\pm2.4$ 	&	 $4895\pm25$ 	&	 $20.2\pm1.4$ 	&	 $5015\pm225$ 	&	 $6.0\pm1.0$\\
SN2005ci& 53621	& 108 &	 $6378\pm65$ 	&	 $82.1\pm1.7$ 	&	 $5421\pm84$ 	&	 $47.7\pm1.3$ 	&	 $2639\pm59$ 	&	 $17.7\pm1.3$ 	&	 $1443\pm34$ 	&	 $12.3\pm2.0$\\
SN2006V& 53822 & 73 &	 $5886\pm238$ 	&	 $38.7\pm3.8$ 	&	 $3721\pm301$ 	&	 $20.7\pm4.5$ 	&	 $2961\pm9$ 	&	 $23.2\pm0.8$ 	&	 $3842\pm151$ 	&	 $7.8\pm1.0$\\
SN2006au& 53850 & 56 &	 $7434\pm180$ 	&	 $59.3\pm1.8$ 	&	 $5958\pm178$ 	&	 $39.9\pm1.4$ 	&	 $4519\pm16$ 	&	 $20.6\pm1.4$ 	&	 $3820\pm115$ 	&	 $6.3\pm1.0$\\
SN2009E & 54922 & 97 &	 $5572\pm187$ 	&	 $20.0\pm2.1$ 	&	 $1706\pm185$ 	&	 $2.8\pm0.4$ 	&	 $1570\pm26$ 	&	 $15.6\pm0.6$ 	&	 $2230\pm6$ 	&	 $21.0\pm1.0$\\
SN2009mw& 55267 & 91	&	 $7296\pm163$ 	&	 $87.8\pm2.0$ 	&	 $5792\pm164$ 	&	 $50.5\pm1.3$ 	&	 $3015\pm23$ 	&	 $29.2\pm0.7$ 	&	 $2496\pm6$ 	&	 $3.8\pm0.0$\\
PTF12gcx& 56134 & 50	&	 $7715\pm306$ 	&	 $84.7\pm8.3$ 	&	 $7795\pm156$ 	&	 $23.3\pm1.1$ 	&	 $4446\pm104$ 	&	 $16.9\pm2.2$ 	&	 $4776\pm85$ 	&	 $8.4\pm1.0$\\
PTF12kso& 56238 & 61	&	 $9369\pm119$ 	&	 $108.7\pm1.7$ 	&	 $6801\pm117$ 	&	 $47.4\pm1.2$ 	&	 $3822\pm59$ 	&	 $11.6\pm0.8$ 	&	 $4139\pm159$ 	&	 $2.8\pm2.0$\\
OGLE-14& 56974 & 101	&	 $8849\pm331$ 	&	 $47.2\pm2.8$ 	&	 $7119\pm326$ 	&	 $29.4\pm1.3$ 	&	 $4917\pm48$ 	&	 $8.7\pm0.9$ 	&	 -- 	&	 --\\
SN Refsdal& 57021 & 21	&	 $8400\pm500$\tablefootmark{a} 	&	 -- 	&	 -- 	&	 -- 	&	 -- 	&	 -- 	&	 -- 	&	 --\\
DES16C3cje& 57805 & 127	&	 $1359\pm31$ 	&	 $7.1\pm0.1$ 	&	 $1088\pm51$ 	&	 $7.6\pm0.2$ 	&	 $1370\pm19$ 	&	 $10.3\pm0.0$ 	&	 $875\pm19$ 	&	 $2.4\pm0.0$\\
SN2018cub&	58335 & 114	&	 $8399\pm36$ 	&	 $55.7\pm4.6$ 	&	 $6456\pm216$ 	&	 $39.5\pm1.6$ 	&	 $3587\pm249$ 	&	 $19.3\pm0.5$ 	&	 $3426\pm110$ 	&	 $6.0\pm3.0$\\
SN2018ego& 58373 & 120	&	 $7989\pm242$ 	&	 $58.3\pm8.5$ 	&	 $6179\pm244$ 	&	 $45.7\pm7.0$ 	&	 $4238\pm147$ 	&	 $17.1\pm0.6$ 	&	 -- 	&	 --\\
SN2018imj& 58487 & 106	&	 $6825\pm189$ 	&	 $86.0\pm2.9$ 	&	 $5510\pm59$ 	&	 $49.7\pm2.4$ 	&	 $2977\pm405$ 	&	 $31.3\pm11.4$ 	&	 $2081\pm104$ 	&	 $17.1\pm1.0$\\
SN2018hna & 58483 & 72	&	 $6948\pm142$ 	&	 $21.2\pm1.7$ 	&	 $4047\pm1006$ 	&	 $13.4\pm11.8$ 	&	 $3456\pm22$ 	&	 $16.3\pm1.0$ 	&	 $3354\pm171$ 	&	 $1.4\pm1.0$\\
SN2019bsw & 58562 & 77	&	 $7306\pm171$ 	&	 $79.1\pm1.7$ 	&	 $6060\pm122$ 	&	 $30.4\pm3.4$ 	&	 $3305\pm91$ 	&	 $16.7\pm6.0$ 	&	 $3332\pm80$ 	&	 $13.7\pm2.0$\\
SN2020faa& 59076 & 144	&	 $7546\pm509$ 	&	 $42.7\pm5.8$ 	&	 $5933\pm37$ 	&	 $18.8\pm1.3$ 	&	 $4591\pm74$ 	&	 $10.9\pm0.7$ 	&	 $3077\pm76$ 	&	 $1.0\pm0.0$\\
SN2020oem& 59090 & 103	&	 $5656\pm201$ 	&	 $40.1\pm3.7$ 	&	 -- 	&	 -- 	&	 -- 	&	 -- 	&	 -- 	&	 --\\
SN2020abah & 59195 & 19 &	 $9475\pm141$ 	&	 $116.1\pm9.8$ 	&	 $8088\pm14$ 	&	 $22.8\pm1.8$ 	&	 $4051\pm273$ 	&	 $18.2\pm3.3$ 	&	 $4399\pm420$ 	&	 $5.6\pm2.0$\\
SN2021mju& 59365 & 75	&	 $7048\pm83$ 	&	 $76.9\pm1.4$ 	&	 -- 	&	 -- 	&	 -- 	&	 -- 	&	 $5483\pm1336$ 	&	 --\\
SN2021skm & 59438 & 47 &	 $7911\pm199$ 	&	 $31.0\pm4.7$ 	&	 $5236\pm54$ 	&	 $7.0\pm0.3$ 	&	 $3132\pm82$ 	&	 $8.9\pm3.3$ 	&	 $2284\pm155$ 	&	 $9.9\pm2.0$\\
SN2021wun & 59468 & 42 	&	 $7642\pm438$ 	&	 $50.2\pm1.5$ 	&	 -- 	&	 -- 	&	 $2243\pm331$ 	&	 $19.0\pm6.6$ 	&	 $3619\pm25$ 	&	 $4.3\pm1.0$\\
SN2021aatd & 59592 & 98	 &	 $6099\pm94$ 	&	 $65.9\pm11.8$ 	&	 $5108\pm231$ 	&	 $48.8\pm6.9$ 	&	 $3691\pm56$ 	&	 $15.4\pm3.4$ 	&	 $2454\pm534$ 	&	 $7.0\pm5.5$\\
        \hline
    \end{tabular}
    
\tablefoot{In the second and third columns, the spectra acquisition time in MJD and the SN rest-frame epoch (since the explosion) are reported, respectively.\\
\tablefoottext{a}{The spectrum of SN Refsdal is not optical and its $V_{\rm H_\alpha}$ value is taken from \citet{kelly16}.}
    }
    \label{Tab:Spectra_feature}
\end{table*}

\begin{table*}
\caption{Physical parameters of the \Alike~SN sample derived by \textsc{SuperBAM} procedure.}
\centering
\scriptsize
\begin{tabular}{l|ccc|ccc|ccc}
\hline
SN Name & \multicolumn{3}{c}{Energy $[{\rm foe}]$} & \multicolumn{3}{c}{Ejected Mass $[{\rm M}_\odot]$} & \multicolumn{3}{c}{Initial Radius $[10^{12}\,{\rm cm}]$} \\
 & Pr. & Po. & MC & Pr. & Po. & MC & Pr. & Po. & MC \\
\hline
\hline

SN1987A 	&	$1.4\pm0.9$	&	$1.5\pm0.2$	&	$1.4\pm0.3$	&	$16\pm5  $	&	$18\pm1  $	&	$17\pm2  $	&	$4.4\pm3.6$	&	$3.\pm0.2$      &	$3\pm0.3$	\\
SN1998A 	&	$4.1\pm3.8$	&	$2.08\pm0.87$	&	$2.28\pm1.25$	&	$24\pm14.2$	&	$20.5\pm3.2$	&	$21.3\pm7.6$	&	$4.8\pm4.6$	&	$3.2\pm0.3$	&	$3.2\pm0.4$	\\
SN2000cb 	&	$6.12\pm6.01$	&	$2.85\pm0.88$	&	$2.4\pm1.16$	&	$24.7\pm14.3$	&	$16.8\pm1.9$	&	$14.7\pm4.8$	&	$3.3\pm2.6$	&	$2.3\pm0.2$	&	$2.3\pm0.2$	\\
SN2004ek 	&	$11.63\pm8.98$	&	$2.82\pm0.31$	&	$2.83\pm0.54$	&	$37.3\pm19.4$	&	$22.9\pm1$	&	$22.9\pm2.4$	&	$44.1\pm23.2$	&	$81.7\pm4.1$	&	$82.3\pm9.4$	\\
SN2004em 	&	$20.41\pm13.65$	&	$1.32\pm0.57$	&	$1.38\pm0.69$	&	$50.7\pm18.4$	&	$22.2\pm3.6$	&	$23.9\pm5.6$	&	$53.1\pm22.3$	&	$51.7\pm4.4$	&	$54.7\pm10.4$	\\
SN2005ci 	&	$4.66\pm2.42$	&	$0.41\pm0.19$	&	$0.44\pm0.23$	&	$27.8\pm8$	&	$15.7\pm2.8$	&	$16.3\pm5$	&	$15.1\pm10.3$	&	$14\pm1.3$	&	$13.9\pm1.5$	\\
SN2006au 	&	$9.58\pm6.32$	&	$1.2\pm0.08$	&	$1.2\pm0.12$	&	$24.4\pm7$	&	$10.2\pm0.3$	&	$10.2\pm0.7$	&	$24.5\pm21.2$	&	$13\pm0.7$	&	$13.2\pm1$	\\
SN2006V 	&	$2.19\pm1.28$	&	$0.83\pm0.08$	&	$0.86\pm0.15$	&	$16.6\pm5.3$	&	$7.3\pm0.3$	&	$7.5\pm0.9$	&	$21.9\pm15.3$	&	$16.5\pm1.5$	&	$16.7\pm2.2$	\\
SN2009E 	&	$0.81\pm0.5$	&	$1\pm0.32$	&	$1.03\pm0.41$	&	$15.2\pm5.8$	&	$22.5\pm2.6$	&	$22.4\pm4.9$	&	$3.9\pm2.6$	&	$3.9\pm0.6$	&	$4\pm0.6$	\\
SN2009mw 	&	$3.66\pm2.79$	&	$0.35\pm0.09$	&	$0.35\pm0.11$	&	$22.2\pm8.5$	&	$8.3\pm0.7$	&	$8.3\pm1.4$	&	$7.6\pm5.2$	&	$6.6\pm0.6$	&	$6.5\pm0.9$	\\
PTF12gcx 	&	$4.49\pm4.29$	&	$10.21\pm7.85$	&	$13.06\pm94.15$	&	$20.1\pm14.5$	&	$12.3\pm4.1$	&	$11.5\pm7.6$	&	$1\pm0.7$        &	$0.9\pm0.1$	&	$0.9\pm0.1$	\\
PTF12kso 	&	$3.12\pm3.35$	&	$1.47\pm0.57$	&	$1.58\pm0.68$	&	$18.2\pm11.6$	&	$6.7\pm0.9$	&	$6.9\pm1.4$	&	$29.9\pm29.4$	&	$20.5\pm1.9$	&	$20.5\pm2.1$	\\
OGLE14-73 	&	$19.1\pm8.85$	&	$19\pm3.16$	&	$18.91\pm3.68$	&	$48.9\pm12.8$	&	$58.3\pm3.3$	&	$58\pm5.5$	&	$24.9\pm22.6$	&	$43.9\pm2.2$	&	$44\pm4.8$	\\
SNRefsdal 	&	$9.35\pm8.67$	&	$4.9\pm1.24$	&	$5.1\pm1.57$	&	$26.8\pm17.2$	&	$34.8\pm3.2$	&	$35.5\pm5.5$	&	$3.3\pm2.3$	&	$3.3\pm0.3$	&	$3.3\pm0.4$	\\
DES16C3cje 	&	$4.12\pm3.9$	&	$0.67\pm0.41$	&	$0.74\pm0.47$	&	$56.7\pm35.7$	&	$42.2\pm10.6$	&	$45.2\pm12.5$	&	$60.5\pm40$	&	$56.9\pm6.7$	&	$54.3\pm7.4$	\\
SN2018ego 	&	$5.76\pm2.54$	&	$1.85\pm0.22$	&	$1.89\pm0.5$	&	$21\pm5.4$	&	$13.2\pm0.6$	&	$13.2\pm2.3$	&	$1.3\pm1.4$	&	$0.9\pm0.1$	&	$0.9\pm0.1$	\\
SN2018hna 	&	$4.14\pm2.78$	&	$0.97\pm0.05$	&	$0.99\pm0.12$	&	$21.9\pm6$	&	$13.6\pm0.2$	&	$13.7\pm1.1$	&	$7.6\pm4.6$	&	$3.3\pm0.2$	&	$3.3\pm0.3$	\\
SN2018imj 	&	$4.3\pm3.39$	&	$2.21\pm1.35$	&	$2.16\pm1.52$	&	$23.8\pm10.3$	&	$19.4\pm4.9$	&	$19\pm7.1$	&	$3\pm2.3  $	&	$2.4\pm0.2$	&	$2.3\pm0.2$	\\
SN2018lrq 	&	$30.34\pm29.05$	&	$2.81\pm1$	&	$2.67\pm1.22$	&	$52.1\pm27$	&	$38.6\pm4.9$	&	$36.8\pm10.7$	&	$1.4\pm1.3$	&	$0.9\pm0.1$	&	$0.9\pm0.1$	\\
SN2019bsw 	&	$7.1\pm3.74$	&	$1.01\pm0.53$	&	$0.98\pm0.57$	&	$31.4\pm10.1$	&	$24.7\pm5.1$	&	$23.9\pm7.5$	&	$60.5\pm40$	&	$61.3\pm1.2$	&	$61.3\pm1.3$	\\
SN2020faa 	&	$39.16\pm27.35$	&	$7.37\pm1.75$	&	$7.35\pm1.83$	&	$77\pm30$       &	$50.9\pm4.2$	&	$50.8\pm4.8$	&	$60.5\pm40$	&	$59.1\pm2$	&	$59.4\pm3.6$	\\
SN2020abah 	&	$8.95\pm8.24$	&	$4.11\pm1.41$	&	$4.38\pm1.88$	&	$35\pm20.2$	&	$36.1\pm4.5$	&	$38.4\pm8.8$	&	$9.2\pm10.5$	&	$7.6\pm0.9$	&	$8\pm1.5$	\\
SN2020oem 	&	$3.71\pm4.02$	&	$7.07\pm5.28$	&	$7.06\pm13.67$	&	$23.8\pm16.3$	&	$21.4\pm6.6$	&	$19.7\pm11.6$	&	$1.5\pm1.2$	&	$0.9\pm0.1$	&	$0.9\pm0.1$	\\
SN2021zj 	&	$24.36\pm17.72$\tablefootmark{a}	&	$2.78\pm0.9$	&	$2.87\pm1.02$	&	$53.7\pm25.8$\tablefootmark{a}	&	$12.5\pm1.4$	&	$12.7\pm2.2$	&	$54.1\pm17.6$	&	$62.8\pm2.2$	&	$62.4\pm2.9$	\\
SN2021mju 	&	$24.7\pm20.31$	&	$1.35\pm0.63$	&	$1.36\pm0.72$	&	$41.6\pm16.3$	&	$21\pm3.6$	&	$20.6\pm6$	&	$1.5\pm1.2$	&	$1.0\pm0.1$	&	$1.0\pm0.1$	\\
SN2021skm 	&	$3.2\pm2.75$	&	$3.9\pm2.8$	&	$3.76\pm3.28$	&	$18.7\pm9.6$	&	$13.7\pm4$	&	$13.5\pm6.3$	&	$1\pm0.7 $	&	$0.9\pm0.1$	&	$0.9\pm0.1$	\\
SN2021wun 	&	$1.3\pm1.45$	&	$7.17\pm3.2$	&	$7.41\pm4.47$	&	$19\pm14.3$	&	$39.2\pm6.6$	&	$39.9\pm16.6$	&	$1\pm0.7$	&	$0.9\pm0.1$	&	$0.9\pm0.2$	\\
SN2021aatd 	&	$5.03\pm4.36$	&	$1.47\pm0.46$	&	$1.37\pm0.53$	&	$23.8\pm10$	&	$14.8\pm1.7$	&	$14\pm3	$        &	$14.4\pm14.4$	&	$9.8\pm1.1$	&	$9.6\pm1.3$	\\

\hline
\end{tabular}

\tablefoot{The table is organized with the SN names in the first column, followed by nine columns clustered into triplets. Each triplet corresponds to a primary physical parameter—$E$, $M_{\rm ej}$, and $R_{\rm 0}$—and contains three sub-columns reporting the values obtained via the different validation methods discussed in Sect. \ref{SubSec:test}.\\
\tablefoottext{a}{Different from other SNe in the sample, the $E$ and $M_{\rm ej}$ prior estimates for SN 2021zj were obtained through the pure photometric scaling relations in Eq. \ref{Eq:Scaling_new} (see Sect. \ref{SubSubSec:Prior} for further details).}}
\label{Tab:Modeling}

\end{table*}

\end{appendix}

\end{document}